\documentclass[twocolumn,pre,nofootinbib,showpacs,preprintnumbers,amsmath,amssymb]{revtex4}
\usepackage{graphicx}
\usepackage{dcolumn}% Align table columns on decimal point
\usepackage{bm}
\usepackage{float}
\begin{document}
\title{Nonequilibrium thermodynamics of interacting tunneling transport:
variational grand potential, density-functional formulation, and nature of
steady-state forces}
\author{P. Hyldgaard}
\affiliation{Department of Microtechnology and Nanoscience, 
MC2, Chalmers University of Technology, 
SE--41296 Gothenburg, Sweden}
\date{April 5, 2012}

%######################## ABSTRACT #######################################
\begin{abstract}
The standard formulation of tunneling transport rests on an open-boundary 
modeling.  There, conserving approximations to nonequilibrium Green function 
or quantum statistical mechanics provide consistent but computational 
costly approaches; alternatively, use of \textit{density}-dependent 
\textit{ballistic-transport} calculations [e.g., Phys. Rev. B 
{\bf 52}, 5335 (1995)], here denoted `DBT', provide computationally efficient 
(approximate) atomistic characterizations of the electron 
behavior but has until now lacked a formal justification.  
This paper presents an exact, variational nonequilibrium thermodynamic theory 
for fully interacting tunneling and provides a rigorous foundation 
for frozen-nuclei DBT calculations as a lowest order approximation
to an exact nonequilibrium thermodynamics density functional evaluation.
The theory starts from the complete electron nonequilibrium quantum statistical mechanics and I identify 
the operator for the nonequilibrium Gibbs free energy which, generally, must be 
treated as an implicit solution of the fully interacting 
many-body dynamics.  I demonstrate a minimal 
property of a functional for the nonequilibrium thermodynamic grand potential which 
thus uniquely identifies the solution as the exact nonequilibrium density matrix.  I also 
show that the uniqueness-of-density proof from a closely related 
Lippmann-Schwinger collision density functional theory [Phys.~Rev.~B {\bf 78}, 
165109 (2008)] makes it possible to express the variational nonequilibrium thermodynamic 
description as a single-particle formulation based on universal 
electron-density functionals; the full nonequilibrium single-particle formulation improves 
the DBT method, for example, by a more refined account of Gibbs 
free energy effects. I illustrate a formal evaluation of the zero-temperature 
thermodynamics grand potential value which I find is closely related to the 
variation in the scattering phase shifts and hence to Friedel density oscillations. 
This paper also discusses the difference between the here-presented exact thermodynamics forces
and the often-used electrostatic forces. Finally the paper documents an inherent 
adiabatic nature of the thermodynamics forces and observes that these
are suited for a nonequilibrium implementation of the Born-Oppenheimer 
approximation.
\end{abstract}

%\textbf{Keywords:} 
\maketitle

%######################## INTRODUCTION #######################################
\section{Introduction} 

An understanding of self-organizing functional molecular systems is a
challenge for condensed-matter physics theory.  Molecular self-assembly 
mechanisms \cite{MolOverview,WeissReview,MolAdsorb,SlidingRings}\footnote{An
example of a van der Waals interaction study for larger-scale molecular-overlayer 
assembly is included in \protect\cite{Honeycomb}.}
provide a low-cost fabrication approach\footnote{For example, based on DNA sequencing that 
programs and controls three-dimensional organization of DNA-cotaed gold particles \protect\cite{DNAbuilders}.}
which mimics the 
molecular-recognition principle \cite{MolRec} of Nature. It can open
technological possibilities, for example, atomically precise production of 
electronics \cite{MolEl} and switching \cite{MolSwitch} components.  
The functionality is specified by the quantum-mechanical behavior 
of the electrons which typically respond to nonequilibrium
conditions. It is natural to assume a Born-Oppenheimer approximation 
of sorts and solve the nonequilibrium electron dynamics problem for model systems 
or for frozen nuclei coordinates within systematic approximations to 
the electron quantum-kinetic account (QKA)\footnote{The present paper benefits from connecting 
to several different formulations of the quantum physical behavior of electrons and must
relate to many concepts.  The use of abbreviations has been restricted to terms that are 
both long and used frequently or that enter as labels in equations and may assist an identification 
of the nature of the implied evaluation.  The abbreviations are in alphabetical order: 
\textit{DBT} --- density-dependent ballistic transport, calculated from the effective potential of 
ground-state DFT and assuming independent transmission; \textit{DFT} --- Density Functional Theory; 
\textit{GCE} --- grand canonical ensemble; \textit{LS} --- Lippmann-Schwinger; 
\textit{MSS} --- metallic surface state; 
\textit{ReNQSM} --- exact reformulation of nonequilibrium quantum statistical 
mechanics; \textit{QKA} --- quantum-kinetic account; \textit{SP} --- single-particle.}
\cite{BergmannLebowitz,Fano,KadBaym,NEQQSM,Keldysh,Langreth,MahanQKA,Araki,Grandy}.

Electron QKA methods range from time-evolution 
formulations\footnote{For formulations based on Lindblad functional
see, for example, \cite{LindbladTime}.}
\footnote{For formulations based on renormalization group theory,
see, for example, \cite{NEQRGT}.},
exact reformulations (here termed ReNQSM) of the nonequilibrium quantum statistical mechanics for 
tunneling \cite{Grandy,NEQreform} (including formal \cite{exactKondo,NEQunified,DoyonAndrei,Han,Dutt}, 
time-averaged \cite{NEQaverageEvolution}, and variational \cite{NEQvariational} approaches), 
over constrained entropy maximization descriptions \cite{NgCurConstraint,
HeinonenJohnsonCurConstraint,GodbyCurConstraint}, and to conserving nonequilibrium Green 
function approximations \cite{KadBaym,Keldysh,Langreth,MahanQKA,Caroli,ContinuumCaroli,FormalNEGF,Datta}.
In the context of nanoscale tunneling transport, there exist, for example, formal 
generalizations \cite{Baranger,KondoPRL,YigalNed} of the Landauer-B{\"u}ttiker 
formula \cite{LandauerButtiker}\footnote{For a corresponding wavefunction analysis of noninteracting tunneling,
please see \cite{Pendry}.} to problems with many-body interaction, but 
also warnings \cite{NEQreform,KondoPRL,RateEq,Phon,IntrinsicNEQlimitationA,
IntrinsicNEQlimitationB} of a breakdown for an oversimplified single-particle (SP) interpretation of 
this formal result.  The electron QKA furthermore includes Wigner-distribution calculations, 
for example, as described in an early and clear discussion \cite{Frensley} of the role 
of open boundaries and of a need for a thermodynamical treatment of nonequilibrium tunneling. The 
grand canonical ensemble (GCE) thermodynamics foundation is essential to correctly handle all 
charging \cite{RahmanDavies,Phon} (and hence Gibbs free energy) effects, 
spontaneous- and stimulated-emission effects \cite{QCL}, the
nonequilibrium entropy production \cite{NeqEntropy} and the
entropy-transport effects \cite{Linke} in a tunneling 
structure connecting leads at different chemical potentials.

Electron QKA studies have, until recently, delayed the discussion of the nonequilibrium forces 
which are exerted on the nuclei by transport-induced changes in the electron behavior. 
Nevertheless, the frozen-nuclei (i.e., electron) QKA methods are, by themselves, 
still empirical in the sense that they depend on an a priori characterization of 
the materials structure.  
This is a limitation since it is a central result from the family of electron density functional 
theory (DFT) formulations \cite{HohKohn,KohnSham,Mermin65,RungeGross,ThermTDDFT,InftyTDDFT,
CurDFT,RelativistDFT,LiouvilleDFT,CurTDDFT,SCDFT,Stefanucci,GebCar,Burke,LSCDFT} 
that the electron behavior is uniquely determined by the specific 
atomic configuration.  Also, molecular systems are characterized by 
sparseness \cite{SparseMatter}, i.e., have regions of low electron 
concentrations where dispersive forces acts and where the system components are 
relatively free to adapt their morphology to the specific (transport-dependent) 
environment.  To guide development of a molecular system to a pre-specified functionality, we must 
develop a nonempirical theory of nonequilibrium interaction effects \cite{DoyonAndrei,Han,Dutt,NEQvariational,Frensley,
NedKarstenJohn,NedLee,KondoPRL,NedLeeYigal,EarlyNoise,TunnelingNoise,
FloRun,RungeNoise,RateEq,Phon,acKondo,exactKondo,NedAnttiYigal,MartinDavidPeter,
Wenzel,WackerTransp,PlihalGadzuk,NEQpulsed,HFcurconstraintcalc,ThygesenTransport,
NeatonTun,FlensbergTun,ThygesenTun,LeeuwenTun},
and structural relaxations/excitations \cite{Electromigration,
Sorbello,GadzukTunnelingDesorption,KumarSorbello,ShamElectroMig,SorbelloDasgupta,
SurfaceFriction,DTUtunnelingForces,ChalmersTunneling,PlihalGadzukVib,
FrictionSwitchWithPhon,CurrentExcitation,TodorovSuttonThermodynForces,
Pantelides,TodorovTB1,TodorovTB2,Todorov1,Todorov2,Todorov3,
ElPhonTransport,Verdozzi,RecentBerndt,ForcesWithPhonFriction,vonOppen,
Narayanamurti} in nanoscale transport. 

Traditional, that is, ground-state DFT \cite{HohKohn,KohnSham},
\textit{illustrates} the level of detail and predictive power that we desire from 
a new nonempirical yet efficient nonequilibrium computational theory of interacting
tunneling.  Ground-state DFT rests on an equilibrium (canonical-ensemble) total-energy 
variational principle and regularly delivers accurate characterizations of 
both the electron and atomic structure for dense (hard) matter problems
when using a generalized gradient approximation \cite{GGA} for the 
electron density functional. With the introduction of truly nonlocal functionals 
like the van der Waals density functional method \cite{vdWDF,vdWDFsc} or
other formulations \cite{vdWother} it has become possible to extend the reach of 
nonempirical ground-state DFT also to the broader class of sparse 
matter \cite{SparseMatter,OthervdWreviews} (including molecular structures and 
devices) \textit{in equilibrium}. These developments come in addition to 
many versatile (but not entirely nonempirical) ground-state DFT extensions \cite{vdWsemiempirical}. The 
parameter-free DFT characterization of, for example, molecular-system 
structure, is possible because  ground-state DFT determines adiabatic 
forces which guide relaxations to an optimal atomic configuration.  \textit{However,} it 
is essential to point out that the desired nonempirical theory for nonequilibrium tunneling must involve 
a significant step beyond  ground-state DFT; the nonequilibrium operating conditions voids the  ground-state DFT variational 
principle and the current flow causes the equilibrium Born-Oppenheimer-approximation 
concepts\footnote{A general, concept-based (as opposed to operational) description 
of an assumed adiabatic nature is given in \cite{AdiabaticChap}.} 
(system staying in the evolving ground state while being thermically isolated, 
free of charging) to lose meaning.

The GCE modeling framework of tunneling is fully incorporated in the
\textit{density}-dependent \textit{ballistic-transport} 
method \cite{Lang,DiVentra,TranSiesta,PuskaRisto,gpawNEGF} 
(here termed DBT) inspired by the Landauer-B{\"u}ttiker formula \cite{LandauerButtiker}.
This method utilizes the density functionals of ground-state DFT to define an effective (density-dependent) 
potential for elastic scattering; it also solves for the density, formally through use of the 
SP Lippmann-Schwinger (LS) equation \cite{Lang,LippSchwing,AlbertXIX3}. I note that 
this DBT approach delivers computationally efficient, parameter-free (but approximate) 
system-specific characterizations of the electron behavior. However, the DBT
method has had no real status as a nonempirical theory because it 
has had no rigorous foundation.

The QKA delivers a full GCE thermodynamics description \cite{Fano,Langreth,NEQQSM,MahanQKA,Grandy,Frensley} 
which is inherently exact.  It is natural to seek a QKA recast and thus enable exact nonempirical characterizations 
of both electron behavior and of nuclei relaxations (specified by thermodynamics forces of the exact electron QKA).  
Access to efficient computations of the exact thermodynamics of 
interacting nonequilibrium tunneling would refine the description of the coupling between 
charge-transfer processes and molecular assembly and structure. It would also deepen the understanding 
of defect-induced local-density-of-state changes measured by scanning-tunneling microscopy 
\cite{MolOverview,WeissReview,STMclassicRef,FriedelInteractSurfTest}, 
of inelastic tunneling microscopy \cite{InelasticSTMclassicRef,ChalmersTunneling,PlihalGadzukVib},
of current-induced catalytic processes \cite{GadzukTunnelingDesorption}, of 
the current-induced nanopipette effect\footnote{The nanopipette is an explicit demonstration (by in-situ 
experiments) that electron currents cause matter transport in a nanoscale system,
\cite{NanoPipette}.},
of current-induced phonon generation \cite{Narayanamurti,FrictionSwitchWithPhon,CurrentExcitation,vonOppen}
light emission and lasing \cite{QCL}\footnote{For an example theory study discussing 
a significant bias-induced enhancement of 
electron-electron scattering in a resonant-tunneling
quantum-cascade-laser \protect\cite{QCL} system, please see
\cite{QCLdisc}.}, as well as of molecular switching and 
memristor effects \cite{MolEl,MolSwitch}.  Furthermore, from an understanding of the exact 
nonequilibrium thermodynamics we can discuss a nonequilibrium Born-Oppenheimer 
approximation for a \textit{strictly electron QKA} of 
morphology relaxations in working nonequilibrium tunneling systems (thus supplementing a recent 
analysis for tunneling with an electron-phonon coupling \cite{vonOppen}).
This is because we can identify the nonequilibrium adiabatic nature \cite{AdiabaticChap} of infinitely 
slow relaxations that both ensures an automatic compliance with the Friedel sum rule \cite{FriedelSum}, 
(and hence charge adjustments \cite{Phon} that leads to important Gibbs free energy effects) and
which avoids creation of an under-relaxed entropy content.

%################ THIS PAPER #############################
\textit{This paper} reformulates the electron QKA of nonequilibrium quantum statistical mechanics 
of interacting tunneling and thus (a) derives an \textit{exact} variational 
GCE formulation of the nonequilibrium thermodynamics. The paper furthermore (b)
expresses the thermodynamics as a regular DFT and, for steady-state problems, 
(c) identifies universal density functionals permitting a rigorous SP
formulation. The paper thus (d) establishes the widely used but 
\textit{previously ad-hoc} DBT method for frozen-nuclei electron density 
calculations as a lowest-order yet consistent approximation.  Moreover, 
the paper (e) shows that the variational property of the thermodynamics potential 
reflects a maximization of entropy subject to an automatic and rigorous implementation
of boundary conditions. 
The author is not aware of any previous derivations of the 
DBT method as a rigorous nonempirical theory. In any case, the here-presented derivation of a 
SP framework for efficient DFT calculations of nonequilibrium thermodynamics is useful 
for it opens for systematic improvements of the DBT method by identifying the nature of a 
set of transport-relevant, exact, universal density functionals.  These universal functionals govern 
the nonequilibrium internal-energy exchange and correlation \textit{and} the Gibbs free energy effects.

The nonequilibrium thermodynamics functional theory is a generalization of Mermin's equilibrium 
thermodynamic theory \cite{Mermin65} and takes off from the exact ReNQSM 
descriptions \cite{NEQreform,acKondo,NEQpulsed,NEQreformDetails,DoyonAndrei,Han,Dutt}.
In this paper I motivate and provide a definition of an 
exact nonequilibrium thermodynamic grand potential value and a corresponding functional,
\begin{eqnarray}
\Omega_{\rm col}(t) & \equiv  
& -\frac{1}{\beta}\ln \hbox{Tr}\{e^{-\beta [H(t)-\hat{Y}_{\rm col}(t)]}\},
\label{eq:NEQGrandPotVal}\\
\Omega_{\rm col}[\hat{\rho}(t)] & \equiv & 
%\frac{
{\rm Tr}\{ \hat{\rho}(t) [H(t)-\hat{Y}_{\rm col}(t) + \beta^{-1} \ln \hat{\rho}(t)]\},
%}
%{{\rm Tr}\{\hat{\rho}(t)\}},
\label{eq:NEQGrandPotFunct}
\end{eqnarray}
respectively. Here, $H(t)$ denotes a general time-dependent Hamiltonian and 
$\beta$ is the inverse temperature. The definition and formal results are made possible 
by the identification of an operator $\hat{Y}_{\rm col}(t)$ for the nonequilibrium Gibbs free energy. 
The Gibbs free energy operator is in general emerging, that is, implicitly defined and given through
an operator description of the interacting many-body time evolution.
The nonequilibrium thermodynamic grand potential (\ref{eq:NEQGrandPotFunct}) is a functional of the class of 
possible nonequilibrium (time-dependent) density matrices $\hat{\rho}(t)$.  The subscripts `col' (used for 
general time-dependent tunneling cases) on thermodynamics operators and values, e.g. in 
Eq.~(\ref{eq:NEQGrandPotFunct}), emphasize that the nonequilibrium interaction problem is solved as 
a formal many-body collision problem. That is, it is expressed in the framework of general collision 
theory \cite{LippSchwing,Dyson,Pirenne,Gellmann,DeWitt,LangrethLS,LangrethFriedel,AlbertXIX},
an approach that was also taken in a recent (related) formulation of a LS-based collision-theory 
DFT, below identified as `LS collision DFT' \cite{LSCDFT}. The here-presented exact nonequilibrium 
thermodynamics theory 
rests on a continuum Caroli partition scheme \cite{Caroli,ContinuumCaroli} and it is therefore possible 
to discuss the formal nature of forces on nuclei.\footnote{There does not seem to be many prior
nonequilibrium thermodynamics investigations of the nature of forces in the strictly adiabatic limit, when
the mass of the electrons is taken to be infinitely smaller than that of the ions. The reference
\protect\cite{TodorovSuttonThermodynForces} lists two interesting thermodynamics 
investigations of the nature of forces, for example, in linear response.  Both are limited to 
a study of noninteracting particles and build from a ground-state DFT characterization and analysis of 
left- (`1') and right-moving (`2') states by thermodynamics arguments.  The starting point 
assumes that two chemical potentials $\mu_{i=1,2}$ ensure a given electron occupation $N_1$ and $N_2$ 
of these different classes of states. At fixed $N_1$ and $N_2$, one has a canonical ensemble 
description and forces are then electrostatic in nature and given by the traditional 
Hellmann-Feynman theorem. For an open system, the papers observe that 
$\delta U = \sum_i\mu_i \delta N_i$ provides the correct GCE link between relevant 
extensive quantities (here given in the absence of entropic effects). The papers 
proceed by adding an electrostatic-force work term $\delta W^{\rm es} \equiv 
\mathbf{F}_{\mathbf{R}_i}^{\rm es,NEQ} \dot \delta \mathbf{R}_i$ when characterizing the 
added effects of an infinitesimal coordinate change. However, for any given value of 
the occupation changes $\delta N_{1,2}$, a coordinate change must (by the Friedel sum 
rule) cause concerted changes in both the internal energy and in the chemical 
potentials. Also, in a GCE thermodynamics system one
cannot assume a thermically isolated behavior and the correct determination of the work 
$\delta W$ must therefore equate both changes in the internal energy and the heat $\delta Q$, that
is, $\delta W=\delta U+\delta Q$.  The papers listed in \protect\cite{TodorovSuttonThermodynForces},
cannot be viewed as a conclusive thermodynamics analysis of forces in nonequilibrium tunneling because they
do not state an argument that a GCE evaluation of electrostatic forces (\ref{eq:esForceFirstDef}) 
will capture \textit{all} heat effects, including all nonequilibrium charging and all 
electron redistribution mechanisms \protect\cite{NEQreform}.}

In steady-state the nonequilibrium thermodynamics solutions are given by the many-body LS solution \cite{LippSchwing,AlbertXIX} 
$|\Psi_\xi^{(+)}\rangle$ which are projections \cite{Gellmann} of the full time-evolution of initial states 
$|\Phi_\xi\rangle$. However, while $|\Phi_\xi\rangle$ is an eigenstate of energy $E_\xi$
of $H(t\to -\infty)$, the many-body LS solution (generally) describe evolution at a different 
energy,  $E_\xi+\Delta_\xi$, with shifts $\Delta_\xi$ 
specified from the T-matrix behavior or by the self energy \cite{Pirenne,Gellmann,DeWitt}.
Refs.~\onlinecite{LangrethLS} and \onlinecite{LangrethFriedel} give example usage of the 
formal many-body LS solution in condensed-matter 
physics problems.  We can choose the initial states to also be eigenstates $Y_\xi$ of the initial Gibbs 
free energy $\hat{Y}_d=\hat{Y}_{\rm col}(t\to -\infty)$. In steady state the many-body LS solutions
thus define a formal evaluation of the nonequilibrium density matrix 
\begin{equation}
\hat{\rho}_{\rm LS} \equiv 
\sum_{\xi}|\Psi_\xi^{(+)}\rangle \exp[-\beta (E_\xi+\Delta_\xi-Y_\xi)] \langle \Psi_\xi^{(+)} |.
\label{eq:rhod-expandLS}
\end{equation}
In such steady state problems, the operator for the nonequilibrium Gibbs free energy $\hat{Y}_{\rm col}(t)$ becomes 
identical to the electron redistribution operator `$Y$' of the steady-state ReNQSM \cite{NEQreform}. 
Below, I identify steady-state thermodynamics operators and values by 
subscripts `LS' to emphasize the direct link to the many-body LS solution.

The demonstration of an \textit{extremal nature} of $\Omega_{\rm LS}[\hat{\rho}]>\Omega_{\rm LS}[\hat{\rho}_{\rm LS}]$ 
leads to additional formal and rigorous results.  The foundation in the electron QKA ensures 
that the here-presented thermodynamics theory contains all boundary, entropy, and entropy-flow 
effects in the electron system.  The paper therefore (f) provides an \textit{exact demonstration}
that the thermodynamics forces 
\begin{equation}
\mathbf{F}^{\rm GCE}_{{\rm LS},\mathbf{R}_i} \equiv - \frac{\partial}{\partial \mathbf{R}_i}
\bar{\Omega}_{\rm LS}
\label{eq:NTForceAdiabInit}
\end{equation}
are explicitly conservative and specified by a here-demonstrated generalized 
Hellmann-Feynman theorem,
\begin{equation}
\frac{\partial}{\partial \mathbf{R}_i} \bar{\Omega}_{\rm LS}
= \langle \frac{\partial}{\partial \mathbf{R}_i} \left[ H - \hat{Y}_{\rm LS} \right]
\rangle_{\hat{\rho}_{\rm LS}}.
\label{eq:AdiabForceRelation}
\end{equation}
These thermodynamics forces are formally different from the (nonequilibrium, identified by `NEQ') 
electrostatic-force definition,
\begin{equation}
\mathbf{F}^{\rm es, NEQ}_{\mathbf{R}_i} \equiv
- \langle \frac{\partial H} {\partial \mathbf{R}_i} \rangle_{\rm NEQ \, GCE},
\label{eq:esForceFirstDef}\\
\end{equation}
which have gained usage in GCE modeling, for example, as exploited for 
relaxations in DBT implementations\footnote{Both Ref.~\protect\onlinecite{CurrentForcesWireCalc} and 
Ref.~\protect\onlinecite{TranSiestaForce} report calculations as well as details of efficient 
code implementations of an approximative nonempirical description of tunneling (including
current-induced relaxations) based on use of electrostatic
forces. This electrostatic force evaluation is given by the density variation as can
be calculated in the DBT method \protect\cite{Lang,DiVentra,TranSiesta,PuskaRisto,gpawNEGF}.}
\cite{CurrentForcesWireCalc,TranSiestaForce}.

There exists arguments for use of electrostatic forces in linear response, for
example, as summarized in Refs.~\onlinecite{Sorbello} and \onlinecite{TodorovSuttonThermodynForces}. 
However, there is no rigorous argument\footnote{The use of electrostatic forces (\ref{eq:esForceFirstDef}) 
is sometimes motivated by a referal to Ehrenfest theorem \protect\cite{EhrenfestMerzbacher}.
Section IX explains limitations, also pointed out in Ref.~\protect\onlinecite{Pantelides}, 
that prevent that approach from always giving a conclusive argument in an open system.}
for the general use of electrostatic forces (\ref{eq:esForceFirstDef}) in infinite, 
open nonequilibrium tunneling systems.  I demonstrate that the electrostatic forces (\ref{eq:esForceFirstDef}) 
do agree with the exact thermodynamics forces (\ref{eq:NTForceAdiabInit}) when the electron 
distribution can be entirely described from the full density of states.  At the same time, I 
observe that the latter is not generally a correct assumption under nonequilibrium conditions. In any 
case, the paper (g) represents a rigorous QKA demonstration that electrostatic forces 
(\ref{eq:esForceFirstDef}) with a nonconservative character \cite{Todorov2,Todorov3} 
can never agree with the steady-state thermodynamics forces (\ref{eq:NTForceAdiabInit}). 

I stress that the here-derived steady-state thermodynamics forces (\ref{eq:NTForceAdiabInit}) are useful
as they define an adiabatic nature of relaxations. A sufficiently slow implementation 
of the exact thermodynamics forces (\ref{eq:NTForceAdiabInit}) ensures that the nonequilibrium system will 
emerge with correct and extremal steady-state thermodynamics potential value. The paper thereby (h) 
demonstrates that the resulting system must be described by the uniquely defined nonequilibrium density 
matrix $\hat{\rho}=\hat{\rho}_{\rm LS}$ that represents the steady-state solution for the 
system (Hamiltonian) after the coordinate translation.
A slow deformation around any closed loop can never produce any change in any thermodynamics quantity. 
The paper (i) shows that the thermodynamics forces (\ref{eq:NTForceAdiabInit})
ensure a correct Gibbs free energy content and 
always avoid an under-relaxed entropy content. The conservative thermodynamics forces 
express a robust physical principle, namely entropy optimization tempered by an automatic implementation 
of rigorous nonequilibrium GCE boundary conditions.  In essence, the paper (j) suggests that the exact thermodynamics 
forces (\ref{eq:NTForceAdiabInit}) are suited for use in a nonequilibrium Born-Oppenheimer approximation for steady-state tunneling 
described as an open-boundary GCE system.

While the set of formal results, (a) through (j), represents the main contribution,
the paper also summarizes a SP framework for computing the zero-temperature variation 
of the exact thermodynamics grand potential $\Delta \Omega_{\rm LS}$. 
This is relevant for completing a nonempirical characterization of nonequilibrium tunneling systems 
and, for example, describe the transport which emerges after (adiabatic) relaxation of the morphology.
Since the paper constitutes a rigorous DFT formulation of nonequilibrium thermodynamics, it is sufficient to 
consider an evaluation of the thermodynamics variation for noninteracting particles. 
Robust computational schemes \cite{Lang,DiVentra,TranSiesta,PuskaRisto,gpawNEGF} already 
exist and provide SP LS solutions, i.e., the noninteracting scattering states.  Previous 
determinations of nonequilibrium thermodynamical forces have often focused on the real-space variation of 
such scattering states, making it important to ensure consistency in the limiting 
processes \cite{Gellmann,DeWitt,TodorovSuttonThermodynForces}. Noting that the SP LS solutions are 
local in wavevector space, I summarize an alternative formal approach
defined by the S-matrix formulation \cite{SmatrixStatMech} of the generalized Friedel sum 
rule \cite{LangerAmbegaokar,LangrethFriedel,FriedelSum} and emphasizing computations of 
the phase-shift variation \cite{Albert1D,AlbertPhaseVar,DoniachCite,HarrisFriedelInteract}.
The formal framework attempts to utilize the availability of the DBT solvers and their description 
of scattering phase shifts. It may well be possible to also develop alternative, perhaps more 
robust and efficient, computation strategies.

%################ PAPER ORGANIZATION #############################
The paper is organized as follows. The next section summarizes computations
of the noninteracting thermodynamics grand potential variation, emphasizing 
the key role of Gibbs free energy in GCE thermodynamics problems.
Section III defines the partition scheme while Section IV presents the electron QKA and the
exact reformulation as a variational nonequilibrium thermodynamics theory.  Section V presents the 
set of general nonequilibrium GCE thermodynamics results for steady-state tunneling while Sec.~VI 
defines and discusses state functions and forces for steady-state tunneling.  The linking with 
LS collision DFT \cite{LSCDFT} and the SP formulation as a thermodynamics DFT is presented 
in Sec.~VII while Sec.~VIII outlines a SP framework for calculating the zero-temperature 
thermodynamics variation.  Section IX contains a discussion, while section X contains 
summary and conclusions. In addition, the paper has 5 appendices, introduced in and
supporting the text.

%Appendix A and B summarize the link to a 
%a formal NEGF description and to an S-matrix, i.e., phase-shift, formulation of 
%the scattering problem.  Appendix C contains formal proof for a Mermin-type variational 
%principle including the GHF theorem. Appendix D provide a mathematical argument 
%for compatibility of LSC-DFT and QKA, a link which underpins the exact GCE thermodynamics 
%DFT. Finally, appendix E summarizes rigorous results relating derivatives of 
%the set of (interacting) thermodynamics state-function potentials. 

\section{Open-boundary thermodynamics of noninteracting particles}

The paper begins by summarizing a general framework for calculating the noninteracting
GCE thermodynamics variation \textit{and} with illustrations the key role held 
by the Friedel sum rule in specifying, for example, the Gibbs free energy variation.  
The section presents a nonequilibrium thermodynamics formulation given in terms of the
nonequilibrium Green function techniques \cite{NEQQSMthermodynamics}. I detail the relation to formal 
scattering theory through a recast that emphasizes the independent-particle nature 
of the associated S matrix \cite{DoniachCite,LangerAmbegaokar,ZarambaKohnLS,ScatteringPhysisorption,HarrisFriedelInteract}.
An analysis of the formal S-matrix behavior is also exploited in Ref.~\onlinecite{vonOppen}.
The summary and presentation is given to illustrate feasibility of such calculations and
because textbooks on thermodynamics calculations tend to rapidly proceed to the more difficult
problem of investigating the effects of an actual many-body interaction. The latter is something 
that this paper instead proposes to (eventually) handle through a rigorous DFT SP 
formulation (given also developments of the nonequilibrium density-functionals).
The Gibbs free energy is essential in GCE thermodynamics studies due to the possibility 
and necessity of open infinite systems to accommodate charge adjustments \cite{FriedelSum,LangerAmbegaokar}.

The presentation is supported by appendix A, which summarizes a Green function thermodynamics 
description \cite{NEQQSMthermodynamics}, and appendix B,  which details the S-matrix,
Friedel sum rule \cite{LangerAmbegaokar} and scattering phase-shift analysis \cite{AlbertPhaseVar,Albert1D}.
The formal framework is presented with the intention to explore (in later works) 
the existence of efficient SP LS solvers, an integral part of the DBT 
codes \cite{Lang,DiVentra,TranSiesta,PuskaRisto,gpawNEGF},
for thermodynamics calculations.

\subsection{Friedel sum rule, phase shifts, and thermodynamics} 

With a GCE description of the electron system, there are typically
charge transfers and associated thermodynamics effects that significantly influence
the interactions of classical particles (e.g., defects or nuclei immersed in the 
electron gas).  The charge adjustments is in the GCE specified by the Friedel sum 
rule \cite{FriedelSum,LangerAmbegaokar,LangrethFriedel}, expressed as an integrated change in 
the density of state, and thus a direct cause for changes in the Gibbs free energy. This is true 
in equilibrium and out of equilibrium.

A natural starting point for a discussion of noninteracting tunneling
thermodynamics is the general formulation of the Friedel sum rule \cite{LangerAmbegaokar},
\begin{equation}
\Delta \mathcal{N}(\mu')= \frac{1}{2i\pi}\hbox{Tr}\{\ln[\mathcal{S}(\mu')]\}.
\label{eq:FormalFriedel} 
\end{equation}
The sum rule is here expressed as a relation between the scattering (or S) matrix 
$\mathcal{S}(\mu)$ for single-particle excitations at a given chemical potential $\mu'$ 
and the total number of displaced electrons, $\mathcal{N}(\mu')$. The trace 
includes a factor of 2 for spin; I shall, for simplicity, assume spin degeneracy and
restrict the summation over SP states to per-spin representation. 
I shall also, in this section, focus exclusively on a zero-temperature formulation.  
In a central potential, the eigenstates of 
the S-matrix are just $\exp[2i\delta_{l,m}(\omega)]$ and the traditional Friedel sum rule \cite{FriedelSum} 
follows simply from an evaluation of the trace of the logarithm in the powerful generalization (\ref{eq:FormalFriedel}).
The sole condition for the derivation \cite{LangerAmbegaokar} of (\ref{eq:FormalFriedel}) is that the 
imaginary part of the many-body self energy vanishes, $\Im \Sigma_{\mu'}=0$.  This is true at the actual Fermi level 
$\mu'=\mu_F$ even for an interacting equilibrium many-body system.

Refs.~\onlinecite{ZarambaKohnLS,ScatteringPhysisorption} provide early examples of a broader 
usage of (\ref{eq:FormalFriedel}) for DFT-based interaction studies, namely in elegant SP-LS 
studies of the adsorbate-induced density of state changes $\Delta \mathcal{D}(\omega)$.
Formally one seeks the adsorption-induced changes in the density of states 
\begin{eqnarray}
\Delta \mathcal{D}(\omega) & \equiv & 2 \sum_{\lambda}
\{\delta(\omega-\tilde{\epsilon}_{\lambda})
-\delta(\omega-\epsilon_{\lambda})\}
\label{eq:defDOSchange}\\
& = & \frac{1}{2i\pi}\frac{\partial}{\partial \omega} \hbox{Tr}\{\ln[\mathcal{S}(\omega)]\}.
\label{eq:FormalFriedelDOSresult} 
\end{eqnarray}
Here $\epsilon$ and $\tilde{\epsilon}$ denote the energy levels of the original ($t\to -\infty$) and of 
the emerging (relevant $t\approx 0$) system, respectively.  
The energy differences $\tilde{\epsilon}_\lambda-\epsilon_{\lambda}$
are, in principle, specified by the same formal collision theory arguments \cite{Gellmann,DeWitt} which 
underpin the expansion (\ref{eq:rhod-expandLS}) but here pertains to the SP LS 
solutions \cite{AlbertXIX3}. The level shifts are only infinitesimal when considered \textit{per 
level} but produce, as in all Freidel-phase-shift analysis \cite{FriedelSum}, an integral effect 
which constitutes the full (noninteracting particle) thermodynamics-grand-potential 
variation\footnote{References~\protect\cite{ZarambaKohnLS,ScatteringPhysisorption,HarrisFriedelInteract} 
are examples of equilibrium interaction studies that illustrate the formal equivalence in equilibrium of \textit{either} moving 
to a canonical-ensemble evaluation \textit{or} retaining the equilibrium Gibbs free energy 
term in a Harris scheme \protect\cite{Harris} adapted for the GCE thermodynamical DFT \cite{Mermin65}.
On the one hand, the stated form of the one-electron contribution, Eq.~(4) in 
Refs.~\protect\onlinecite{HarrisFriedelInteract}, 
can be viewed simply as an adjustment of the Fermi level; this canonical-ensemble approach is
discussed and used in Refs.~\protect\onlinecite{FriedelGrimley,FriedelTed,FriedelLK}. 
On the other hand, the expression for interaction energy (as
expressed from one-electron contributions) \textit{also} serves to explicitly include 
the Gibbs free energy term in an open-boundary thermodynamics evaluation. In that form it is consistent 
with a zero-temperature evaluation of Mermin's equilibrium thermodynamics DFT \protect\cite{Mermin65}.
Either way, a careful handling of the implications of Freidel sum rule plays
a vital role in correctly describing charge conservation and the resulting
interaction \protect\cite{HarrisFriedelInteract,FriedelInteractSurfTest,WeissReview}.}
\cite{ZarambaKohnLS,HarrisFriedelInteract},
\begin{eqnarray}
	\Delta \Omega_{\rm LS}
&=& 
\int_{-\infty}^{\mu_F} d\omega \, 
(\omega-\mu_F) \Delta \mathcal{D}(\omega)
\label{eq:GCEandCEresults}\\
& = & -
\int_{-\infty}^{\mu_F} d\mu' \, 
\Delta \mathcal{N}(\mu').
\label{eq:FormalFriedelOmegaresult} 
\end{eqnarray}
The key observation is that the formal result (\ref{eq:FormalFriedel}) applies for all values of $\mu'$ for 
noninteracting particles and that the S-matrix has an explicit on-shell character \cite{SmatrixStatMech,ZarambaKohnLS,AlbertXIX}.
The formal-scattering theory formulation (\ref{eq:FormalFriedelDOSresult}) can be expanded in terms of the 
(on-shell) T-matrix.  The procedure suggests a rapid convergence, for example, in 
a systematic calculation of the effective interaction between an adsorbate and a metal surface \cite{ZarambaKohnLS}.

A text-book analysis \cite{DoniachCite} provides a simple example of the role of the phase-shifts variation in
thermodynamics calculations.  The study of s-wave scatterers is relevant for a discussion of 
adsorbate-induced local-density-of-state changes in the metallic surface state (MSS) of 
Cu(111) \cite{FriedelInteractSurfTest,WeissReview}.
There one can have a simple (yet nonperturbative) T-matrix behavior \cite{HarrisFriedelInteract,HarrisFriedelTrioInteract}
$T^{\delta_{\rm F}}_0(\omega) \propto \exp[i2\delta_0(\omega)-1]$ (appendix A) which is completely 
characterized by the experimentally observed Fermi-level phase-shift value \cite{HarrisFriedelInteract}, $\delta_{\rm F}=\delta_0(\omega=\mu_{\rm F})
\approx \pm \pi/2$. 
By assumption, the s-wave phase shift is the only eigenvalue of $\mathcal{S}(\omega)$, and inclusion of the
s-wave scattering adsorbate causes a nonperturbative change in both the density of state and in the
the equilibrium thermodynamics potential \cite{DoniachCite}
\begin{eqnarray}
	\Delta \mathcal{D}_{\rm MSS,1s}^{\delta_{\rm F}}(\omega) & = &  \frac{2}{\pi}\;\frac{\partial}{\partial \omega}\; \delta_0(\omega),
\label{eq:EQ1sDeltaDos}\\
\Delta \Omega_{\rm MSS,1s}^{\delta_{\rm F}} & = &  - \frac{2}{\pi}\int_{-\infty}^{\mu_{\rm F}} d\omega \, \delta_0(\omega).
\label{eq:EQ1sThermodynamicsPotential}
\end{eqnarray}
The text book result (\ref{eq:EQ1sThermodynamicsPotential}) is consistent with the formal thermodynamics result 
given by Eqs.~(\ref{eq:FormalFriedel}) and (\ref{eq:FormalFriedelOmegaresult}). 

\subsection{A finite-bias computational framework} 

For calculations of the thermodynamics behavior of nonequilibrium \textit{noninteracting} tunneling it 
is natural to pursue a nonequilibrium Green function/scattering-state formulation which is effectively
suggested in Ref.~\onlinecite{NEQQSMthermodynamics}. 
Under nonequilibrium conditions,  we have left and right leads which are described 
by different chemical potentials $\mu_{\rm L}>\mu_{\rm R}$.  The leads $P={\rm L/R}$ are formally
of infinite volume $\mathcal{V}_{\rm L}$ and $\mathcal{V}_{\rm R}$.  Using the extensive nature of SP LS scattering 
solutions that arise from either leads one can define and evaluate the nonequilibrium noninteracting thermodynamics grand potential 
from a sum of partial terms
\begin{eqnarray}
\Delta \Omega_{\rm LS}(\mu_{\rm L},\mu_{\rm R})  & = & 
\Delta \Omega_{\rm LS}^{{\rm L}}(\mu_{\rm L})
+\Delta \Omega_{\rm LS}^{{\rm R}}(\mu_{\rm R}),
\label{OmegaLSsplit}\\
\Delta \Omega_{\rm LS}^{P={\rm L/R}}(\mu_P) & = &  - \int^{\mu_P}_{-\infty} d\mu'\; \Delta \mathcal{N}_{P}(\mu').
\label{eq:OmegaLS:leadEvaluation}
\end{eqnarray}
Here the value of $-\Delta \mathcal{N}_P/\mathcal{V}_P$ (for $P={\rm L/R}$) takes the role of a
change in partial pressure given by the change in electron density (with, for example, the onset of tunneling).
In a simple adaption of Ref.~\onlinecite{NEQQSMthermodynamics} it also follows that this partial
pressure is given by well-defined components of the less-than Green function.

As further detailed in the following section, I consider a system which is initially disconnected
`$d$' but in which tunneling arises upon the adiabatic turn on. The connected system
is generally identified by a superscript $(0)$ as it is noninteracting.  Relevant initial-system 
eigenstates $|\lambda_P\rangle$ of energy $\epsilon_{\lambda_P}$  are those which belong to a 
lead, $P={\rm L/R}$.  In the noninteracting tunneling systems, these initial eigenstates give
rise to distinct \cite{Gellmann,DeWitt} SP LS solutions 
$|\tilde{\lambda}_P\rangle=|\psi^{(0)}_{{\rm LS},\lambda_P}\rangle $ of 
energy $\tilde{\epsilon}_{\lambda_P}$. The resulting eigenstate set is (for noninteracting particles)
orthonormal \cite{DeWitt} and complete; I do not here explicitly consider effects of possible bound states \cite{Gellmann}.
The states $|\tilde{\lambda}_P\rangle$ are eigenstates of the connected system, and the initial and resulting
electron distribution can conveniently be described lead `$P$' and state-specific contributions:
\begin{eqnarray}
	g_<^{(0)} (\tilde{\lambda}_P,\omega) & = & \delta(\omega-\tilde{\epsilon}_{\lambda_P}) f_{\mu=\mu_P}(\omega)
	\label{eq:glt0defLSeigenstate}\\
	g_<^{d} (\lambda_P,\omega) & = & \delta(\omega-\epsilon_{\lambda_P}) f_{\mu=\mu_P}(\omega)
	\label{eq:gltdeigenstate}
\end{eqnarray}
where, at a given chemical potential $\mu$,
\begin{equation}
	f_\mu(\omega)=\frac{1}{1+\exp[\beta(\omega-\mu)]}
\end{equation}
denotes the Fermi-Dirac distribution function.
An evaluation of the level shifts $\tilde{\epsilon}_{\lambda_P}-\epsilon_{\lambda_P}$ determines the partial pressures and 
GCE thermodynamics \cite{NEQQSMthermodynamics} via \textit{lead-projected} density of state changes
\begin{equation}
	\Delta \mathcal{D}_{P = {\rm L/R}}(\omega) = 2 \sum_{\lambda_P} [\delta(\omega-\tilde{\epsilon}_{\lambda_P})
	-\delta(\omega-\epsilon_{\lambda_P})], 
	\label{eq:deltaDshiftEvaluation}
\end{equation}
and corresponding integrated density of state changes
\begin{eqnarray}
	\Delta \mathcal{N}_{P}(\mu') & = & \int_{-\infty}^{\mu'} \Delta \mathcal{D}_P(\omega) d\omega
	\label{eq:deltaNshiftEvaluation}\\
	& = & 2 \sum_{\lambda_P} \int \frac{d\omega}{2\pi} \; 
	[g_<^{(0)}(\tilde{\lambda}_P,\omega)-g_<^d(\tilde{\lambda}_P,\omega)].
	\label{eq:deltaNshiftEvaluationNEGF}
\end{eqnarray}
Section VIII.D suggests use of the formal specification 
(\ref{eq:deltaNshiftEvaluationNEGF}) to determine also the interacting 
nonequilibrium thermodynamics variation through the DFT SP formulation.

\begin{figure}
\begin{center}
\includegraphics[width=0.47\textwidth]{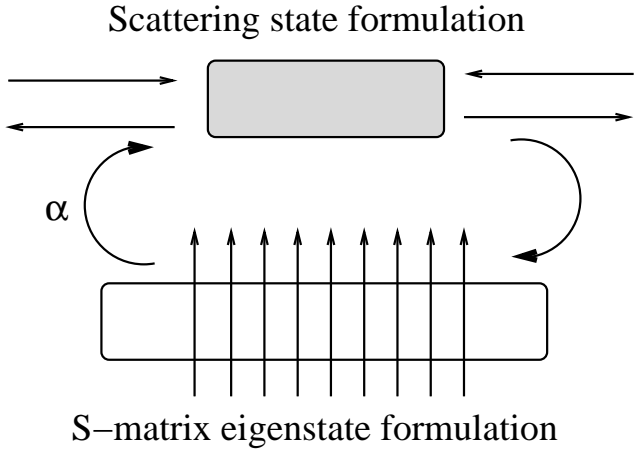}
\caption{
Schematics of the S-matrix decomposition analysis that is used here for
a formal expression and interpretation of general thermodynamics grand 
potential changes.  The essential step is identifying the unitary transformation
(curved arrows) $\boldsymbol{\alpha}=\langle j | \lambda_{\rm L/R} \rangle $ which relates the 
mutually independent scattering of lead-specific initial states $|\lambda_{\rm L/P}\rangle$
(upper panel) with the eigenstates $|j\rangle$ of the S-matrix (lower panel). The S-matrix 
decomposition provides a formal evaluation of the thermodynamics potential value 
in terms of effective (projected) phase shifts.
}
\end{center}
\end{figure}

Figure 1 shows schematics of the S-matrix eigenstate analysis, Appendix B, that leads to
an alternative formulation and interpretation of (\ref{eq:OmegaLS:leadEvaluation}) based
on scattering phase shifts \cite{FriedelSum,AlbertPhaseVar,Albert1D}.
The approach exploits the fact that for noninteracting particles, the S-matrix, and hence the Friedel
sum rule (\ref{eq:FormalFriedel}) already contains a complete mapping of the \textit{mutually independent} SP dynamics.
I note that the S-matrix is specified by the T-matrix behavior, i.e., by the independent nature of 
the SP LS eigenstates \cite{SmatrixStatMech,ZarambaKohnLS} and that there is no difference between the 
Green-function framework (above) and the scattering-phase formulation (below). 

In essence the analysis uses the generalized Friedel sum rule formulation (\ref{eq:FormalFriedel}) 
to sort the SP contributions according to the lead from which the 
scattering states emerge. This is done simply by restricting the trace to initial states in 
either leads, appendix B. The key step is expressing the S-matrix eigenstates $|j\rangle$ of eigenvalue 
$\exp[2i\delta_j(\omega)]$ in original states `$|\lambda_{P={\rm L/R}}\rangle$' and obtain a unitary 
matrix $\boldsymbol{\alpha}=\langle j | \lambda_{P}\rangle $ for the representation change. For each initial 
state $|\lambda_P\rangle$ one can extract the independent SP dynamics from the noninteracting 
S-matrix, a step which identifies lead- and state-specific phase shifts
\begin{eqnarray}
	\delta_{\lambda_P}(\omega) & \equiv & \sum_j \delta_j(\omega)  |\langle j| \lambda_{P} \rangle |^2\\
	& = & \frac{1}{2 i} \langle \lambda_{P}| \ln[\mathcal{S}(\omega)] | \lambda_{P}\rangle.
	        \label{eq:levelspecificphaseshift}
\end{eqnarray}
The lead-and-state specific contribution to the (tunneling-induced) change in density of state can thus be expressed in a familiar Friedel-type analysis \cite{FriedelSum}
\begin{equation}
\delta(\omega-\tilde{\epsilon}_{\lambda_P}) - \delta(\omega-\epsilon_{\lambda_P})  \leftrightarrow 
\frac{1}{2\pi} \frac{\partial}{\partial\omega} \delta_{\lambda_P}(\omega).
\label{eq:NEQdoschangeVsPhaseShift}
\end{equation}

The S-matrix analysis (\ref{eq:NEQdoschangeVsPhaseShift}) implies in turn an evaluation of (\ref{eq:deltaNshiftEvaluation}) 
which simply involves restricting the trace in the Friedel sum rule (\ref{eq:FormalFriedel}) to initial-system eigenstates 
in either of the leads. Also, one may from (\ref{eq:levelspecificphaseshift}) define corresponding frequency dependent
lead-specific phase shifts $\delta^{\rm eff}_{P={\rm L/R}}(\omega)$ that satisfy
\begin{equation}
	\delta^{\rm eff}_{P={\rm L/R}}(\omega) 
	\sum_{\lambda_P} \delta(\omega - \epsilon_{\lambda_P}) 
	= \sum_{\lambda_P} \delta(\omega - \epsilon_{\lambda_P}) 
	\delta_{\lambda_P}(\omega).
\end{equation}
The effective phase shifts reflect the density of states in the leads.  
An alternative formal determination of 
lead-specific integrated density of state changes is then
\begin{equation}
\Delta \mathcal{N}_{{\rm P=L/R}}(\mu_{\rm P}) = \frac{2}{\pi}
	\delta_{P}^{\rm eff}(\mu_P).
\label{eq:NleadSpecLS:phaseEvaluation}
\end{equation}
Similarly, the S-matrix eigenstate decomposition for noninteracting particles thus leads
to a scattering phase shift expression for the lead-specific contributions
\begin{equation}
\Delta \Omega_{\rm LS}^{P={\rm L/R}}(\mu_{\rm P}) = -\frac{2}{\pi}\int^{\mu_{\rm P}}_{-\infty} d\omega \,
	\delta_{P}^{\rm eff}(\omega),
\label{eq:OmegaLS:phaseEvaluation}
\end{equation}
to the nonequilibrium thermodynamics grand potential (\ref{OmegaLSsplit}).

\subsection{Role of Gibbs free energy in surface interactions} 

The single-s-wave scatterer result (\ref{eq:EQ1sThermodynamicsPotential}) points to the 
central importance of Gibbs free energy changes. I shall focus the discussion on the
adsorption interaction effect that is observable in
scanning-tunneling microscopy  investigations \cite{FriedelInteractSurfTest,WeissReview}. The importance is 
made explicit by noting that the equilibrium Gibbs free energy, here evaluated as
\begin{eqnarray}
	\mu_{\rm F} \Delta \mathcal{N}_{\rm MSS,1s}^{\delta_{\rm F}}(\mu_{\rm F}) & = &  \mu_{\rm F} \int^{\mu_{\rm F}} d\omega \Delta 
\; \mathcal{D}_{\rm MSS,1s}^{\delta_{\rm F}}(\omega) \nonumber\\
& = & \mu_{\rm F} \frac{2}{\pi} \delta_0(\mu_F),
\label{eq:phaseshiftIDOSchange}
\end{eqnarray}
is specified by the Friedel sum rule (\ref{eq:FormalFriedel}). 
The scanning-tunneling microscopy observations \cite{FriedelInteractSurfTest} of Fermi level (s-wave) phase shifts
$\delta_{\rm F}\approx \pm \pi/2$ is a direct indication that the adsorbate causes 
strong changes in the Gibbs free energy. 

The importance of Gibbs free energy variation in general open-infinite problems is further illustrated by 
calculations \cite{FriedelGrimley,FriedelTed,FriedelLK,HarrisFriedelInteract,HarrisFriedelTrioInteract,SlidingRings}  
of indirect electronic interaction arising between adsorbates on a noble metal surface and mediated by the 
MSS \cite{WeissReview}. 
Previous presentations have summarized the GCE evaluation \cite{Mermin65,ZarambaKohnLS,HarrisFriedelInteract} of the 
resulting MSS mediated 
interaction \cite{HarrisFriedelInteract,HarrisFriedelTrioInteract,SlidingRings,FriedelInteractSurfTest,WeissReview}.
Here I provide a simple discussion of the underlying GCE thermodynamics behavior.

The analysis of the interaction problem involves (appendix A) a determination of
additional scattering-induced changes in the integrated density of state (\ref{eq:FormalFriedel}). 
This calculation proceeds through the use of Lloyds' formula \cite{Lloyds} and the 
result \cite{HarrisFriedelInteract,NEQreformWithTime}
is given relative to the single-s-wave 
characterization, (\ref{eq:phaseshiftIDOSchange}) and (\ref{eq:EQ1sThermodynamicsPotential}).
From the determination of the integrated density of state changes
(\ref{eq:IntDOSchangesMSSexplicit}) one directly extracts the noninteracting-particle result 
both for the internal and GCE Gibbs free energy variations
\begin{eqnarray}
	\Delta U_{\rm MSS,2s}^{\delta_{\rm F}} (d; \mu_{\rm F}) & = & 
	\int_{-\infty}^{\mu_{\rm F}} d\omega' 
\, \omega' \frac{\partial}{\partial \omega} \Delta \mathcal{N}_{\rm MSS,2s}^{\delta_{\rm F}}(d; \omega'),
\label{eq:U2sFormal}\\
\Delta Y_{\rm MSS,2s}^{\delta_{\rm F}}(d; \mu_{\rm F})
& = & \mu_{\rm F} \Delta \mathcal{N}_{\rm MSS,2s}^{\delta_{\rm F}}(d; \mu_{\rm F}).
\label{eq:Y2sFormal}
\end{eqnarray}
Use of formal Fourier transform analysis \cite{Lighthill} establishes the (leading) asymptotic behavior of 
both of these terms as oscillatory but with a $1/d$ decay. 
The net (zero-temperature) interaction is given by the variation in thermodynamics grand potential
\begin{eqnarray}
\Delta	\Omega_{\rm MSS,2s}^{\delta_{\rm F}} (d; \mu_{\rm F}) 
	& \equiv & \Delta U_{\rm MSS,2s}^{\delta_{\rm F}} (d;\mu_{\rm F}) - \Delta Y_{\rm MSS,2s}^{\delta_{\rm F}}(d;\mu_{\rm F})
\label{eq:Omega2sFormal}\\
& \sim & - \epsilon_{\rm F} 
\left(\frac{2\sin(\delta_{\rm F})}{\pi}\right)^2 \, 
\frac{\sin(2q_{\rm F}d+2\delta_{\rm F})}{(q_{\rm F}d)^2},
\label{SPsumExpandFriedel}
\end{eqnarray}
where $\epsilon_{\rm F}$ denotes the difference between $\mu_{\rm F}$ and the
bottom of the MSS band. The last expression (\ref{SPsumExpandFriedel}) states the asymptotic 
behavior.

The thermodynamics grand potential interaction results \cite{HarrisFriedelInteract,SlidingRings},
the asymptotic oscillatory $1/d^2$ decay, and the phase in (\ref{SPsumExpandFriedel}), correspond 
well with experimental observations \cite{FriedelInteractSurfTest,WeissReview} for adatoms 
and admolecules on Cu(111).  However, I stress that this correct interaction behavior only 
arises upon a cancellation of terms with a much larger $1/d$ leading order behavior. This 
observation illustrates the essential role that charging, i.e., the Gibbs free energy 
variation and adherence to the Friedel sum rule, plays in understanding the GCE systems.

\subsection{Role of Gibbs free energy in nonequilibrium tunneling} 

In nonequilibrium tunneling the Gibbs free energy is no less important since it describes
charge adjustments and thus thermodynamics changes arising with the
onset of tunneling.  It thus affects the nature of nonequilibrium thermodynamics forces that guides
\textit{strictly adiabatic} transformations, i.e., forces that are  applicable (and needed) 
in the limit when the masses of the ions can be seen as infinitely greater than 
the mass of the electrons.

\begin{figure}
\begin{center}
\includegraphics[width=0.47\textwidth]{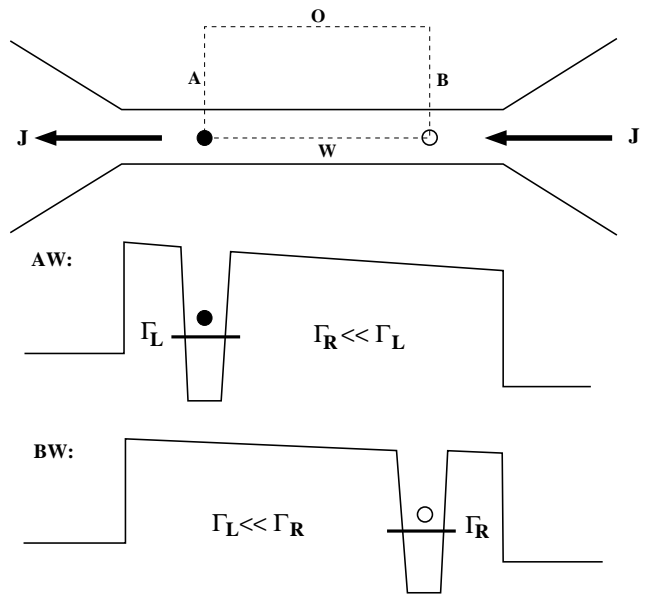}
\caption{
Schematics of possible (adiabatic) translations (inside the current-carrying wire, W, 
or outside, along path legs A, O, B) of the resonant level position in a lower-dimensional 
resonant tunneling system \textit{and} identification of a nonequilibrium Gibbs free energy effect 
(associated with resonant-level charging) that requires a thermodynamics-force analysis. 
The charging or discharging is independent of whether the resonant level is moved inside 
or outside of the wire.  The nonequilibrium Gibbs free energy (as well as the total nonequilibrium thermodynamics 
grand potential) variation can not be captured by forces which are proportional to the current 
density vector $\mathbf{J}$ (arrow in top panel) but will generally arise (via changes in the 
scattering phase shifts) also when the resonant level (or defect or ion) is moved 
perpendicularly to $\mathbf{J}$, along path legs A and B.}
\end{center}
\end{figure}

The top panel of Fig.~2 shows a schematics of a type of resonant-tunneling 
problem in which it is relevant to study both (nonadiabatic) electromigration \cite{Sorbello} 
\textit{and} nonequilibrium adiabatic relaxations from an exact thermodynamics analysis. I stress that there 
is only a limited zone of overlap in the nature and applicability of the here-derived 
adiabatic thermodynamics forces and the electrostatic forces investigated in the standard 
electromigration literature, for example, summarized in Ref.~\onlinecite{Sorbello}.  
In general, the electromigration literature \cite{Sorbello} splits the nonequilibrium forces into 
a wind-force contribution $\mathbf{F}_w$, defined by a electron-nuclei momentum transfer,
and a direct force $\mathbf{F}_d$. The wind force is proportional to the current 
density vector $\mathbf{J}(\mathbf{r})$, but is never relevant for a discussion of 
strictly adiabatic transformations (i.e., when one assumes an infinite ratio of 
the nuclei mass to the electron mass).  The direct force $\mathbf{F}_d$ is generally 
assumed to be electrostatic in nature, and typically assumed to be proportional
to $\mathbf{J}(\mathbf{r})$, although there are also observations that the
electrostatic nature can be more complex \cite{Todorov1}. 
There are arguments that an electrostatic nature of $\mathbf{F}_d$ is relevant 
in linear response \cite{Sorbello,TodorovSuttonThermodynForces}. 
However, there is no rigorous demonstration 
for a general equivalence of electrostatic and thermodynamics forces, Sec.~VI.C.  

An example of nonequilibrium Gibbs free energy effect, also shown in Fig.~2, is motivated to
help illustrate the differences between thermodynamics forces and nonconservative
electrostatic forces that are proportional to $\mathbf{J}(\mathbf{r})$.  The example
shows that there is no formal conflict with the Sorbello's gedanken experiment 
(explained, for example,  in Refs.~\onlinecite{Todorov1,Todorov2}) \textit{because} 
that analysis is not complete in the discussion of strictly adiabatic thermodynamics changes.  
The Sorbello gedanken experiment represents an argument that the nonequilibrium forces should be 
nonconservative \textit{if} they are proportional to the current density. A force relation 
(a proportionality) to the current density vector is seen as the fundamental reason that 
the electrostatic forces can sometime acquire a nonconservative nature \cite{Todorov1,Todorov2}.
However, the assumption that $\mathbf{F}\propto \mathbf{J}(\mathbf{r})$
\textit{can not} generally be relevant --- and I find that the Sorbello gedanken experiment 
is not conclusive --- for an analysis of adiabatic thermodynamics forces. This follows 
because such forces omit a full treatment of the nonequilibrium Gibbs free energy variation.

The middle and bottom panels of Fig.~2 summarize an analysis of the thermodynamics changes 
(defined by level occupation changes) that must arise in simple resonant tunneling systems when adiabatically moving 
the resonant level from a position `AW' to another position `BW' (positions which are
intersections of the set of path legs shown in the top panel). These are changes that arise when
the resonant level goes from a charged state (filled circle) to an discharged states (empty 
circle) and creates different regimes of tunneling \cite{RahmanDavies,Phon}. The nonequilibrium interactions 
in such systems has for example been explored in Refs.~\onlinecite{KondoPRL,RateEq} and the changes
in occupation cause differences in the nature and strength of the many-body interactions \cite{Phon,QCLdisc}.
A strictly adiabatic transformation along different paths must still 
produce the same (uniquely defined) final steady-state solution and, in particular,
identical changes in the thermodynamics grand potential and Gibbs free energy values, Sec. VI.D. 

I stress that the above-described adiabatic change in Gibbs free energy arises exclusively 
in the presence of a finite applied bias. I observe that the nonequilibrium Gibbs free energy change will 
arise (from phase-shift changes) also when moving the resonant level position perpendicularly to 
the current flow (along path legs A and B).  Fig.~2 thus identifies a nonequilibrium Gibbs free energy and 
thermodynamics grand-potential change which will not (generally) turn up in a path integral of 
the electromigration-type of forces [that satisfy $\mathbf{F}\propto \mathbf{J}(\mathbf{r})$].  

I conclude that an analysis of exact thermodynamics forces for strictly adiabatic transformations 
may sometimes be necessary to supplement the standard electromigration analysis \cite{Sorbello}.

\section{Partition scheme and Hamiltonian}

The partition scheme of Caroli {\it et al,} Ref.~\onlinecite{Caroli}, 
allows a description of tunneling which is first principle and 
predictive (i.e., parameter-free, atomistic).
This is true when the Caroli partition scheme
is extended from the original tight-binding framework to the continuum
formulation \cite{ContinuumCaroli,LSCDFT}.
The emphasis on an atomistic approach (i.e., with the description 
of the electron dynamics having a parametric dependence of the
nuclei positions) is important because we desire a thermodynamic theory 
specific to the materials nature.  Avoiding an empirical modeling (set, e.g., by 
tight-binding parameters) implies that electron behavior is completely 
specified by the position of the nuclei and by whatever external field one chooses to apply. 
Subject to a suitable nonequilibrium implementation of the Born-Oppenheimer approximation, one 
can therefore relax atomic structure and morphology to provide a parameter-free 
and fully predictive thermodynamic description of interacting 
nonequilibrium tunneling starting from an electron Hamiltonian.

\subsection{Initially disconnected equilibria system}

Using atomic units throughout, the total kinetic energy can be expressed 
as a formal second-quantization operator
\begin{eqnarray}
K & = & 
\sum_s \int d\mathbf{r}' \, \int d\mathbf{r} \,
\hat{\psi}_{s}^{\dagger}(\mathbf{r}')
\frac{1}{2} \langle \mathbf{r}'| \hat{k}^2 | \mathbf{r}\rangle
\hat{\psi}_{s}(\mathbf{r}),
\label{Kdef}\\
\langle \mathbf{r}'| \hat{k}^2 | \mathbf{r}\rangle
&  = & - \int d\mathbf{r}_1 
\frac{\partial}{\partial \mathbf{r}'} \delta(\mathbf{r}'-\mathbf{r}_1)
\frac{\partial}{\partial \mathbf{r}_1} \delta(\mathbf{r}_1-\mathbf{r}).
\label{KmatDef}
\end{eqnarray}
This formulation reduces to the traditional second-quantization form 
\begin{equation}
K= \sum_s \int d\mathbf{r} \,
\hat{\psi}_{s}^{\dagger}(\mathbf{r})
\left(- \frac{1}{2} \nabla^2 \right) \hat{\psi}_{s}(\mathbf{r}),
\label{Kdeftrad}
\end{equation}
with a handling of boundary conditions. 

The Feuchtwang/Caroli partition scheme \cite{Caroli,ContinuumCaroli,Pendry} for 
nonequilibrium Green function and/or wavefunction studies of tunneling assumes that the system is disconnected 
at $t\to -\infty$, split into three sections, and having a kinetic energy described by
$K-\delta K$. The tunneling term $\delta K$ is defined below but is adiabatically
turned on, 
\begin{eqnarray}
K(t) & = &  (K-\delta K) + \delta K(t)\\
\delta K(t) & \sim & \delta K \exp(\eta t)
\end{eqnarray}
so that the full kinetic energy (\ref{Kdeftrad}) emerges at $t=0$. The $t\to 0$ splitting 
into disconnected sections allows thermodynamics to be unambiguously defined in each of the components, 
a left lead `$L$' ($z<z_L$), a center region `$C$' ($z_L<z<z_R$), and a right lead `$R$'. 
This Caroli partition scheme has, for example, been used to derive a generalization of 
the Landaur-B{\"u}ttiker-type current formula to interacting tunneling \cite{KondoPRL,YigalNed}.
The total kinetic energy $K$ of the system is written as a sum, $K\equiv K_L+K_R+K_C+\delta K$, 
where $K_{L,R,C}$ have supports which are confined to the respective, separate components. 

The continuum (and thus atomistic) Caroli partition scheme \cite{ContinuumCaroli,Pendry,LSCDFT} 
assumes that each of the disconnected subsections $H_d=\sum_{J=L,C,R} H_J$ is in 
equilibrium at different chemical potentials $\mu_{L/C/R}$ and that the three spatial
regions of $H_d$ together can represent the full spatial variation of any given potential 
$v(\mathbf{r},t)$.  At $t\to -\infty$, the system is assumed described by the 
static external potential $v_d(\mathbf{r})$ and corresponding operator 
\begin{equation}
V_d=\int d\mathbf{r} \, v_d(\mathbf{r})\, \hat{n}(\mathbf{r}).
\label{eq:V0def}
\end{equation}
where $\hat{n}(\mathbf{r})\equiv\sum_s{\psi}_{s}^{\dagger}(\mathbf{r})
\hat{\psi}_{s}(\mathbf{r})$ denotes the electron-density operator.

The initial system in the continuum (and hence atomistic) 
Caroli scheme is simply
\begin{equation}
H_d =  
\sum_{P=L/R/C} K_P +V_{d}.
\label{eq:Hdterm}
\end{equation}
I use $\{c^\dagger_{\lambda_{P=L/R/C}} \}$  to denote 
the set of operators which creates an electron in a $H_d$ SP
eigenstate (of energy $\epsilon_{\lambda, i}$) with basis in the left, right, 
or center regions, respectively.  
The quadratic Hamiltonian (\ref{eq:Hdterm}) can be expressed 
\begin{equation}
H_d =  
\sum_{J=L/R/C} \sum_{\lambda} \epsilon_{\lambda,J} c^{\dag}_{\lambda,J}c_{\lambda,J}.
\label{eq:HdtermOpexpand}
\end{equation}
In a second-quantization formulation, the tunneling term can then be expressed
\begin{eqnarray}
\delta K & = & 
\sum_s \int d\mathbf{r} \,
\hat{\psi}_{s}^{\dagger}(\mathbf{r})
\left[- \frac{1}{2} \nabla^2 + v_d(\mathbf{r}) \right] \hat{\psi}_{s}(\mathbf{r})
\nonumber\\
& & -\sum_{J=L/R/C} \sum_{\lambda} \epsilon_{\lambda,J} c^{\dag}_{\lambda,J}c_{\lambda,J}.
\label{KdeltaTradOp}
\end{eqnarray}
This is a traditional starting point in nonequilibrium Green function calculations. 

\begin{figure}
\begin{center}
\includegraphics[width=0.47\textwidth]{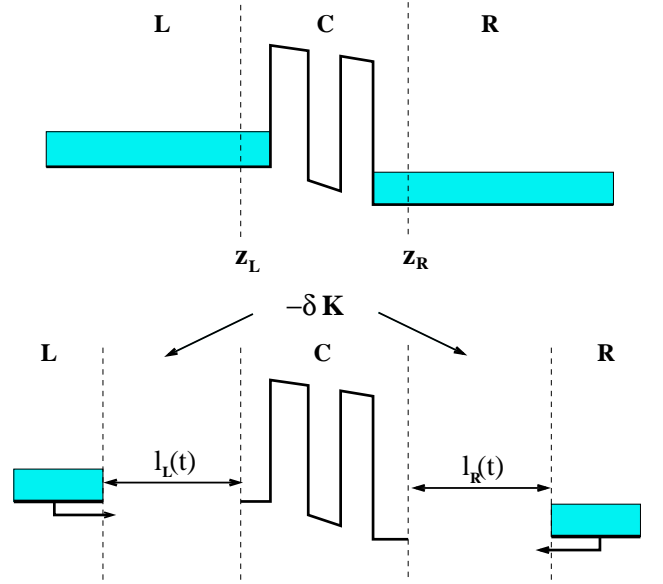}
\caption{
Schematics of possible \textit{interpretation} of the continuum Caroli partition scheme. In the
interpretation one assumes that the three sections (a left `L' and right `R' lead, and a center 
tunneling region `C') are split at $z_L$ and $z_R$ (shown in top panel) and well separated 
(bottom panel) at $t\to -\infty$, thus ensuring a well-characterized equilibrium-thermodynamics 
starting point for the quantum-kinetic account. The starting kinetic energy differs by $\delta K$ 
from that of the connected system that characterizes the system at $t=0$ (top panel). The kinetic-energy
difference $\delta K$ is adiabatically turned on together with changes in the single-particle
potential $\delta V$ and a general electron-electron interaction term $W$. I stress that the actual
partition scheme (formulated in the text) is independent of this schematic interpretation; 
it is simply a well-defined and standard second-quantization starting point for computing properties 
of interacting nonequilibrium tunneling \protect\cite{KondoPRL,YigalNed,LSCDFT}.}
\end{center}
\end{figure}

Figure 3 suggests that this Caroli framework can be interpreted
as representing a system which at $t\to -\infty$ (and around the divisions $z_{\rm L/R}$)
has voids of high fictitious barriers of width $l_{\rm L/R}\to \infty$.
At a finite time, the width $l_{\rm L/R}(t)$ of the fictitious barriers decreases
(and eventually vanishes at $t=0$) so as to produce the actual form of the adiabatic 
turn on of the tunneling term $\delta K(t)\sim \delta K\exp(\eta t)$.
%It is, in principle, also possible to use the formal second-quantization expression 
%(\ref{Kdef}) for a formal delta-function formulation
%\begin{eqnarray}
%\delta K & =  &\lim_{\delta z\to 0} \sum_{P=L/R}
%\sum_s 
%\int_{z_P-\delta z/2}^{z_P+\delta z/2} d\mathbf{r} \times\nonumber\\
%&& 
%\int d\mathbf{r}' \, 
%\hat{\psi}_{s}^{\dagger}(\mathbf{r}')
%\frac{1}{2} \langle \mathbf{r'}| \hat{k}^2 | \mathbf{r} \rangle 
%\hat{\psi}_{s}(\mathbf{r}).
%\label{eq:deltaKdeforig}
%\end{eqnarray}
%However, such a formulation is only precise up to a discussion of boundary conditions, 
%[a fact which is also true for Eq.~(\ref{KmatDef})].  In any case, 
I stress that the 
precise formulation of $\delta K$ or interpretations is irrelevant for the formal 
results presented here and for all parts of the previous LS collision DFT analysis\footnote{The 
explicit form given for the tunneling term $\delta K$
in \protect\cite{LSCDFT} is not correct. However,
that fact is completely irrelevant for the formal exploration of 
the tunneling based on the Caroli partition, $K=(K-\delta K)+\delta K$ (an equation 
which formal scattering theory solves to all orders in $\delta K$). For example, in the
uniqueness of density proof presented in the appendix of Ref.~\protect\onlinecite{LSCDFT}, 
the specific form for $\delta K$ is completely void of consequence since the difference 
between two relevant Hamiltonians will always exclusively depend on a possible 
difference in the external potentials.}
\cite{LSCDFT}.

The precise formulation and expression for $\delta K$ (including the choice of $z_{L/R}$) 
is also of no consequence for practical nonequilibrium thermodynamics calculations. This is because I here
demonstrate that the relevant thermodynamics can be expressed through the phase shifts of
the SP LS solutions \cite{Lang,AlbertXIX3}, a quantity for which there exists readily 
available efficient solvers.  In fact, I show that the SP standard DBT 
codes \cite{Lang,DiVentra,TranSiesta,PuskaRisto,gpawNEGF} (for frozen-nuclei 
problems) already approximate the exact SP LS solutions and, moreover, can be
adapted also for systematic improvements (given refinement of the nonequilibrium 
density-functional description).

A simple Gibbs-weighting operator $\hat{X}_d$ and associated unnormalized
density matrix,
\begin{equation}
\hat{\rho}_d \equiv  
\exp[-\beta\hat{X}_d],
\label{eq:rhoXddef}
\end{equation}
characterizes the disconnected ($t\to -\infty$) system at inverse 
temperature $\beta$. 
The initial density matrix is a direct product of L/C/R equilibrium density 
matrices and the initial Gibbs weighting 
can therefore directly be expressed as a difference 
\begin{equation}
\hat{X}_d\equiv H_d-\hat{Y}_d
\label{eq:XdDef}
\end{equation}
between $H_d$ and an operator for the $t \to -\infty$ 
Gibbs free energy,
\begin{equation}
\hat{Y}_d\equiv \sum_{J=L/R/C} 
\mu_J \hat{c}^{\dagger}_{\lambda,J} \hat{c}_{\lambda,J}.
\label{eq:YdSumDef}
\end{equation}

It is possible (and likely helpful for nonequilibrium thermodynamics 
calculations) to define the initial disconnected equilibria system $H_d$ 
as composed of isolated ground-state DFT systems.  One may, for example, adapt the 
embedding formulation \cite{EmbeddingDFT} from the original application 
for surfaces and to the present case of (infinitely extended) barriers around
$z_{L/R}$. In such an embedding approach 
it may furthermore be convenient to assume that the form of $v_d(\mathbf{r})$ for 
$\mathbf{r}\to \pm \infty$ reduces to that of a jellium model, i.e., describing
a uniform background potential $\phi_{L/R}$ for $z\to +/- \infty$.  
The values of $\phi_{L}$ and $\phi_{R}$ must be chosen to ensure charge 
neutrality of the initial, disconnected Hamiltonian
at $t\to -\infty$ (and at $z\to \pm \infty$ in general).  

The continuum Caroli partition scheme ensures that we have initially
a disconnected-equilibria configuration and therefore a well defined thermodynamics starting point.
The disconnected-equilibria ($t \to -\infty$) density matrix
\begin{equation}
\frac{\hat{\rho}_d}
{\hbox{\rm Tr}\{\hat{\rho}_d\}}= 
\frac{\exp[-\beta(H_d-\hat{Y}_d)]}
{\hbox{\rm Tr}\{\exp[-\beta(H_d-\hat{Y}_d)]\}}
\label{eq:rhoddef}
\end{equation}
is independent of the value of the applied bias $\phi_{\rm bias} 
\equiv \phi_L-\phi_R = \mu_L-\mu_R$. I stress that the interacting tunneling system will 
evolve to gain the correct density distribution and that the relation $\phi_L - \phi_R = \mu_L-\mu_R$ 
is not used at relevant times $t\approx 0$.  The initial density matrix depends 
exclusively on the initial electron concentration in the leads and the pair of values for lead chemical potentials, 
and, when relevant, on the initial occupation (formally described by chemical potential
$\mu_C$) of the central island `C' \cite{LSCDFT}.

\subsection{The interacting tunneling system}

The continuum Caroli partition scheme assumes an adiabatic turning on of the fixed 
tunneling term $\delta K$, of a given many-body interaction $W$ (for example, the full 
electron-electron interaction) and of a static electron-scattering potential 
$v_{\rm sc}(\mathbf{r})$ which includes the effects of the applied bias and of 
the set of atomic potentials.  The scheme can also allow for an additional time-dependent 
potential $\phi_{\rm g}(\mathbf{r},t>t_g)$, then describing a gate operation 
starting at some finite time $t_g$.  The total (time-dependent) collision potential 
is 
\begin{equation}
v_{\rm col}(\mathbf{r},t)= v_{\rm sc}(\mathbf{r})
+ \phi_{\rm g}(\mathbf{r},t).
\label{eq:vcoldef}
\end{equation}
This single-particle potential defines a quadratic contribution 
\begin{equation}
V_{\rm col}(t)  =  
\int d\mathbf{r} \, v_{\rm col}(\mathbf{r},t) \, \hat{n}(\mathbf{r})
\label{eq:Vterm}
\end{equation}
to the Hamiltonian.

In several cases in the presentation of a thermodynamic theory
for tunneling, the focus will be on the steady-state regime. Here the
collision term simply describes the effect of the time-independent 
scattering potential
\begin{equation}
V_{\rm col}(t) \longrightarrow V_{\rm sc} = 
\int d\mathbf{r} \, v_{\rm sc}(\mathbf{r}) \, \hat{n}(\mathbf{r}).
\label{eq:VSCterm}
\end{equation}

For a continuum description (which allows atomistic calculations) of electron
dynamics in nonequilibrium interacting tunneling, I investigate the full quantum-kinetics of 
the electrons in the connected (time-dependent and interacting) Hamiltonian
\begin{eqnarray}
H(t) & = & H_d + a_\eta(t)\, H_{1,{\rm col}}(t)
\label{eq:Hamilton}\\
H_{1,{\rm col}}(t) & = & \{\delta K + W +V_{\rm col}(t)-V_d\}, 
\label{eq:HamiltonH1}
\end{eqnarray}
The adiabatic turn on is described by a factor $a_\eta(t)$ which for all practical 
purposes\footnote{It is possible to link the LS collision DFT \protect\cite{LSCDFT} 
with the present nonequilibrium thermodynamic account because one can make compatible 
choices in the details of adiabatic turn on and thus satisfy even stringent mathematical
conditions, appendix D.}
can be taken as $\exp(\eta t), \eta \to 0$; the original LS 
analysis~\cite{LippSchwing} of the many-body collision problems used $\exp(-\eta |t|), 
\eta = 0^{+}$. 

Furthermore, to aid the discussion of an exact, effective SP formulation (for efficient 
calculations of the nonequilibrium density, thermodynamics grand potential, and forces) in steady state,
I also introduce the quadratic Hamiltonian
\begin{eqnarray}
H_V^{(0)} & \equiv & H_d +H_{1,V}^{(0)},
\label{eq:H0def}\\
H_{1,V}^{(0)} & \equiv & - V_d + \delta K + V,
\label{eq:H10def}
\end{eqnarray}
where the potential term $V$ is described by a time-independent scattering potential, $V=V_{\rm sc}$.

\subsection{Relation to the quantum kinetics account}

Both the original Caroli scheme and the presently used continuum 
Caroli partition scheme \cite{ContinuumCaroli,LSCDFT} define a many-body collision 
problem \cite{LippSchwing} for nonequilibrium tunneling. Traditional many-body collision 
problems involve an approach of initially free particles to a central 
interacting collision region; this approach can conveniently be expressed 
by an adiabatic turn on, for example, as used in the formal LS 
formulation \cite{LippSchwing}, and in subsequent analysis by, for example,
Dyson \cite{Dyson}, Pirenne \cite{Pirenne}, Gellmann and Goldberger \cite{Gellmann}, 
and DeWitt \cite{DeWitt}.

Ref.~\onlinecite{LangrethLS} provides an excellent introduction and discussion of 
application of formal collision theory to condensed matter theory problems.
The trick of a soft turn on of the perturbations was soon applied 
also for (infinite) condensed matter problems to drive the system from 
a well-defined initial configuration to the physically relevant 
case of an interacting many-body system. The adiabatic turn on (and hence 
the inherent collision nature) is essential for the 
quantum-kinetics account of nonequilibrium dynamics~\cite{Langreth} as well as 
specific Green function calculations of interacting tunneling for it provides 
such calculations with a robust starting point.  

The steady-state Gibbs free energy operator coincides with the operator (originally
termed `$\hat{Y}$') that the steady-state ReNQSM uses to describe electron 
redistribution in the nonequilibrium density matrix 
\begin{equation}
\hat{\rho}_{\rm LS} \equiv \exp[-\beta (H-\hat{Y}_{\rm LS})].
\label{eq:rhoReNQSM}
\end{equation}
This redistribution operator can under general conditions be interpreted \cite{NEQreform,exactKondo,Dutt} 
\begin{equation}
\hat{Y}_{\rm LS}=\sum_{\lambda,J=L/R/C} \mu_J \hat{\psi}_{LS,\lambda,J}^{\dagger} \hat{\psi}_{LS,\lambda,J},
\label{eq:HershfieldLSinterpretation}
\end{equation}
in terms of a family of operators, $\{\hat{\Psi}^\dagger_{LS,\lambda,J}\}$
for many-body-LS single-quasiparticle excitations of energy $\epsilon_{\lambda_{P=L/R/C}}$. 
For most transport problems, when bound states are not relevant, one can ignore the 
evolution which originates from the smaller number of states in the center regime.  For simplicity in notation
and discussion I will therefore  restrict the summation for the connected system [and in 
(\ref{eq:HershfieldLSinterpretation})] simply to the scattering contributions originating 
from leads, $P={\rm L,R}$. 

In any case, the operators (\ref{eq:HershfieldLSinterpretation}) express the adiabatic evolution
(described by a many-body Liouville operator $\mathcal{L}_{H}\hat{A} \equiv [H,\hat{A}]$) 
of the initial (independent) states subject to full many-body 
interaction \cite{NEQreform,exactKondo},
\begin{equation}
\hat{\psi}^{\dagger}_{LS,\lambda_P} = \frac{i\eta}{\epsilon_{\lambda_P}+\mathcal{L}_H+i\eta} 
\, \hat{c}^{\dagger}_{\lambda_P}.
\label{eq:LSopSolution}
\end{equation}
This operator-based ReNQSM analysis can be generalized \cite{DoyonAndrei,NEQunified,NEQaverageEvolution,Han,Dutt,NEQreformDetails,NEQreformWithTime} 
and it is possible to express not only the general Gibbs free 
energy operator $\hat{Y}_{\rm col}(t)$ but also the variational LS expression for 
the many-body T-matrix \cite{LSCDFT} in terms of single-quasi-particle excitations.

\section{Nonequilibrium quantum statistical mechanics}

The formal structure of the exact nonequilibrium thermodynamic theory is 
obtained in the spirit of Mermin's 1965 formulation of 
equilibrium thermodynamics \cite{Mermin65}.
In the absence of any applied bias one regains the GCE
system for which Mermin succeeded in formulating an 
equilibrium-thermodynamic DFT.

\subsection{Collision nature of transport}

The general strategy for studying many-body collision problems 
is to formally solve for projections of the full time-evolution operator
\begin{equation}
i\frac{\partial \mathcal{U}(t,-\infty)}{\partial t} = H(t) \mathcal{U}(t,-\infty).
\label{eq:genQM}
\end{equation}
This operator describes, for example, the evolution of a many-body collision state 
\begin{equation}
|\Psi_\xi(t)\rangle
= \mathcal{U}(t,-\infty)|\Phi_\xi\rangle
\label{eq:MBsolution}
\end{equation} 
originating and evolving from an initial state $|\Phi_\xi \rangle $.
I shall assume that such initial states are simultaneous eigenstates 
of the disconnected-system operators $H_d$ and $\hat{Y}_d$ (of eigenvalues
$E_\xi$ and $Y_\xi$, respectively) and I expand the unnormalized,
initial density matrix 
\begin{equation}
\hat{\rho}_d \equiv 
\sum_{\xi}|\Phi_\xi\rangle \exp[-\beta (E_\xi-Y_\xi)] \langle \Psi_\xi |.
\label{eq:rhod-expand}
\end{equation}

For a steady state collision problem one can explicitly construct a
nonequilibrium density matrix which incorporates the full complexity of the 
many-body LS solution \cite{LippSchwing} (\ref{eq:rhod-expandLS}). The adiabatic 
turn on, including scattering and interactions, causes energy shifts 
$\Delta_\xi$ which are consistently handled within the general 
LS solution by simultaneously adjusting \cite{Pirenne,Gellmann,DeWitt}
\begin{eqnarray}
\delta V & \to & \delta V - \Delta_\xi |\Phi_\xi\rangle \langle\Phi_\xi|, 
\label{eq:Vshift}\\
E_\xi & \to & E_\xi +\Delta_\xi.
\label{eq:Eshift}
\end{eqnarray}
The resulting LS solution 
\begin{equation}
| \Psi_\xi^{(+)} \rangle = \frac{i\eta}{E_\xi-H+i\eta}|\Phi_\xi\rangle
\label{eq:LSmbEquation}
\end{equation} 
constitutes an explicit construction of an interacting eigenstate
of energy $E_{\xi}+\Delta_\xi$ for steady-state problems. The
energy shifts \cite{Pirenne} $\Delta_\xi$ are themselves determined by 
the state evolution \cite{Gellmann,DeWitt,LangrethLS}. In condensed matter
problems such shifts are essential and permit applications
also to problems where the perturbation includes a periodic
potential (and the dynamics must reflect the band structure that arise) \cite{Gellmann}.
The inclusion of the integrated effects of (infinitesimal) energy shifts provides
a consistent GCE determination of the indirect electronic interactions of adsorbates
on surfaces \cite{HarrisFriedelInteract}.

Steady state solutions
arise in the collision problems when the collision potential reduces
to the time-independent form (\ref{eq:VSCterm}) and when the system 
dynamics is completely specified by $H(t=0)$. The many-body LS solution
(\ref{eq:LSmbEquation}) constitutes a projection of the time-evolution 
operator $\mathcal{U}(t,-\infty)$, since it satisfies
\begin{equation}
\exp(-i(E_\xi+\Delta_\xi) t) | \Psi_\xi^{(+)} \rangle 
 =  \mathcal{U}(t,-\infty) | \Phi_\xi\rangle
\label{eq:LSformaltime}
\end{equation}
in the limit of an infinitely slow adiabatic turn on, $\eta \to 0^{+}$.
Combining Eqs.~(\ref{eq:rhod-expand}) and (\ref{eq:LSformaltime}) yields
an (unnormalized) steady-state nonequilibrium collision density 
matrix (\ref{eq:rhod-expandLS}).

The explicit construction (\ref{eq:rhod-expandLS}) of the steady-state 
nonequilibrium density matrix $\hat{\rho}_{\rm LS}$ provides a direct link of the 
nonequilibrium thermodynamic theory (and of ReNQSM) to formal many-body LS 
solutions \cite{LippSchwing}. Closely related links are expressed 
in the seminal steady-state ReNQSM papers, Ref.~\onlinecite{NEQreform,exactKondo}, 
and in several recent extensions and applications that study
steady-state interacting tunneling transport \cite{NEQunified,NEQaverageEvolution,DoyonAndrei,Han,Dutt,NEQvariational}.

\subsection{Collision picture of interacting tunneling}

Standard QKA methods, including nonequilibrium Green function techniques, permit 
calculations of operator expectation values under nonequilibrium conditions and
must therefore provide a consistent treatment of the collision nature 
(\ref{eq:MBsolution}).  The tested versatility of the many-body LS equation 
for steady state problems motivates an effort to keep a corresponding 
collision picture for solving general (time-dependent or steady-state) 
interacting tunneling through a nonequilibrium collision density matrix 
$\hat{\rho}_{\rm col}(t)$.

Traditionally, the formal starting point for calculating the expectation value
of an operator $\hat{O}(t)$ in a nonequilibrium system is \cite{Langreth} 
\begin{equation}
\langle \hat{O}\rangle_{\rm col} (t)
\equiv 
\frac{\hbox{Tr}\{\hat{\rho}_d \, \hat{O}_\mathcal{H}(t)\}}
{\hbox{Tr}\{\hat{\rho}_d \}}.
\label{eq:neqqsmdef}
\end{equation}
The subscript `col' emphasizes that a Caroli collision picture underpins the definition.
The evaluation (\ref{eq:neqqsmdef}) rests on the Heisenberg picture which represents 
a canonical transformation of operators,
\begin{equation}
\hat{O}(t) \longrightarrow \hat{O}_\mathcal{H}(t)\equiv  
\mathcal{U}^{\dagger}(t,-\infty) \hat{O}(t) \mathcal{U}(t,-\infty).
\label{HeisenbergPicture}
\end{equation}
The Heisenberg picture has advantages for 
time-independent problems (for it leaves the many-body 
eigenstates fixed), but it is not a natural description 
in the presence of the adiabatic turn on used in the collision 
problems. Nevertheless, the original $t\to -\infty$ system ($H_d$) 
has a well-defined separated-equilibrium thermodynamics $\hat{\rho}_d$.
The formulation (\ref{eq:neqqsmdef}) still serves to uniquely define 
the expectation value because the QKAs emphasize accurate (conserving) 
descriptions of the time-evaluation $\mathcal{U}(t,\infty)$.

For development of a nonequilibrium thermodynamic theory it is more 
advantageous to make the collision nature (\ref{eq:MBsolution}) explicit
in the operators for thermodynamic quantities. I work 
below with a collision picture defined by operator transformations
\begin{equation}
\hat{A}(t) \longrightarrow \hat{A}_{\rm col}(t)\equiv
\mathcal{U}(t,-\infty) \hat{A}(t) \mathcal{U}^{\dagger}(t,-\infty)
\label{CollisionPicture}
\end{equation}
where $\hat{A}(t)$ denotes a time-dependent operator in the Schr{\"o}dinger
picture. This is done simply because it offers a simpler formulation and interpretation; 
the here-presented description is formally exact. This approach is suggested
by the formal steady-state LS solution \cite{LippSchwing} and it is
here used for a general time-dependent tunneling case.
In such a collision picture we obtain an explicit construction of the exact nonequilibrium 
collision density matrix 
\begin{eqnarray}
\hat{\rho}_{\rm col}(t) & \equiv & 
\mathcal{U}(t,-\infty) 
\hat{\rho}_d
\mathcal{U}^{\dagger}(t,-\infty)
\nonumber\\
& = & 
\sum_\xi
|\Psi_\xi(t)\rangle 
\exp[-\beta (E_\xi-Y_\xi)] 
\langle \Psi_\xi (t)|.
\label{eq:rhoColdefine}
\end{eqnarray}
The time evaluation of the density matrix is
\begin{equation}
i\frac{d\hat{\rho}_{\rm col}}{dt} = [H(t),\hat{\rho}_{\rm col}(t)].
\label{InverseHeisenbergEvolveRho}
\end{equation}
The formal time-evolution of other operators are also simpler than 
in the Heisenberg picture because of the opposite ordering of time-evolution operators in (\ref{eq:rhoColdefine}). 
In using the collision picture for thermodynamic operators (\ref{CollisionPicture}) 
care is taken to verify that the resulting formulation remains exact and fully equivalent 
with the traditional formulations of the QKA \cite{Langreth}. For the collision density matrix 
(\ref{eq:rhoColdefine}) the formal equivalence
\begin{equation}
\langle \hat{O}\rangle_{\rm col} (t)
 \equiv 
 \frac{\hbox{Tr}\{\hat{\rho}_d \, \hat{O}_{\mathcal{H}}(t)\}}
 {\hbox{Tr}\{\hat{\rho}_d \}}
 = \frac{\hbox{Tr}\{\hat{\rho}_{\rm col}(t) \, \hat{O}(t)\}}
{\hbox{Tr}\{\hat{\rho}_{\rm col}(t)\}}.
\label{eq:ObarHneqqsmdef}
\end{equation}
follows simply from a cyclic permutation of operators.

\subsection{Thermodynamical description of nonequilibrium tunneling}

The collision density matrix (\ref{eq:rhoColdefine}) can be 
expressed in terms of an evolving Gibbs-weighting factor $\hat{X}_{\rm col}(t)$. 
I start from the disconnected-equilibrium Gibbs weighting 
\begin{equation}
\hat{X}_{\rm col}(t\to -\infty ) = \hat{X}_d\equiv H_d-\hat{Y}_d,
\label{eq:Xddef}
\end{equation}
and express the evolution 
\begin{eqnarray}
\hat{X}_{\rm col} & = &
\mathcal{U}(t,-\infty) \hat{X}_d \mathcal{U}^{\dagger}(t,-\infty),
\label{eq:GWdefine}\\
\hat{\rho}_{\rm col}(t) 
& = &
\exp[-\beta \hat{X}_{\rm col}(t)],
\label{eq:rhoGWdefine}
\end{eqnarray} 
thus identifying (\ref{eq:GWdefine}) as the emerging Gibbs weighting.

The introduction of a nonequilibrium Gibbs weighting factor 
motivates, in turn, a definition of the thermodynamic grand potential
\begin{eqnarray}
\Omega_{\rm col}(t) & \equiv & -\frac{1}{\beta}
\ln \hbox{Tr}\{\hat{\rho}_{\rm col}(t)\}
\nonumber\\
& = & -\frac{1}{\beta} \ln \hbox{Tr}\{\exp[-\beta \hat{X}_{\rm col}(t)]\}.
\label{eq:OmegNEQcolDef}
\end{eqnarray}
The exact ReNQSM \cite{NEQreform,exactKondo,NEQunified} 
shows that $H-Y_{\rm LS}$ serves as a Gibbs weighting 
factor in the steady state problem. For the general tunneling 
problem I define
\begin{equation}
\hat{Y}_{\rm col}(t)\equiv H(t)-\hat{X}_{\rm col}(t),
\label{eq:Ydef}
\end{equation} 
corresponding to a reformulation the normalized collision density matrix
\begin{equation}
\frac{\hat{\rho}_{\rm col}(t)}{\hbox{Tr}\{\hat{\rho}_{\rm col}(t)\}} 
= \exp[\beta [\Omega_{\rm col}(t)-H(t)+\hat{Y}_{\rm col}(t)]].
\label{eq:EffectiveWeightingEvaluation}
\end{equation}
A formal evaluation of the entropy $S_{\rm col}(t)$ in the exact 
collision density matrix can therefore be expressed \cite{Fano}
\begin{eqnarray}
S_{\rm col}(t) &\equiv & 
- \langle \ln (\hat{\rho}_{\rm col}) \rangle_{\rm col}
+ \langle \ln (\hbox{Tr}\{\hat{\rho}_{\rm col}\}) \rangle_{\rm col}
\nonumber\\ 
& = & -\beta \Omega_{\rm col}(t)+ \beta \langle H(t)\rangle_{\rm col}
-\beta \langle \hat{Y}_{\rm col}(t)\rangle_{\rm col}.
\end{eqnarray}

I consequently make the natural identification of $\hat{Y}_{\rm col}(t)$ 
as the operator for the nonequilibrium Gibbs free energy and I stress that the 
operator can be seen as emerging, with a time-evolution given by 
the LS collision picture (\ref{CollisionPicture}).  This interpretation is 
consistent with the boundary condition, 
\begin{equation}
\hat{Y}_{\rm col}(t\to -\infty)= \hat{Y}_d=\sum_{\lambda,J=L/R/C} 
\mu_J 
\hat{c}^{\dagger}_{\lambda,J}
\hat{c}_{\lambda,J}, 
\end{equation}
which obviously provides an explicit formulation of the Gibbs free energy 
in the original, disconnected-equilibria system.  I note that, 
irrespectively of the interpretation, the operator 
$\hat{Y}_{\rm col}(t)$ is uniquely specified by the collision 
nature of the Gibbs weighting operator $\hat{X}_{\rm col}(t)$ or, 
equivalently, of the exact collision density matrix (\ref{eq:rhoGWdefine}).  

\subsection{Nonequilibrium Gibbs free energy}

An evaluation of the exact collision density matrix (\ref{eq:rhoGWdefine}) 
requires an explicit determination of the operator for nonequilibrium Gibbs free energy, 
$\hat{Y}_{\rm col}(t)$.  This was (with the interpretation stated above) a central
achievement of the exact steady-state ReNQSM, proceeding through an 
explicit construction $\hat{Y}_{\rm LS}$ order by order in the formal 
perturbation term \cite{NEQreform}. Here I start instead with the 
collision nature of the problem, to obtain a generalization 
to general tunneling and to clarify the explicit connection to 
a full nonequilibrium thermodynamic theory.

By the emerging nature of the Gibbs free energy operator (\ref{eq:Ydef}), we can
directly extract formal time evolution
\begin{equation}
i\frac{d \hat{Y}_{\rm col}(t)}{dt} = [H(t),\hat{Y}_{\rm col}(t)] + i\frac{dH(t)}{dt},
\label{eq:YtimeDeriv}
\end{equation}
a differential equation which, of course, must be solved subject to the
(collision) boundary condition $Y_{\rm col}(t\to -\infty)=\hat{Y}_d$.
For the set of operators $\{\hat{c}^{\dagger}_{\lambda_{P=L/R/C}}\}$ which 
(at $t\to -\infty$) describes creation of single-particle $H_d$ eigenstates 
(of single-particle energies $\epsilon_{\lambda_{P=L/R/C}}$,)
we can formally establish the time-evolution in the collision picture 
\begin{equation}
\hat{c}^{\dagger}_{\lambda_P} \longrightarrow {\hat{\psi}}_{\lambda_P}^{\dagger}(t) = 
\mathcal{U}(t,-\infty)\hat{c}^{\dagger}_{\lambda_P} \mathcal{U}^{\dagger}(t,-\infty).
\label{eq:createSPsolve}
\end{equation}
In turn, the formal solution (\ref{eq:createSPsolve}) provides an explicit 
determination of the operator for the emerging Gibbs free energy operator
\begin{equation}
	\hat{Y}_{\rm col}(t)=\sum_{\lambda_{P=L/R}} 
\mu_P \hat{\psi}_{\lambda_P}^{\dagger}(t) \hat{\psi}_{\lambda_P}(t).
\label{eq:YCOLconstruct}
\end{equation}

The exact nonequilibrium Gibbs free energy value is calculated from (\ref{eq:YCOLconstruct})
and from the uniquely specified nonequilibrium solution density matrix,
\begin{equation}
Y_{\rm col}(t) = 
\frac{\hbox{Tr}\{\hat{Y}_{\rm col}(t) \hat{\rho}_{\rm col}(t) \}}
{\hbox{Tr}\{\hat{\rho}_{\rm col}(t)\}}.
\label{eq:GibbsEnergyContentWithTimeDef}
\end{equation}

For problems with a time-independent scattering potential $v_{\rm sc}(\mathbf{r})$,
and when an actual steady-state transport results, the emerging Gibbs free energy 
$\hat{Y}_{\rm sc}$ must coincide exactly with the ReNQSM redistribution operator,
\begin{equation}
\hat{Y}_{\rm sc}=\hat{Y}_{\rm LS}.
\end{equation}
The time evaluation is formally described by
\begin{equation}
\hat{Y}_{\rm sc}(t) = \mathcal{U}(t,-\infty)\hat{Y}_d \mathcal{U}^{\dagger}(t,-\infty),
\label{eq:Yscresult}
\end{equation}
but it is alone the many-body Hamiltonian [$H\equiv H(t=0)$] and the open boundary conditions which
determine the properties of this emerging Gibbs free energy. The formal time dependence is 
therefore limited to the adiabatic turn on itself,
\begin{equation}
\hat{Y}_{\rm sc} \approx \hat{Y}_d + \exp(\eta t) [\hat{Y}_{\rm sc}(t=0)-\hat{Y}_d].
\end{equation}
There is in steady-state a cancellation of the time-dependence in the creation and destruction 
operators in (\ref{eq:Yscresult}). As a consequence,  the resulting Gibbs free energy is characterized 
by a time evolution
\begin{equation}
i\frac{d\hat{Y}_{\rm sc}}{dt} \propto \eta,
\label{eq:Yscderiv}
\end{equation}
which is identical to that derived for the steady-state electron-redistribution operator 
$\hat{Y}_{\rm LS}$ derived and analyzed by Hershfield and Schiller \cite{NEQreform,exactKondo}.

\subsection{Total-internal energy and state renormalization}

When dealing with a fully time-dependent problem we obtain the exact total internal energy 
directly from the nonequilibrium density-matrix solution
\begin{equation}
U_{\rm col}(t) = 
\frac{\hbox{Tr}\{H \hat{\rho}_{\rm col}(t) \}}
{\hbox{Tr}\{\hat{\rho}_{\rm col}(t)\}}.
\label{eq:totalInternalEnergyContentWithTimeDef}
\end{equation}
This expression has just a trivial denominator, 
$\hbox{Tr}\{\hat{\rho}_{\rm col}(t)\}=\hbox{Tr}\{\hat{\rho}_d\}$, because of the unitary 
character of the time-evolution operator, and the conserving nature of the nonequilibrium solution
density matrix $\hat{\rho}_{\rm col}(t)$.

It is instructive to discuss an apparent --- but not actual --- difference which arises in 
the evaluation of steady-state thermodynamic quantities depending on whether one retains the 
full nonequilibrium time dependence or whether one exploits the ReNEQSM [and the connection to formal many-body LS 
solution via (\ref{eq:rhod-expandLS})]. The apparent differences are most clearly evident in 
a discussion of a system-specific evaluation of the steady-state 
total internal energy, 
\begin{equation}
U_{\rm LS} = 
\frac{\hbox{Tr}\{H\, \hat{\rho}_{\rm LS} \}}
{\hbox{Tr}\{\hat{\rho}_{\rm LS}\}}.
\label{eq:totalInternalEnergyContentSteadyStateDef}
\end{equation}
Because the solution density matrix $\hat{\rho}_{\rm LS}$
commutes \cite{NEQreform} with the fully interacting Hamiltonian ($H$),
it is tempting (but wrong) to conclude that the set of initial $t\to -\infty$ 
disconnected equilibria eigenstates $E_\xi$ 
provides a complete specification of $U_{\rm LS}$. 

For a steady-state tunneling problem there can be no difference between
the formally exact evaluation (\ref{eq:totalInternalEnergyContentWithTimeDef})
as a time-dependent problem and the also exact steady-state 
description (\ref{eq:totalInternalEnergyContentSteadyStateDef}). 
An apparent difference arises only because the formal many-body LS solution to 
the collision problem implies a projection (\ref{eq:LSformaltime}) of the 
full dynamics onto the eigenstates of the ($t\approx 0$) 
Hamiltonian \cite{LangrethLS,LippSchwing,Gellmann}.  In a formal 
scattering theory description it is, of course, essential to include a description
of both the level shifts \cite{Pirenne,Gellmann} and a possible collision-state 
renormalization \cite{LangrethLS,DeWitt}.

\section{Variational nonequilibrium thermodynamic grand potential}

The introduction provides a summary of the central enabling result of this paper, namely the 
description of the nature and properties of the exact nonequilibrium thermodynamic grand potential. Appendix
C contains a formal proof of the variational properties, essentially adapting the analysis
of Mermin \cite{Mermin65}. Here I motivate and detail the recast of the exact QKA
as a variational thermodynamics theory while also emphasizing central results.

\subsection{Notation and motivation}

The evaluation of expectation values [for a general operator $\hat{A}(t)$] 
is viewed as a functional 
\begin{equation}
\langle \hat{A}(t) \rangle_{\hat{\rho}(t)} 
\equiv \frac{\hbox{Tr}\{\hat{\rho}(t) \hat{A} \}}{\hbox{Tr}\{\hat{\rho}(t)\}}
\label{eq:GenRhoExpectDef}
\end{equation}
of a general nonequilibrium, time-dependent density matrix $\rho(t)$.  

For a general nonequilibrium density matrix (electron distribution) there
exists a natural functional evaluation of the entropy
\begin{equation}
S[\hat{\rho}(t)]  = 
-\langle \ln \hat{\rho}(t) \rangle_{\hat{\rho}(t)}
+\langle \ln [\hbox{Tr}\{\hat{\rho}(t)\}] \rangle_{\hat{\rho}(t)}.
\label{eq:Sfunctional}
\end{equation}
Similarly, the evaluations of the nonequilibrium internal 
energies and Gibbs free energy,
\begin{eqnarray}
U_{\rm col}[\hat{\rho}(t)]& = & \langle H(t) \rangle_{\hat{\rho}(t)},
\label{eq:Einternal}\\
Y_{\rm col}[\hat{\rho}(t)]& = & \langle \hat{Y}_{\rm col}(t) \rangle_{\hat{\rho}(t)},
\label{eq:Yenergy}
\end{eqnarray}
are set directly by the Hamiltonian $H(t)$ and by the emerging Gibbs free
energy operator $\hat{Y}_{\rm col}(t)$.

The nonequilibrium evaluations of entropy, internal energy, and Gibb free energy 
functionals suggest, in turn, a functional form for the nonequilibrium grand potential:
\begin{equation}
\Omega_{\rm col}(t) [\hat{\rho}(t)]=
-\frac{1}{\beta}S[\hat{\rho}(t)] + U_{\rm col}[\hat{\rho}(t)] - Y_{\rm col}[\hat{\rho}(t)].
\label{eq:OmegaFunctConstruct}
\end{equation}
This expression is a generalization 
(\ref{eq:NEQGrandPotFunct}) of Mermin's (equilibrium) result to a 
nonequilibrium thermodynamic grand potential for collision problems. The 
formulation (\ref{eq:OmegaFunctConstruct}) is relevant 
for tunneling systems described by the continuum Caroli partition scheme.

For a compact presentation of derivatives and forces
it is convenient to view the nonequilibrium thermodynamic grand potential functional 
(\ref{eq:NEQGrandPotFunct}) as a special (collision) 
instance of the more general functional
\begin{equation}
\Omega_{\tilde{X}(t)}[\rho(t)]\equiv 
\langle \tilde{X}(t) + \beta^{-1} \ln \hat{\rho}(t)
- \beta^{-1} \ln [\hbox{Tr}\{\hat{\rho}(t)\}] \rangle_{\hat{\rho}(t)}.
\label{GenNEQGrandPot}
\end{equation}
Obviously, it holds that
\begin{equation}
\Omega_{\rm col}[\hat{\rho}(t)] = 
\Omega_{\tilde{X}(t)\equiv H(T)-\hat{Y}_{\rm col}(t)}[\hat{\rho}(t)].
\end{equation}

\subsection{Variational thermodynamics} 

A simple generalization of Mermin's analysis (appendix C) demonstrates that the 
nonequilibrium thermodynamics grand potential 
(\ref{eq:NEQGrandPotFunct}) is extremal at the exact nonequilibrium collision 
density matrix $\hat{\rho}_{\rm col}(t)$:
\begin{equation}
\Omega_{\rm col}[\hat{\rho}(t)]  > 
\Omega_{\rm col}[\hat{\rho}_{\rm col}(t)] =  \Omega_{\rm col}(t).
\label{eq:OmegaExtremal}
\end{equation}
That is, the grand potential functional acquires the expected minimum 
(\ref{eq:NEQGrandPotVal}) as the extremal value and thus serves both to identify the
exact nonequilibrium grand potential value (\ref{eq:NEQGrandPotVal}) and the exact 
nonequilibrium density matrix for interacting tunneling problems. A Mermin-type variational 
principle (\ref{eq:OmegaExtremal}) for nonequilibrium thermodynamic theory is, of course, 
expected given the ReNQSM nature, i.e., the reformulation nonequilibrium tunneling as
an effective equilibrium statistical-mechanics problem \cite{NEQreform,exactKondo,NEQunified,NEQaverageEvolution,DoyonAndrei,Han,Dutt}.
The (generalized) Mermin thermodynamics variational principle \cite{Mermin65,Araki} supplements
the ground state, action, and the many-body LS T-matrix principles. Variational principles
are indispensable in DFT formulations.

The variational property (\ref{eq:OmegaExtremal}) is proved (in appendix C) for 
members, $\hat{\rho}$,  of the class of normalized nonequilibrium density matrices with a 
statistical weighting given by a Hermitian operator 
$\hat{X}(t)$ and normalization $N_{\hat{X}(t)}$,
\begin{eqnarray}
	\hat{\rho} & \in & \{ N_{\hat{X}(t)}
	\hat{\rho}_{\hat{X(t)}}\},\\ 
\hat{\rho}_{\hat{X(t)}}  
&\equiv & \exp[-\beta \hat{X}(t)],
\label{eq:rhoXdefine}\\
N_{\hat{X }(t)} & = & \hbox{Tr} \{
	\hat{\rho}_{\hat{X(t)}}\}.
\end{eqnarray}
The conservative nature of the QKA is, of course, sufficient
to ensure that the normalization is independent of time; when 
using the formal many-body LS solution for steady state problems one must
explicitly ensure normalizations \cite{Gellmann,DeWitt,LangrethLS}.

The density-matrix class $\{N_{\hat{X}(t)}\hat{\rho}_{\hat{X}(t)}\}$
encompasses all possible nonequilibrium density matrices $\hat{\rho}_{\rm col}(t)$ 
that emerge (with conserving dynamics) in a general 
collision problem, Eqs.~(\ref{eq:GWdefine}) and (\ref{eq:rhoGWdefine}).
The extremal property is therefore sufficient to uniquely identify the correct nonequilibrium 
collision density matrix (\ref{eq:rhoColdefine}) for a given (interacting) tunneling 
problem specified as continuum Caroli partition scheme.

From the Mermin-type analysis one immediately finds a stationary property of
\begin{equation}
\Omega_{\tilde{X}(t)}\equiv \Omega_{\tilde{X}(t)}[\hat{\rho}_{\tilde{X}(t)}]
\end{equation}
which can be formulated from the functional expression
\begin{eqnarray}
\left. \frac{\partial \Omega_{\hat{X}_0(t)+\lambda \hat{\Delta}(t)}}
{\partial \lambda}\right|_{\lambda=0}
 & = &
\langle \left. \frac{\partial}{\partial \lambda} \hat{\rho}_{\hat{X}_0(t)+\lambda \hat{\Delta}(t)} 
\right|_{\lambda=0} \rangle_{\hat{\rho}_{\hat{X}_0(t)}}
\nonumber\\
& = & \langle \hat{\Delta}(t) \rangle_{\hat{\rho}_{\hat{X}_0(t)}}.
\label{eq:metaGenHFtheorem}
\end{eqnarray}
As in the corresponding equilibrium case \cite{Mermin65}, this property
is central to demonstrating the variational and extremal properties of the
thermodynamic grand potential at the solution density matrix $\hat{\rho}_{\rm col}(t)$.

\subsection{Extremal property as a tempered maximization of entropy}

The nonequilibrium solution density matrix $\hat{\rho}_{\rm col}(t)$ is unique and can be identified 
by the minimum of the variational thermodynamical grand potential functional. 
From the construction of the thermodynamic potential it follows that
\begin{equation}
\frac{1}{\beta} \frac{\delta}{\delta \hat{\rho}} S[\hat{\rho}] = - \frac{\delta}{\delta \hat{\rho}} 
\left\{ \Omega_{\rm col}(t) [\hat{\rho}] - (U_{\rm col}(t)[\hat{\rho}] - Y_{\rm col}(t)[\hat{\rho}]) \right\},
\label{eq:derivSwithChange}
\end{equation}
and that, following a system perturbation, $\hat{\rho}(t)\neq 
\hat{\rho}_{\rm col}(t)$, we expect a relaxation because it corresponds to
a maximization of electron entropy. This maximization is naturally limited or tempered 
by a functional for the total-internal-energy and Gibbs-free energy difference, 
$U_{\rm col}(t)[\hat{\rho}] - Y_{\rm col}(t)[\hat{\rho}]$. 

The maximum-entropy character of the here-presented theory 
makes it clear that there are close relations to the range of constrained maximum entropy 
formulations \cite{NgCurConstraint,HeinonenJohnsonCurConstraint,GodbyCurConstraint}.
However, the are also important differences. While the constrained maximum entropy 
approaches use Lagrange multipliers to specify selected expectation values
(for example, the current flow), the explicit need for constraints is bypassed in the 
ReNQSM and in this variational thermodynamic description. Specifically, I stress
that the approach does not impose a value for the internal energy or Gibbs free
energy and that $U_{\rm col}(t)[\hat{\rho}] - Y_{\rm col}(t)[\hat{\rho}]$
is itself uniquely specified by the boundary conditions and by the actual time evolution. 

\subsection{Generalized Hellmann-Feynman theorem}

The fact that the exact nonequilibrium solution can be viewed as a maximization of entropy 
provides a rationale for introducing generalized thermodynamic forces 
\begin{equation}
\mathbf{F}^{\rm GCE}_{{\rm col},{\hat{\rho}}} \equiv - \frac{\partial}{\partial \hat{\rho}} \Omega_{\rm col}(t)[\hat{\rho}].
\end{equation}
The more specific thermodynamic-force definition
\begin{equation}
\mathbf{F}^{\rm GCE}_{{\rm col},{\mathbf{R}_i}}(t) \equiv - \frac{\partial}{\partial \mathbf{R}_i} 
\Omega_{\rm col}(t)
\end{equation}
is given by a parametric dependence of the density matrix (as will be
further motivated for steady-state problems in the following section). 
The combined electron-nuclei system is here simply assumed to
further the maximization of the entropy content (in the electron system).

A generalized Hellmann-Feynman theorem exists to simplify the determination of such
thermodynamic forces, here as discussed for time-dependent tunneling.
As an explicit expression of (\ref{eq:metaGenHFtheorem}), one
can specify the perturbation
\begin{equation}
\hat{\Delta}(t) = \frac{\partial}{\partial \mathbf{R}_i} \hat{X}_{\rm col}(t) \cdot \delta \mathbf{R}_i
\end{equation}
and evaluate the changes corresponding to a derivative of the thermodynamic potential along a 
coordinate-line element $\delta \mathbf{R}_i$:
\begin{equation}
\delta \mathbf{R}_i \cdot \frac{\partial}{\partial \mathbf{R}_i} \Omega_{\rm col}(t) =
\delta \mathbf{R}_i \cdot \langle \frac{\partial \hat{X}_{\rm col}}{\partial \mathbf{R}_i} 
\rangle_{\hat{\rho}=\rho_{\rm col}(t)}.
\end{equation}
This provides a demonstration of a nonequilibrium generalized Hellmann-Feynman, 
\begin{equation}
\frac{\partial}{\partial \mathbf{R}_i} \Omega_{\rm col}(t) =
\langle \frac{\partial}{\partial \mathbf{R}_i} \left[ H(t) - \hat{Y}_{\rm col}(t) \right]
\rangle_{\rho_{\rm col}(t)},
\label{eq:GenHFtheorem}
\end{equation}
for a thermodynamic description of tunneling.

\section{Steady-state thermodynamics and adiabatic transformations}

The formulation in terms of a nonequilibrium grand potential functional (above) is
by construction fully equivalent to the exact solution in the QKA \cite{Langreth}.
This formal results can also be used to investigate tunneling described by a
time-independent potential and subject to the assumption that a steady-state 
solution (described by the many-body LS equation) has emerged. There does exist an 
argument that a steady-state solution will emerge under general 
conditions\footnote{Tunneling in a system described by a time-independent scattering potential 
need not always have stationary electron dynamics. As also discussed in Ref.~\protect\onlinecite{LSCDFT}, 
a time-dependent tunneling behavior can arise through an intermittent behavior, 
for example, in an ordinary avalanche diode.} \cite{DoyonAndrei}.
The following steady-state analysis provides a framework for a discussion
of adiabatic relaxations and a nonequilibrium Born-Oppenheimer approximation. 

\subsection{State functions for steady-state tunneling}

If all nuclei coordinates are changed at an infinitely slow rate along a given 
deformation path, it follows that the system can always
be assumed to be described by the exact steady-state density matrix solution
$\hat{\rho}_{\rm LS}(\{\mathbf{R}_i\})$. As indicated, this solution
is then a function of the set of classical-coordinate values as they
vary along the deformation path. The density matrix solution 
specifies in turn an exact evaluation of corresponding thermodynamics state functions
\begin{eqnarray}
\bar{U}_{\rm LS}(\mu_L, \mu_R; \{\mathbf{R}_i \})
&\equiv & U_{\rm LS}[\hat{\rho}_{\rm LS}],
\label{eq:EStatFunct}\\
\bar{Y}_{\rm LS}(\mu_L, \mu_R; \{\mathbf{R}_i \})
&\equiv & Y_{\rm LS}[\hat{\rho}_{\rm LS}],
\label{eq:YStatFunct}\\
\bar{S}_{\rm LS}(\mu_L, \mu_R; \{\mathbf{R}_i \})
&\equiv & S[\hat{\rho}_{\rm LS}],
\label{eq:MSStatFunct}\\
\bar{\Omega}_{\rm LS}(\mu_L, \mu_R; \{\mathbf{R}_i \})
&\equiv & \Omega_{\rm LS}[\hat{\rho}_{\rm LS}],
\label{eq:OmStatFunct}
\end{eqnarray}
namely for the internal energy, for the Gibbs free energy, for the entropy,
for the nonequilibrium grand potential, respectively.  

\subsection{Steady-state thermodynamic forces}

From the state functions and using the generalized Hellmann-Feynman of the variational nonequilibrium thermodynamics
theory, one immediately obtains the thermodynamics forces 
\begin{equation}
\mathbf{F}^{\rm GCE}_{{\rm LS},\mathbf{R}_i} \equiv - \frac{\partial}{\partial \mathbf{R}_i}
\bar{\Omega}_{\rm LS}(\mu_L,\mu_R; \{\mathbf{R}_i\}).
\label{eq:NTForceAdiabRestate}
\end{equation}
I find (as expected) that the GCE conditions invalidates a traditional Hellmann-Feynmann evaluation
\begin{eqnarray}
\frac{\partial}{\partial \mathbf{R}_i} \bar{U}_{\rm LS}
& = & 
\langle \frac{\partial H}
{\partial \mathbf{R}_i} \rangle_{\hat{\rho}_{\rm LS}} +
\nonumber\\
& & 
\frac
{\hbox{Tr}\{(H-\bar{U}_{\rm LS}) \frac{\partial}{\partial\mathbf{R}_i}
e^{-\beta (H-\hat{Y}_{\rm LS})}\}}
{\hbox{Tr}\{e^{-\beta (H-\hat{Y}_{\rm LS})}\}}
\\
& \neq & 
\langle \frac{\partial}{\partial \mathbf{R}_i} H \rangle_{\hat{\rho}_{\rm LS}}.
\label{eq:noHFeint}
\end{eqnarray}
There is a corresponding complication in a GCE evaluation of the
Gibbs free energy
\begin{eqnarray}
\frac{\partial}
{\partial \mathbf{R}_i} 
\bar{Y}_{\rm LS}
& = & \langle \frac{\partial
\hat{Y}_{\rm LS} 
}{\partial \mathbf{R}_i} 
\rangle_{\rho_{\rm LS}} + \nonumber\\
& & \frac
{\hbox{Tr}\{(\hat{Y}_{\rm LS}-\bar{Y}_{\rm LS}) \frac{\partial}{\partial\mathbf{R}_i}
e^{-\beta (H-\hat{Y}_{\rm LS})}\}}
{\hbox{Tr}\{e^{-\beta (H-\hat{Y}_{\rm LS})}\}}
\\
& \neq & 
\langle \frac{\partial}{\partial \mathbf{R}_i} \hat{Y}_{\rm LS} \rangle_{\hat{\rho}_{\rm LS}}.
\label{eq:noHFgfe}
\end{eqnarray}
and the exact steady-state generalized Hellmann-Feynman force description (\ref{eq:NTForceAdiabRestate}) arises only upon cancellations.

The steady-state thermodynamics forces (\ref{eq:NTForceAdiabRestate}) are useful for they
serve to trace out changes in the thermodynamics grand potential value among different 
possible nuclei configurations \textit{as well as} concerted changes in other extensive thermodynamics
functionals (\ref{eq:EStatFunct})--(\ref{eq:MSStatFunct}).  I also stress that integrating the thermodynamics
grand potential changes from the force expression (\ref{eq:NTForceAdiabRestate}) --- and using the extremal
character of the thermodynamics grand potential --- will in turn uniquely identify the solution
nonequilibrium density matrix at the final coordinates.  Sec.~VI.D contains an argument for using these 
steady-state thermodynamics forces in a nonequilibrium Born-Oppenheimer approximation.

\subsection{Relation to electrostatic forces}

\textit{I observe,} that the thermodynamics forces (\ref{eq:NTForceAdiabRestate}) are formally 
different from the electrostatic forces often used in nonequilibrium studies.  By a natural assumption of
ergodicity it follows that the steady-state description must also reflect 
fluctuations \cite{EarlyNoise,TunnelingNoise,RungeNoise} and that the tunneling
systems must be treated as a statistical ensemble (even at $\beta \to 0$).
Electrostatic forces will only be interesting in the context of a GCE thermodynamics if (\ref{eq:esForceFirstDef}) 
provides an evaluation which is representative for this ensemble. However, the source of an additional force contribution 
with a statistical origin (for example, the electron-redistribution effects arising in open 
systems \cite{NEQreform,exactKondo,HarrisFriedelInteract,NEQunified}) is directly evident in the 
exact ReNQSM result for an effective (unnormalized) steady state nonequilibrium density matrix \cite{NEQreform}
(\ref{eq:rhoReNQSM}).  This exact nonequilibrium density matrix depends on the operator $\hat{Y}_{\rm LS}$ for 
interaction-induced redistributions of electrons, i.e., the emerging Gibbs free energy operator.
Accordingly, it is important that the forces that effectively act on this ensemble (of steady-state nonequilibrium 
tunneling systems) reflect the sensitivity of this electron redistribution on the nuclei 
position.\footnote{The steady-state ReNQSM \protect\cite{NEQreform,NEQreformDetails} 
contains an explicit construction of the operator $\hat{Y}_{\rm LS}$ order-by-order in the perturbation term
(that includes the external potential). This construction strongly suggests that it is essential to retain
an explicit dependence on the nuclei positions in the resulting operator $\hat{Y}_{\rm LS}$ (under nonequilibrium conditions).}

In a GCE thermodynamic description, the classical nuclei still respond partly to electrostatic 
forces but the adiabatic relaxations (of classical nuclei) will \textit{also} reflect 
a tendency to seek thermodynamically favorable electron distributions \cite{NEQreform};  
these distribution changes are specified in part by a correct evaluation of the Gibbs 
free energy effects. The analysis of Ref.~\onlinecite{NEQreform} shows that there is no 
change in the Gibbs free energy operator $\hat{Y}_{\rm LS}$ in equilibrium. However, that does 
not generally apply under general nonequilibrium conditions where the operator $\hat{N}_{L(R)}$ for counting 
particles in either lead can no longer be treated as as a conserved property.  An assumption that 
$\partial \hat{Y}_{\rm LS}/\partial \mathbf{R}_i$ vanish would significantly simplify the formulation 
of density functionals (Sec. VII) but such assumption is not motivated by the steady-state ReNQSM \cite{NEQreform}.

\textit{I also observe,} that the Green function framework (Section II.B and appendix A) and the S-matrix 
framework (section II.B and appendix B) for calculations of finite-bias noninteracting thermodynamics suggest that 
we must generally expect differences in the electrostatic and thermodynamics forces.
Appendix A provides a formal nonequilibrium Green function demonstration that the exact adiabatic thermodynamics forces 
(\ref{eq:NTForceAdiabRestate}) does reduce to GCE electrostatic forces \textit{when} the changes in 
the full electron distribution would be completely specified by the changes in full density of states.  
However, that is an assumption which does not generally hold outside equilibrium (for example as discussed 
in Refs.~\onlinecite{KondoPRL}, \onlinecite{RateEq}, and \onlinecite{Phon}).

Similarly, a S-matrix analysis (appendix B) links the electrostatic forces and the thermodynamics forces to two
different components of a scattering phase-shift characterization. An assumption that changes in 
the full density of states determine the full electron distribution implies that the
GCE thermodynamics potential is specified by 
\begin{equation}
\Delta \mathcal{N}_{\rm tot}(\omega) = \frac{2}{\pi}
[\delta_{\rm L}^{\rm eff}(\omega) + \delta_{\rm L}^{\rm eff}(\omega)]
\equiv \frac{2}{\pi} \delta_{\rm tot}^{\rm eff}(\omega).
\label{eq:NtotLS:phaseEvaluation}
\end{equation}
The partial-pressure determination \cite{NEQQSMthermodynamics} of the nonequilibrium systems yields instead lead-specific 
thermodynamics grand potential components given by $\delta_{\rm L}^{\rm eff}(\omega)$ for $\mu_{\rm L} \gg \mu_{\rm R}$.
Appendix B
contains an example analysis for a one-dimensional transport description, based on Ref.~\onlinecite{Albert1D}. In this
model study one can identify  the lead-projected [full, unprojected] phase-shift component $\delta_{\rm L}^{\rm eff}(\omega)$ 
[$\delta_{\rm tot}^{\rm eff}(\omega)$] as the phase shift for free-particle reflection into the emitter lead 
[transmission into the collector].  There is no general nature of the S-matrix behavior which would ensure 
equivalence of thermodynamics forces (given by the reflection phase shift) and electrostatic forces 
(given by the transmission phase shift).

There are thus several arguments for a formal and actual difference between thermodynamics and electrostatic
forces under nonequilibrium conditions (beyond linear response). The most important argument for a difference is, 
of course, this fact: while there are argument for a nonconservative nature of the electrostatic forces, 
this paper shows that the adiabatic thermodynamics forces will always remain explicitly conservative. 

\subsection{Nature of steady-state adiabatic forces}

The steady-state thermodynamics forces (\ref{eq:NTForceAdiabRestate}) 
(like the general variation grand potential) expresses a 
controlled entropy maximization 
\begin{equation}
\frac{1}{\beta} \frac{\delta}{\delta \hat{\rho}} S[\hat{\rho}] = - \frac{\delta}{\delta \hat{\rho}} 
\left\{ \Omega_{\rm LS} [\hat{\rho}] - (U_{\rm LS}[\hat{\rho}] - Y_{\rm LS}[\hat{\rho}]) \right\}.
\label{eq:derivSwitchChange}
\end{equation}
The optimization arises subject to rigorous boundary conditions, namely, the QKA results for the internal-energy and Gibbs-free-energy contents.  
This suggests an intrinsic value for modeling relaxations.

One can, for example, use the exact thermodynamics forces (\ref{eq:NTForceAdiabRestate})
to discuss the thermodynamics stability of a steady-state tunneling system in the limit of infinitely slow deformations.
The initial state (before the transformation around a loop) is uniquely characterized by the solution density matrix 
$\hat{\rho}_{\rm LS}$ and potential value $\bar{\Omega}_{\rm LS}$. Following any transformation around a closed loop
we will in general obtain a new density matrix $\hat{\rho}^f$ and a new thermodynamic potential value 
$\Omega_{\rm LS}^f$. However, with the underlying state-function assumption of infinitely slow deformation 
it follows that $\Omega_{\rm LS}^{f} = \Omega_{\rm LS}[\rho^f]= \bar\Omega_{\rm LS}$, and
that we have must returned to the exact same density matrix, $\hat{\rho}^f\equiv \hat{\rho}_{\rm LS}$.
Not surprisingly, the conserving nature of the thermodynamic forces ensures that all state functions
must attain the exact same value as the initial state functions, 
\begin{equation}
\delta \bar{S}_{\rm LS}=\delta \bar{U}_{\rm LS}=\delta \bar{Y}_{\rm LS} = \delta \bar{\Omega}_{\rm LS} \equiv 0.
\end{equation}
Appendix E provides additional formal results on relations among variations of the set of 
thermodynamics state functions.

The extremal property of $\bar{\Omega}_{\rm LS}$ ensures, in fact, that the exact steady-state thermodynamics 
forces (\ref{eq:NTForceAdiabRestate}) are inherently adiabatic in nature, i.e., valid and needed
in the limit where we can assume an infinite ratio of the nuclei-to-electron masses. This adiabatic nature 
follows because a sufficiently slow implementation will automatically avoid all thermodynamics excitations.

The basis in a physics principle and the automatic ability to avoid all thermodynamics excitations
suggest that the here derived conservative thermodynamics forces are well suited for use in a nonequilibrium Born-Oppenheimer approximation.

\section{Variational thermodynamics for noninteracting tunneling}

The paper develops a DFT formulation with a corresponding a SP framework for efficient 
computations.  This involves a self-consistency loop where SP determinations of the 
electron density is used to update the universal functional description. It is therefore
important to demonstrate that effective SP calculations (of the density) are uniquely 
defined within the (full nonequilibrium interacting) variational thermodynamics theory.

\subsection{The independent-particle dynamics}

For noninteracting particles, described by the quadratic Hamiltonian
$H^{(0)}_V$, it is clear that the set of SP LS solutions \cite{exactKondo}
\begin{eqnarray}
|\psi_{{\rm LS},\lambda_P}^{(0)}\rangle & = & (\hat{\psi}_{{\rm LS},\lambda_P}^{(0)})^{\dag} |0\rangle
\label{eq:LSipSolWav}
\\
(\hat{\psi}_{{\rm LS},\lambda_P}^{(0)})^{\dag} & = & \frac{i\eta}{\epsilon_{\lambda_P}+\mathcal{L}_{H_V^{(0)}} +i\eta}
\label{eq:LSipSolOp}
\end{eqnarray}
coincides with the SP LS solutions introduced and discussed in Sec.~II and in Appendix A. 
It is also clear that these 
SP LS states constitutes  the basis of what is formally the many-body solution for 
a steady-state tunneling problem \cite{NEQreform,exactKondo,DoyonAndrei,NEQunified,NEQaverageEvolution,Han,Dutt,NEQreformDetails,NEQreformWithTime}.
The states (\ref{eq:LSipSolWav}) express the (steady-state) solution that emerge from
an initial $H_d$ eigenstate $|\phi_{\lambda_P}\rangle$ in formal LS collision theory \cite{AlbertXIX}.
There can be a finite shift \cite{Gellmann} of the energy $\epsilon_{\lambda_P}$ if the perturbation 
introduces an infinitely extended periodic potential \cite{Pirenne,DeWitt}; as always in formal 
scattering theory, there will be infinitesimal energy shifts leading to changes 
in the integrated density of states \cite{FriedelGrimley,FriedelTed,HarrisFriedelInteract} that are 
stipulated by the Friedel sum rule \cite{FriedelSum}. I do not (yet) express these energy shifts so 
as to keep the nomenclature simple in the following analysis.  

The states and the their creation operators, 
(\ref{eq:LSipSolWav}) and \ref{eq:LSipSolOp}, retain
a genuine \textit{independent-particle} nature and hence the emerging Gibbs free energy, 
denoted $\hat{Y}_V^{(0)}$ is also quadratic.  It is also clear that these SP solutions must 
enter both the nonequilibrium solution density matrix and the thermodynamics value.  In particular, it 
is possible to obtain a separable form of the nonequilibrium solution density matrix
\begin{eqnarray}
\hat{\rho}_{\rm LS}^{(0)} & = & \Pi_{\lambda_P} 
\hat{\rho}_{{\rm LS},\lambda_P}^{(0)}, \\
\hat{\rho}_{{\rm LS},\lambda_P}^{(0)} & = &
\exp[-\beta(\epsilon_{\lambda_P}-\mu_P)\hat{n}_{\lambda_P}],\\
\hat{n}_{\lambda_P} & = & 
(\hat{\psi}_{{\rm LS},\lambda_P}^{(0)})^{\dag}
\hat{\psi}_{{\rm LS},\lambda_P}^{(0)},
\end{eqnarray}
and of the corresponding nonequilibrium thermodynamical grand potential.

\subsection{Variational link to the single-particle
Lippmann-Schwinger solution}

For noninteracting transport it is sufficient to limit the variational space (for the nonequilibrium solution density matrix) to that 
produced by a corresponding set of mutually orthogonal trial functions $\tilde{\psi}_{\lambda_P}(\mathbf{r})
= \langle \mathbf{r} | \tilde{\psi}_{\lambda_P}\rangle$. These trial functions approximate the form a SP 
scattering state from $H_d$ eigenstates of energy $\epsilon_{\lambda_P}$. No assumption is made for the normalization of these 
trial scattering wavefunctions whereas I use $\{\hat{\tilde{\psi}}^{\dag}_{\lambda_P}\}$ to denote creation 
operators for such states with a suitable standard (delta-function) normalization
\begin{equation}
\hat{\tilde{\psi}}^{\dag}_{\lambda_P} |0\rangle = 
\frac{| \tilde{\psi}_{\lambda_P}\rangle} 
{\langle \tilde{\psi}_{\lambda_P}| 
\tilde{\psi}_{\lambda_P}\rangle^{1/2}}.
\end{equation}
I also introduce $\hat{\tilde{n}}_{\lambda_P}= \tilde{\psi}_{\lambda_P}^{\dag} \tilde{\psi}_{\lambda_P}$
and define the trial density matrix solution
\begin{eqnarray}
\hat{\tilde{\rho}} & = & \Pi_{\lambda_P} 
\hat{\tilde{\rho}}_{\lambda_P}, \\
\hat{\tilde{\rho}}_{\lambda_P} & = &
\exp[-\beta(\epsilon_{\lambda_P}-\mu_P)\hat{\tilde{n}}_{\lambda_P}],
\nonumber\\
& = & 1 + \hat{\tilde{n}}_{\lambda_P} \left(\exp[-\beta(\epsilon_{\lambda_P}-\mu_P)]-1\right).
\end{eqnarray}
I note that the difference between the internal and Gibbs free energies is a SP operator
which can be expressed in terms of the trial wavefunctions 
\begin{eqnarray}
H^{(0)}_V-\hat{Y}_V^{(0)} & = & 
\sum_{\lambda',P'}
\sum_{\lambda_P} 
\frac{\langle \tilde{\psi}_{\lambda',P'} | (H^{(0)}_V-\hat{Y}^{(0)}_V)| \tilde{\psi}_{\lambda_P}\rangle}
{
\langle \tilde{\psi}_{\lambda',P'}| \tilde{\psi}_{\lambda',P'}\rangle^{1/2} \,
\langle \tilde{\psi}_{\lambda_P}| \tilde{\psi}_{\lambda_P}\rangle^{1/2}
}\nonumber\\
& & \times \hat{\tilde{\psi}}_{\lambda',P'}^{\dag}
\hat{\tilde{\psi}}_{\lambda_P}.
\end{eqnarray}

The variation form of the grand potential value is also separable
\begin{eqnarray}
\Omega_V^{(0)}[\hat{\tilde{\rho}}] & = & \sum_{\lambda_P} 
\Omega_{V,\lambda_P}^{(0)}[\hat{\tilde{\rho}}_{\lambda_P}],
\label{eq:SeparableOmega}\\
\Omega_{V,\lambda_P}^{(0)}[\hat{\tilde{\rho}}_{\lambda_P}] & = & 
\Omega^{(0)}_{V,\lambda_P}
+ 
\Delta \Omega_{V,\lambda_P}^{(0)} [\hat{\tilde{\rho}}_{\lambda_P}],
\\
\Omega^{(0)}_{V,\lambda_P} 
& = & - \frac{1}{\beta} \ln \hbox{Tr} \{\hat{\tilde{\rho}}_{\lambda_P}\}\nonumber\\
& = & -\frac{1}{\beta} \ln \left(1+\exp[-\beta(\epsilon_{\lambda_P}-\mu_P)]\right),
\label{eq:Omega0value}
\\
\Delta \Omega^{(0)}_{V,\lambda_P} [\hat{\tilde{\rho}}_{\lambda_P}] & =& 
\frac{
\hbox{Tr}\{\hat{\tilde{\rho}}_{\lambda_P}
[ H_V^{(0)}-\hat{Y}^{(0)}_V + (1/\beta) \ln(\hat{\tilde{\rho}}_{\lambda_P})] \} 
}
{
\hbox{Tr}\{\hat{\tilde{\rho}}_{\lambda_P}\}
}.
\label{eq:DeltaOmega0functional}
\end{eqnarray}
I observe that the components of the grand-canonical potential functional has the minimum
\begin{equation}
\Omega_{V,\lambda_P}^{(0)}[\hat{\tilde{\rho}}_{\lambda_P}=\hat{\rho}_{{\rm LS},\lambda_P}] =
\Omega_{V,\lambda_P}^{(0)}.
\end{equation}
Below, in the spirit of Ref.~\onlinecite{LLipDisc}, I investigate the SP thermodynamics
by considering the variational contributions from each of the orthogonal trial scattering 
eigenstates, $\{\lambda_P\}$, at a time.

It is straightforward to evaluate the terms in the variational separable grand potential functional,
\begin{eqnarray}
\Delta \Omega_{V,\lambda_P}^{(0)}[\hat{\tilde{\rho}}] & = & 
(\epsilon_{\lambda_P}-\mu_P) 
\langle \hat{\tilde{n}}_{\lambda_P} \rangle_{\hat{\tilde{\rho}}_{\lambda_P}}
\left(
C[\tilde{\psi}_{\lambda_P}] -1 
\right), \\
C[\tilde{\psi}_{\lambda_P}] & = & 
\frac
{\langle
\tilde{\psi}_{\lambda_P} |
(H_V^{(0)} -\hat{Y}_V^{(0)})
| \tilde{\psi}_{\lambda_P} 
\rangle}
{(\epsilon_{\lambda_P}-\mu_P) 
\langle 
\tilde{\psi}_{\lambda_P} |
\tilde{\psi}_{\lambda_P} 
\rangle}, 
\label{eq:Cdefine}
\\
\langle \hat{\tilde{n}}_{\lambda_P} \rangle_{\hat{\tilde{\rho}}_{\lambda_P}} & = & 
\frac
{
\hbox{Tr}\{
\hat{\tilde{\rho}}_{\lambda_P} \hat{\tilde{n}}_{\lambda_P}\}
}
{
\hbox{Tr}\{
\hat{\tilde{\rho}}_{\lambda_P}\}
}  \nonumber \\
 & = & \frac
{\exp[-\beta(\epsilon_{\lambda_P}-\mu_P)]}
{1+\exp[-\beta(\epsilon_{\lambda_P}-\mu_P)]} \nonumber \\
& = & \langle \hat{n}_{\lambda_P} \rangle_{\hat{\rho}_{\rm LS}^{(0)}}.
\end{eqnarray}
It follows that the variational nature of the nonequilibrium thermodynamical grand potential can be expressed
\begin{eqnarray}
	\frac{\delta \Omega^{(0)}_V[\hat{\tilde{\rho}}]}{\delta \tilde{\psi}^{*}_{\lambda_P}(\mathbf{r})} & = & 
	\frac{\delta \Omega^{(0)}_{V,\lambda_P}[\hat{\tilde{\rho}}_{\lambda_P}]}
{\delta \tilde{\psi}^{*}_{\lambda_P}(\mathbf{r})}=0,\\
D[\tilde{\psi}_{\lambda_P}]^{-1} 
\frac{\delta \Omega^{(0)}_{V,\lambda_P}[\hat{\tilde{\rho}}_{\lambda_P}]}
{\delta \tilde{\psi}^{*}_{\lambda_P}(\mathbf{r})} & = & 
\langle \mathbf{r} | (H_V^{(0)}-\hat{Y}_V^{(0)}) | \tilde{\psi}_{\lambda_P} \rangle \nonumber\\
& & - C[\tilde{\psi}_{\lambda_P}]  (\epsilon_{\lambda_P}-\mu_P)\langle \mathbf{r} | \tilde{\psi}_{\lambda_P} \rangle,
\label{eq:variationaSPformulation}
\\
D[\tilde{\psi}_{\lambda_P}] & = & \frac
{
\langle \hat{n}_{\lambda_P}\rangle_{\hat{\rho}_{\rm LS}^{(0)}}
}
{\langle 
\tilde{\psi}_{\lambda_P} |
\tilde{\psi}_{\lambda_P} 
\rangle}.
\end{eqnarray}
Since $C[\tilde{\psi}_{\lambda_P}]=1$ (at the solution), it is possible to
express the variational condition (\ref{eq:variationaSPformulation}) simply as
the condition
\begin{equation}
(H_V^{(0)}-\hat{Y}_V^{(0)}) 
|\tilde{\psi}_{\lambda_P} \rangle
= (\epsilon_{\lambda_P}-\mu_P) 
|\tilde{\psi}_{\lambda_P} \rangle.
\end{equation}
This equation is obviously solved by the formal SP LS solution, (\ref{eq:LSipSolWav})
and (\ref{eq:LSipSolOp}).
 
\subsection{Calculation of single-particle properties}

From the variational thermodynamics description of the noninteracting particles we can represent
the full variation of the \textit{noninteracting} Green functions and expectation values of
all SP operators. I exemplify this computational power
with a determination of the electron-density variation.

The initial disconnected system is fully characterized through the retarded Green function
operator $\hat{g}_r^{d}(\omega)=\sum_{J={\rm L/R/C}} \hat{g}_r^{dJ}(\omega)$, given by
Eqs.~(\ref{grddef}) and (\ref{grJdef}), and by the initial electron occupation.
Analytic properties relates the retarded and advanced Green functions [$\hat{g}_a(\omega)=\hat{g}_r^*(\omega)$]
and permits an simple expression also of disconnected equilibria spectral function
(\ref{DEQspectral}). Each of the separate sections are equilibrium and the initial system is thus 
described by an electron-distribution or less-than Green function operator \cite{NEQQSM}
\begin{eqnarray}
	\hat{g}_<^{d}(\omega) & = & \hat{g}_{<,\mu=\mu_J}^{dJ={\rm L/R/C}}(\omega),\\
	\hat{g}_{<,\mu}^{dJ}(\omega) & = &
	f_{\mu}(\omega) i [ \hat{g}_r^{dJ}(\omega)-\hat{g}_a^{dJ}(\omega) ]
\nonumber\\
& = & 
f_\mu(\omega) \sum_{\lambda_J} |\lambda_J \rangle\langle \lambda_J| \delta(\omega-
\epsilon_{\lambda_J} ).
\label{DEQdistrib}
\end{eqnarray}
Here `$\lambda_J$' identifies an eigenstate of energy
$\epsilon_{\lambda_J}$ where resides in the leads or in the center.
This definition is consistent
with the one introduced and used in Section II.B. I have in (\ref{DEQdistrib}) chosen to 
explicitly state the dependence on the (lead-specific) chemical potential, to stress the 
relation with the formulation in Ref.~\onlinecite{NEQQSMthermodynamics}.

After the adiabatic turn on we have the system described by the SP LS solutions  (\ref{eq:LSipSolWav}). The retarded
Green functions for the connected noninteracting system $H^{(0)}$ is given by the Dyson equation \cite{Dyson} as well
as a formal Green function evaluation based on an analysis \cite{Gellmann} of the (SP) LS solution:
\begin{eqnarray}
	\hat{g}^{(0)}_r(\omega) & = &  \sum_{P={\rm L/R}} \hat{g}^{(0),P}_r(\omega),
\label{gr0def}
	\\
	\hat{g}^{(0),P}_r(\omega)
	& = & \sum_{\lambda}\frac{
|\psi_{{\rm LS},\lambda_P}^{(0)}\rangle\langle \psi^{(0)}_{{\rm LS},\lambda_P}|
} 
{\omega-\tilde{\epsilon}_{\lambda,{\rm P}}+i\eta}. 
\label{gr0Pdef}
\end{eqnarray}
The scattering-state poles $\tilde{\epsilon}_{\lambda_P}$ are given by infinitesimal level shifts
relative to original disconnected-system eigenvalues $\epsilon_{\lambda_P}$.  The formulation 
(\ref{gr0def}) applies in transport cases when we can ignore bound states \cite{Gellmann,DeWitt}.
I will not consider possible charging-hysteresis cases here \cite{RahmanDavies}.

From the formal scattering theory results there is a also a link to a full nonequilibrium Green 
function determination for
noninteracting particles.  The Green-theorem boundary terms enters in the 
evaluation of the operator for the noninteracting less-than Green function
\begin{equation}
\hat{g}^{(0)}_<(\omega) = \sum_{P=L/R} [1+\hat{g}^{(0)}_r(\omega) H_{1,V}^{(0)}]
\hat{g}_<^P(\omega)[1+H_{1,V}^{(0)}\hat{g}^{(0)}_a(\omega)].
\label{GFLSequation}
\end{equation}
Inserting (\ref{DEQdistrib}) in (\ref{GFLSequation}) yields 
\begin{eqnarray}
	\hat{g}^{(0)}_<(\omega) & = & \sum_{P={\rm L/R}} \hat{g}^{(0),P}_{<,\mu'=\mu_P}(\omega),\\
	\hat{g}^{(0),P}_{<,\mu'}(\omega) & = & 2\pi f_{\mu'}(\omega) \sum_{\lambda_P} 
	| \psi^{(0)}_{\rm LS,\lambda_P}\rangle\langle \psi^{(0)}_{\rm LS,\lambda_P}| \delta(\omega - \tilde{\epsilon}_{\lambda_P}),
\end{eqnarray}
a sum where each terms corresponds exactly to the contribution from the individual (independent) SP LS solution. 
Again the description here is consistent with that used in Section II.C. The connection 
between the SP LS solutions and the noninteracting Green functions is, of course, well known.

To calculate many SP properties it is practical to seek a real-space representation; here we 
exemplify a determination of the electron-density variation in tunneling. From the SP LS solutions, expressed 
through (\ref{GFLSequation}), one defines
\begin{equation}
g^{(0)}_{<}(\mathbf{r},\mathbf{r}',\omega) \equiv \langle \mathbf{r} | \hat{g}_<^{(0)} (\omega) | \mathbf{r}' \rangle,
\label{gltconstruct}
\end{equation}
and one immediately obtains a determination of, for example, the electron-density variation
\begin{equation}
n(\mathbf{r}) = \int d\omega g_<^{(0)}(\mathbf{r},\mathbf{r},\omega).
\label{fulldensityResNEGF}
\end{equation}
It is sometimes natural to contrast this density variation with the corresponding result
$n^{d}(\mathbf{r})$ for the initial, disconnected system and consider
\begin{equation}
\Delta n(\mathbf{r}) = n(\mathbf{r}) - n^{d}(\mathbf{r}).
\label{DeltadensityResNEGF}
\end{equation}

For simple model systems defined by a double-barrier structure the
tunneling-induced density change (\ref{DeltadensityResNEGF}) includes
a variation in  nonequilibrium resonant-level charging \cite{RahmanDavies,Phon} and
nonequilibrium Friedel density oscillations.\footnote{The Friedel density oscillations are very pronounced in studies of 
	tunneling in two-dimensional systems and evident, for example, in Fig.~10 of 
	Ref.~\protect\onlinecite{PuskaRisto}a and Fig.~2 of Ref.~\protect\onlinecite{PuskaRisto}b.}
With a true \textit{noninteracting-electron} modeling 
(i.e., in the absence of actual many-body scattering) the electron-density change 
$\Delta n(\mathbf{r})$ also includes a position-independent 
component which corresponds to SP transmission.

The robust SP LS solver strategies employed in standard DBT codes \cite{Lang}
codes \cite{Lang,DiVentra,TranSiesta,PuskaRisto,gpawNEGF} excel at
calculation the density variation (\ref{fulldensityResNEGF}) for any given SP potential.
Calculations of the effective density variation (\ref{DeltadensityResNEGF}) is, 
of course, a core step also in an exact nonequilibrium thermodynamics DFT.

\section{Density functional theory of interacting tunneling}

The Mermin-type proof of the variational property (\ref{eq:OmegaExtremal}) of 
the nonequilibrium grand potential does not require, nor by itself provide 
uniqueness of density in nonequilibrium tunneling. 
This is in contrast to the case of thermal properties in equilibrium \cite{Mermin65}.
Nevertheless, the previously formulated LS collision DFT \cite{LSCDFT} and 
open-system time-dependent DFT analysis \cite{ZhengTDDFT} still provide a proof for 
uniqueness of density for collision problems, Appendix D. 

The uniqueness of density proof is here used to define a SP 
formulation for exact calculations of interacting nonequilibrium thermodynamics based on universal 
energy functionals. The formally exact theory contains the set of DBT 
descriptions as a first consistent but also un-quantified approximation.  
The here-presented thermodynamics DFT provides a formal framework for 
systematic improvements beyond DBT.

\subsection{Uniqueness of density in nonequilibrium thermodynamics}

The unitary evolution provides an explicit construction for the mapping
\begin{equation}
\mathcal{N}: v_{\rm col}(\mathbf{r},t)
\longrightarrow n(\mathbf{r},t) =\langle \hat{n}(\mathbf{r}) \rangle_{\rm col}(t).
\label{eq:ExplicitNmap}
\end{equation}
The LS collision DFT proof for uniqueness of density, Ref.~\onlinecite{LSCDFT}, explores 
the time-evolution in the collision picture to prove that $\mathcal{N}$ is unique, 
i.e., that for every $v_{\rm col}(\mathbf{r},t)$ there can be at most one 
time-dependent density. Also, the case of steady-state tunneling resulting from 
the adiabatic turn on is just a particular form of time-dependent behavior and 
the uniqueness of density extends also to cases when the collision term 
reduces to a time-independent scattering potential $v_{\rm sc}(\mathbf{r})$.  
Under the assumption that a steady-state behavior emerges from 
$v_{\rm sc}(\mathbf{r})$ the mapping $\mathcal{N}$ is then instead 
completely described by the steady-state density matrix (\ref{eq:rhoReNQSM}):
\begin{equation}
\mathcal{N}: v_{\rm sc}(\mathbf{r})
\longrightarrow n(\mathbf{r}) =\langle \hat{n}(\mathbf{r}) \rangle_{\rm sc}.
\label{eq:ExplicitNmapSC}
\end{equation}
It follows either by the boundary conditions discussed in Ref.~\onlinecite{NEQreform}
or directly by differentiating the mapping (\ref{eq:ExplicitNmap}) that the
resulting density can have no more than an adiabatic time dependence,
$n(\mathbf{r},t) \sim n(\mathbf{r})\exp(\eta t)$.

This unique mapping between density and external potential allows an explicit construction of density 
functionals for the thermodynamics of nonequilibrium tunneling. I observe that it is possible to choose a detailed form of 
the adiabatic turn on such that it is compatible, even in a stringent mathematical sense, with both the 
requirement of LS collision DFT \cite{LSCDFT} and with the present nonequilibrium thermodynamic description, Appendix D.
A robust, universally applicable, linking of the LS collision DFT with the nonequilibrium 
thermodynamics also requires an assumption of a V-representability\footnote{For traditional time-dependent 
DFT calculations, where one can assume a finite extension of the system and hence system-specific potential, 
there exists a proof of uniqueness of density, \protect\cite{Vrepresentability}.},
i.e., that for every nonequilibrium electron density there exists an external potential for which the 
solution density matrix produces that density.\footnote{It is possible, but not clear, that the proof of the 
regular time-dependent DFT 
V-representatibility \protect\cite{Vrepresentability} also extents to transport calculations 
based on infinite-system  time-dependent DFT \protect\cite{Stefanucci,ZhengTDDFT}, to the LS collision DFT \protect\cite{LSCDFT} 
formulation, and thus to here-presented construction of a rigorous GCE thermodynamics DFT. 
However, the support of the system-specific SP potential term can in any case be made much larger 
than the extension of the actual laboratory and much larger than any four-probe measurement setup. 
I therefore here follow a common practice and simply assume that there exists at least 
an approximative V-representatibility.}

A solution nonequilibrium density matrix corresponding to a given density $n'(\mathbf{r},t)$ can 
be viewed as a general density functional
\begin{equation}
\hat{\rho}_{n'}(t) =\hat{\rho}[n'(\mathbf{r},t)] \equiv \hat{\rho}_{\rm col, V' \leftarrow n'}(t).
\label{eq:rhiofnfunctionaldef}
\end{equation}
Below I focus the discussion on steady-state transport in a time-independent scattering potential.

The formally exact description of the overall dynamics in nonequilibrium tunneling
also requires the introduction of a number of other density functionals\footnote{Here and 
below I do not express the LS foundation in subscripts because I instead 
	seek to emphasize the link to the density, namely via the specification $n \to V \to 
	Y_V$ and thus, eventually, to the effective SP scattering potential $V_{\rm eff}$.}

\begin{eqnarray}
F[n] & = &
\frac{\hbox{Tr} 
\{ \hat{\rho}_n [K+W+(1/\beta) \ln(\hat{\rho}_n/\hbox{Tr}\{\hat{\rho}_n\})] \} } 
{ \hbox{Tr} \{ \hat{\rho}_n \}},
\label{UnivFTerm}\\
Y_V[n] & \equiv & 
Y_{\rm LS}[n]  = 
\frac{\hbox{Tr} 
\{ \hat{\rho}_n \hat{Y}_{V}\} }
{ \hbox{Tr} \{ \hat{\rho}_n \} }.
\label{YVTerm}
\end{eqnarray}
Here, $\hat{Y}_V$ denotes the steady-state operator for the Gibbs free energy.
The variational character of the thermodynamical potential can then be expressed
\begin{eqnarray}
\Omega_V[n] & = & V[n] + F[n] - Y_V[n],
\label{eq:nonUniversalPot}\\
\frac{\partial \Omega_V[n]}{\partial n} & = & 
\frac{\partial \Omega_V^{(0)} [n]}{\partial n} + 
\frac{\partial}{\partial n} \left[ \Omega_V[n] - \Omega_V^{(0)}[n] \right]
\nonumber\\
& = & 0.
\label{eq:nonUniversalPotVariance}
\end{eqnarray}

The correction to the noninteractiong contribution in (\ref{eq:nonUniversalPotVariance}) is \textit{formally} just a term which
reflects an (effective-potential-type) extension to the SP dynamics discussed in Sec.~VII.  However, Eq.~(\ref{eq:nonUniversalPotVariance})
is not directly useful for a DFT formulation of efficient calculations of nonequilibrium GCE thermodynamics. As the subscript `$V$' 
emphasize, this is a formulation which would still require us to succeed with a complicated explicit 
construction \cite{exactKondo,DoyonAndrei,Dutt} of the Gibbs free energy operator $\hat{Y}_V$ 
that is specific for $V$ and enters (\ref{YVTerm}). Explicit solution for
$\hat{Y}_V$ are possible \cite{exactKondo,NEQaverageEvolution,DoyonAndrei,Han,Dutt}
for special interacting-transport cases and reveals exciting details of interacting nonequilibrium dynamics. However, an approach 
based on (\ref{eq:nonUniversalPotVariance}) does not appear directly viable for setting up efficient SP calculations 
within a nonequilibrium thermodynamics DFT. Fortunately, there exists a work around.

\subsection{Relation to standard ballistic transport codes}

The goal is frozen-nuclei calculations of the electron density (and of transmission
and thermodynamics properties) in nonequilibrium tunneling within a generalization of the DBT 
framework \cite{Lang,DiVentra,TranSiesta,gpawNEGF}.
I consider only steady-state transport in a time-independent scattering potential and I introduce
both a system-specific potential-energy and (mean-field Coulomb) Hartree interaction terms
\begin{eqnarray}
V[n] & = & \int d \mathbf{r} v_{\rm sc}(\mathbf{r}) n(\mathbf{r}),
\label{eq:VscfunctDif}\\
\Phi_{\rm H}[n] & = & \frac{1}{2} \int d\mathbf{r} \int d\mathbf{r}' \frac{n(\mathbf{r})n(\mathbf{r}')}{|\mathbf{r}-\mathbf{r}'|}.
\label{CoulombTerm}
\end{eqnarray}
Both are here expressed formally as density functionals.

Of central importance for the usefulness of
the DBT method is undoubtably the emphasis on using a density functional expression
to approximate effects of the many-body interaction. The DBT chooses to approximate these by using the 
exchange-correlation functional term (identified by `GS') of ground-state DFT, that is, 
approximations to the universal functional for equilibrium systems: 
\begin{eqnarray}
E_{\rm xc}[n] & = & G_{\rm GS}[n]-G_{\rm GS}^{(0)}[n],
\label{Excterm}\\
G_{\rm GS}[n] & = & \langle \Psi_0 | K+W |  \Psi_0 \rangle -\Phi_{\rm H}[n],
\label{UnivGTerm}\\
G_{\rm GS}^{(0)}[n] & = & \langle \Psi_0^{(0)} | K |  \Psi_0^{(0)} \rangle.
\label{UnivGnullTerm}
\end{eqnarray}
specified as a difference between expectation values of the fully interacting 
ground state $|\Psi_0\rangle$ and of the noninteracting ground state $|\Psi^{(0)}_0\rangle$.
From these functional the DBT constructs an effective scattering potential
\begin{equation}
v_{\rm DBT}[n] (\mathbf{r}) \equiv  v_{\rm sc} (\mathbf{r}) + 
\frac{\delta \Phi_H}{\delta n(\mathbf{r})} + 
\frac{\delta E_{\rm xc}}{\delta n(\mathbf{r})}. 
\label{eq:DBTFTeffectivePot}
\end{equation}

The fact that solutions of the DBT dynamics are expressed in terms of SP LS solutions 
ensures that the DBT method (for electron-density calculations) is inherently consistent, 
automatically satisfying conservation laws (for any given density functional choice).  
The DBT method incorporates the essential 
Gibbs-free energy effects as it is consistent with the charge adjustments stipulated by 
using Freidel sum rule on dynamics in the effective potential (\ref{eq:DBTFTeffectivePot}).
In effect, the DBT method thus consists not only of constructing an effective potential
$v_{\rm DBT}[n]$ but also, implicitly, of constructing and using the associated SP Gibbs free energy operator, 
\begin{equation}
\hat{Y}_{V_{\rm DBT}} (\mathbf{r}) \equiv  \sum_{\lambda_P} \mu_P
(\hat{\psi}_{V_{\rm DBT},\lambda_P}^{(0)})^{\dag}
\hat{\psi}_{V_{\rm DBT},\lambda_P}^{(0)},
\label{eq:DBTFTeffectiveY}
\end{equation}
from SP creation operators $(\hat{\psi}_{{\rm V,DBT},\lambda_P}^{(0)})^{\dag}$ for the set of DBT 
SP LS solution states.

A determination of the Gibbs free energy operator (and Gibbs free energy effects) is
essentially already available at the DBT level. This opportunity exists because
the DBT codes emphasize efficient calculations of the SP scattering states.

\subsection{Universal functionals and formally-exact single-particle descriptions}

While the grand thermodynamical potential functional (\ref{eq:nonUniversalPot}) 
contains a non-universal component (beyond $V[n]$), it is still possible to achieve 
an effective single-particle formulation in terms of universal density functionals.  
The development of a formally exact SP formulation shows that it contains 
the DBT method \cite{Lang,DiVentra,TranSiesta} as a lowest-order approximation
(for frozen coordinates).

I observe that since the scattering potential is a functional of the density, 
Eq.~(\ref{eq:VscfunctDif}), it follows that the LS operator solution can also be expressed as 
a density functional 
\begin{equation}
\hat{\psi}_{{\rm LS},\lambda_P}^{\dag}[n] = \frac{i\eta}{\epsilon_{\lambda_P}+\mathcal{L}_{H_{V\leftarrow n}} + i\eta} 
c^{\dag}_{\lambda_P}.
\label{eq:psifuncfDef}
\end{equation}
From this functional form of the many-body LS operator solution, it is possible 
to obtain density derivatives of the variational thermodynamics grand potential. 
I define
\begin{eqnarray}
\hat{Y}_n & = & \sum_i \mu_i 
\hat{\psi}^{\dag}_{{\rm LS},\lambda_P}[n]
\hat{\psi}_{{\rm LS},\lambda_P}[n],
\label{UnivYopTerm}\\
Y[n] &= & 
\frac{\hbox{Tr} 
\{ \hat{\rho}_n \hat{Y}_{n}\} }
{ \hbox{Tr} \{ \hat{\rho}_n \} },
\label{UnivYTerm}\\
Y'[n] & = & 
\frac{\hbox{Tr} 
\{ \hat{\rho}_n (\partial \hat{Y}_{n}/\partial n) \} }
{ \hbox{Tr} \{ \hat{\rho}_n \} }.
\label{UnivYprimeTerm}
\end{eqnarray}
I note that 
\begin{equation}
\frac{\partial}{\partial n} Y_V[n] = 
\frac{\partial}{\partial n} Y[n] - Y'[n], 
\end{equation}
expresses the derivative of the Gibbs free energy in terms of universal functionals.

I seek a SP formulation where solving for fictitious SP scattering wavefunctions 
$\psi_{{\rm eff},\lambda_P}(\mathbf{r})$ (moving in some effective scattering  potential $V_{\rm eff}$) 
constitutes a complete specification of both the electron density and the thermodynamical 
potential $\Omega_V$. The individual SP scattering state is
a SP LS solution which emerges from an $H_d$ eigenstate $\lambda_P$ (of energy $\epsilon_{\lambda_P}$).
Assuming a suitable normalization of such solutions leads to a simple determination of the electron density variation
\begin{equation} 
n(\mathbf{r}) = \sum_{P,\lambda_P} |\psi_{{\rm eff},\lambda_P}(\mathbf{r})|^2
\, \langle \hat{n}_{\lambda_P}\rangle_{\hat{\rho}_n}
\label{eq:EffSPdensityevaluate}
\end{equation} 
where $\langle \hat{n}_{\lambda_P}\rangle_{\hat{\rho}_n}$ is the level-occupation expection value evaluated from $\hat{\rho}_n$.
To provide a rigorous SP formulation from universal functionals I introduce a set of (generalized) exchange-correlation energy terms
\begin{eqnarray}
F_{\rm xc}[n] & = & F[n]-F^{(0)}[n],
\label{UnivFxcTerm}\\
Y_{\rm xc}[n] & = & Y[n]-Y^{(0)}[n],
\label{UnivYxcTerm}\\
Y'_{\rm xc}[n] & = & Y'[n]-Y'^{,(0)}[n].
\label{UnivYxcprimeTerm}
\end{eqnarray}
Here, as before, superscripts `$(0)$' identify functional values evaluated at $W=0$.
It follows that the fictitious dynamics is uniquely specified by an effective equation of motion
\begin{eqnarray}
\frac
{\partial \Omega_V^{(0)}}{\partial \psi_{{\rm eff},\lambda_P}^*(\mathbf{r})} & = &
-\frac{\partial [\Omega_V-\Omega_V^{(0)}]}{\partial n}
\psi_{{\rm eff},\lambda_P}(\mathbf{r})\\
& = &
-\left(\frac{\delta F_{\rm xc}[n]}{\delta n(\mathbf{r})}
-\frac{\delta Y_{\rm xc}[n]}{\delta n(\mathbf{r})} + Y'_{\rm xc}[n]\right)
\psi_{{\rm eff},\lambda_P}(\mathbf{r}).
\end{eqnarray}
Since we have assumed a suitable normalization, the equation of motion for an occupied
SP LS solution wavefunction can be reformulated,
\begin{eqnarray}
& & \langle \mathbf{r} | (H_V^{(0)} - \hat{Y}_V^{(0)}) | \psi_{{\rm eff},\lambda_P} \rangle +\nonumber\\ 
& & \left( \frac{\delta F_{\rm xc}[n]}{\delta n(\mathbf{r})}
+Y_{\rm xc}'[n]
-\frac{\delta Y_{\rm xc}[n]}{\delta n(\mathbf{r})} \right)
\langle \mathbf{r} | \psi_{{\rm eff},\lambda_P} \rangle\nonumber\\
& = & C[\psi_{{\rm eff},\lambda_P}] (\epsilon_{\lambda_P}-\mu_P)
\langle \mathbf{r} | \psi_{{\rm eff},\lambda_P} \rangle.
\label{eq:variationalFunctSPformulation}
\end{eqnarray}
The real-valued functional $Y'_{\rm xc}[n]$ can be viewed either as a constant potential shift
or as a shift in the eigenvalue $\epsilon_{\lambda_P}$, for example, as expected when an infinite
scattering potential causes a change in band-structure \cite{Pirenne,Gellmann}.

The SP LS scattering-wave solutions (for dynamics in an effective SP potential 
$V_{\rm eff}$ that gives rise to the SP Gibbs-free energy operator $\hat{Y}_{\rm V_{\rm eff}}$) is:
\begin{equation}
\langle \mathbf{r} | (H_{V_{\rm eff}}^{(0)} - \hat{Y}_{V_{\rm eff}}^{(0)}) | \psi_{{\rm eff},\lambda_P} \rangle 
 =  \left(\epsilon_{\lambda_P}-\mu_i\right)
\langle \mathbf{r} | \psi_{{\rm eff},\lambda_P} \rangle.
\label{eq:effectiveSPformulation}
\end{equation}
Given an assumption of V-representability, it follows that the terms of the fictitious SP formulation 
(\ref{eq:effectiveSPformulation}) and the variational evaluation (\ref{eq:variationalFunctSPformulation}) 
must coincide. This provides a framework to identify the uniquely defined effective SP potential 
\begin{equation}
V_{\rm eff}[n] \equiv 
\int d\mathbf{r} \,  v_{\rm eff}[n](\mathbf{r}) n(\mathbf{r}).
\end{equation}

At the same time, and unlike in ground-state DFT and in DBT, it is not so that a comparison of the dynamics 
specified by the variational principle leads directly to an explicit determination of a new effective potential,
\begin{equation}
	V_{\rm eff}[n] \neq V[n]+\frac{\delta}{\delta n}F_{\rm xc}[n].
\end{equation}
To search for an effective SP potential for description of many-body interactions effects in nonequilibrium tunneling
we must instead require an implicit approach with concerted adjustments of both the potential form and 
of the associated Gibbs free energy term. Future works will explore approaches for simplifying this
implicit search for the effective SP scattering potential (going beyond the DBT approach).

Nevertheless, we are able to now discuss the nature of DBT method for calculations
of electron density as a lowest-order approximation to the above-discussed, exact 
GCE thermodynamical DFT. It is clear that the DBT method is consistent: it is built
from SP LS scattering states and it follows that the description (in itself) is compatible 
with the Friedel sum rule.  The DBT method captures Gibbs free energy effects [to the extent 
that these arise in the DBT choice of the effective potential (\ref{eq:DBTFTeffectivePot})] through 
the implicit construction of the quadratic operator $\hat{Y}_{V_{\rm DBT}}$,  Eq.~(\ref{eq:DBTFTeffectiveY}).
Within the DBT calculation method one can also define the DBT solution density matrix as a density
functional $\hat{\rho}_n^{\rm DBT}$ and, in turn a density functional for the DBT approximation
for the Gibbs free energy in actual potential $V$:
\begin{equation}
Y_{V}^{\rm DBT}[n] = \frac
{\hbox{Tr}\{\hat{\rho}_n^{\rm DBT}\hat{Y}_{V_{\rm DBT}}\}}
{\hbox{Tr}\{\hat{\rho}_n^{\rm DBT}\}}.
\label{eq:YVbtdDef}
\end{equation}

The DBT method does constitute an approximation in that it does 
not express the thermodynamics in terms of the actual functionals $F_{\rm xc}[n]$, $Y_{xc}[n]$, 
and $Y'_{\rm xc}[n]$. A systematic improvement of the DBT method can be made by also consistently 
including the nonequilibrium many-body scattering effects which are expressed in the functional-derivative difference
\begin{eqnarray}
\frac{\delta}{\delta n(\mathbf{r})}
\left(F_{\rm xc}[n]-E_{\rm xc}[n] -\Phi_H[n]\right) - 
& & \nonumber\\ 
\frac{\delta}{\delta n(\mathbf{r})}
\left(Y_{\rm xc}[n]-Y_{V}^{\rm DBT}[n]+Y_V^{(0)}[n]\right) + 
\nonumber \\
Y'_{\rm xc}[n]  
& \neq & 0.
\end{eqnarray}

\subsection{nonequilibrium thermodynamics in interacting tunneling.}

With the universal functionals, the SP formulation, and the generalized Hellmann-Feynman, it is possible
to formally setup rigorous DFT calculations of the exact steady-state thermodynamics variation and of 
the associated conservative, adiabatic forces.  The nonequilibrium Green function or the phase shift formulations 
(sec.~II.B) are viable frameworks for a DFT SP studies of fully interacting nonequilibrium tunneling. 

Such formal DFT calculations provides a determination of the change in
thermodynamics grand potential value but only given relative to the 
(fixed, infinite) value which characterize the initially disconnected system $H_d$. This is natural for 
the open model systems are extensive (infinite) and no other determination would make sense. 
I also note that the disconnected equilibria system $H_d$ might well be given by a ground-state DFT 
calculation of those manifestly equilibrium $t \to -\infty$ subsystems.
This implies that it is possible to 
obtain a starting point which is exactly characterized (through the Kohn-Sham eigenstates 
and eigenvalues). It also implies that one can have a starting point where
the expected density changes (and hence the changes in 
effective scattering potential) will remain essentially confined to the central tunneling region. 
While the transmission, in principle, implies that the density changes extend to infinity, 
this density component will not generally affect the resulting effective potential far from
the central tunneling region. This is an observation which has been explicitly tested in a 
typical DBT study \cite{TranSiesta}.

Computations of the thermodynamics grand potential based on the Green function 
formulation \cite{NEQQSMthermodynamics} (appendix A) can build on
significant experience.  Formal scattering theory evaluations, based 
on a rapidly convergent expansion in the T-matrix behavior, was emphasized already in 
the early DFT history.  The approach was, for example, used in studies of the van der Waals binding and 
kinetic energy repulsion of noble gas atoms on a metal surface \cite{ZarambaKohnLS,ScatteringPhysisorption}\footnote{For a 
recent summary and comparison with van der Waals density functional calculations, please see,
\protect\cite{H2lee}.}.
This type of scattering-system calculations are also actively pursued for nonequilibrium systems 
within the ReNSQM framework and calculations are available even for the more complex 
interacting nonequilibrium tunneling problems \cite{exactKondo,DoyonAndrei,NEQunified,NEQaverageEvolution,Han,Dutt}.
In addition, the phase-shift analysis (appendix B) suggests that DBT computations can
provide a shortcut, at least in simpler cases where it is possible to complete a
S-matrix eigenstate decomposition.

\section{Discussion} 

The rigorously demonstrated \textit{existence} of an exact, variational thermodynamics grand potential is 
important in itself for it leads to an exact SP computational framework [results (a) through (d)] of, for example, the electron density.  
At that stage, the theory constitute an alternative to adaptations \cite{Stefanucci,ZhengTDDFT} of 
 time-dependent DFT \cite{RungeGross} to a GCE modeling framework of tunneling although the functionals are 
different, constructed from a thermodynamics rather than an action principle.  

The fact that the nonequilibrium thermodynamics grand potential is even extremal allows it to deliver additional conclusions 
[results (e) through (j)]; it is relevant to seek a rigorous computation 
also of the variational quantity $\Delta \Omega_{\rm LS}$ itself. 

\subsection{Thermodynamics versus electrostatic forces}

It is interesting that the here-derived exact thermodynamics forces (\ref{eq:NTForceAdiabInit})
\textit{formally and generally differs} 
\begin{equation}
\mathbf{F}^{\rm es, NEQ}_{\mathbf{R}_i} \neq \mathbf{F}^{\rm GCE}_{{\rm LS},\mathbf{R}_i}
\label{eq:GeneralDiffer}
\end{equation}
from the often-used GCE electrostatic forces,
(\ref{eq:esForceFirstDef}).  The electrostatic-forces expressions (\ref{eq:esForceFirstDef}) are 
specified by frozen-nuclei calculations of the nonequilibrium electron-density variation and they are 
often -- but need not be -- used to thus predict nonequilibrium atomic relaxations from DBT or nonequilibrium quantum statistical mechanics 
calculations \cite{CurrentForcesWireCalc,TranSiestaForce,
Pantelides,TodorovSuttonThermodynForces,TodorovTB1,TodorovTB2,Todorov1,Todorov2,Todorov3,Verdozzi,ForcesWithPhonFriction}.

\textit{I observe} that the use of electrostatic forces lacks a formal foundation
beyond (equilibrium and) linear response.  Like Ref.~\onlinecite{Pantelides}, I point out that use of
Ehrenfest theorem \cite{EhrenfestMerzbacher} cannot be used for a conclusive determination of
forces by electrons on the nuclei for infinite open systems. One can succeed with an Ehrenfest-based
force evaluation (and derive electrostatic forces)
\begin{equation}
i\frac{d}{d t}\langle \frac{\partial}{\partial \mathbf{R}_i} \rangle|_{\rm GS\ or\ finite} =
\langle \frac{\partial H}{\partial \mathbf{R}_i} \rangle_{\rm GS\ or\ finite},
\label{eq:esForceEhrenfest}
\end{equation}
of course, for finite systems \cite{EhrenfestMerzbacher} or in a standard ground-state DFT 
formulation\footnote{Ehrenfest theorem can be used for an
equilibrium canonical-ensemble system by first assuming an 
infinite repetition and by then employing a periodic-cell Green's theorem, 
\protect\cite{EQgreengtheorem}.}.
A force evaluation based directly on Ehrenfest theorem 
is also available for studies using traditional, finite-extention canonical-ensemble
 time-dependent DFT \cite{EhrenfestCitation,BurkeDisc} and when the nonequilibrium transport 
problem is recast as a finite (essentialy closed) problem \cite{GebCar,Burke}, i.e., when a 
time-dependent field causes induction (balanced by a coupling to a phonon bath) in a 
horse-shoe setup.  However, for the standard open-boundary modeling of tunneling 
an analysis in terms of GCE thermodynamics remains essential.

An attempt at a direct evaluation of the Ehrenfest theorem is indeterminate in a GCE description
of tunneling. This is because such systems, by nature,  must have infinite extension 
to avoid carrier depletion.  Ehrenfest evaluation of forces (for SP dynamics) rests on using 
Green's theorem 
\begin{equation}
\int_V (u \Delta v -v \Delta u) d\mathbf{r} = \int_S (u\nabla v - v\nabla u) \cdot d \mathbf{\tau} 
\end{equation}
and on using an assumed cancellation of the surface terms (in `S') at 
infinity \cite{EhrenfestMerzbacher}. There is no argument directly available for a cancellation
for scattering states.  In fact, inclusion of surface terms is essential for both the formal 
SP collision formulation \cite{AlbertXIX3}, for a consistent nonequilibrium Green function evaluation of 
noninteracting tunneling transport\footnote{The role of the Green-theorem surface terms (in 
a consistent determination of noninteracting Green's functions for 
tunneling \protect\cite{KondoPRL}) is not always expressed but they are
nevertheless essential \protect\cite{RateEq} (until one includes 
actual many-body scattering and associated dephasing effects everywhere). In the
nonequilibrium Green function descriptions, we traditionally seek to separate out the 
single-particle terms (here denoted $\delta KV$) and many-body terms (described by 
a self-energy $\sigma$).  In the absence of many-body interactions (or dephasing) in 
the leads, the Green-theorem surface terms forces us to work with 
$g_{<}=[1+g_r (\delta KV+\sigma_r)]g_<^d[(\delta KV+\sigma_a) g_a+1]+g_r\sigma_< g_a$, 
not just the last term (that vanishes when $\sigma=0$).}
\cite{RateEq}, and for a LS collision description 
of an open many-body interaction problem \cite{LippSchwing,AlbertXIX29}. The evaluation must be pursued in 
the context of a Wronskian \cite{Albert1D} formulation, i.e., with adherence to the flux conservation 
that formal scattering theory automatically provides \cite{AlbertXIX}.

\textit{This paper shows} (appendix A) that the exact thermodynamics forces (\ref{eq:AdiabForceRelation}) 
reduce to electrostatic-force definition (\ref{eq:esForceFirstDef}) when the electron distribution 
can be entirely described from the full density of states. I note that this condition is
not generally satisfied outside equilibrium.  Before proceeding, I also
emphasize that the here-presented formal and exact QKA recasts rigorously demonstrates 
that the steady-state thermodynamics forces for adiabatic transformation can never 
reduce to the nonconservative electrostatic forces discussed and explored, 
for example, in Refs.~\onlinecite{Todorov2,Todorov3}.

Many existing descriptions of tunneling forces starts with the definition
(\ref{eq:esForceFirstDef}) and the assumption that since electrostatic forces
are applicable in equilibrium and in linear response \cite{Sorbello,TodorovSuttonThermodynForces} 
it should also hold under general nonequilibrium conditions.
The assumption of the validity of electrostatic forces (\ref{eq:esForceFirstDef}) enters, for example, 
as expressed by a choice of Lagrangian \cite{Verdozzi}, or when motivation is sought from Ehrenfest 
theorem \cite{EhrenfestMerzbacher} followed by an Ehrenfest approximation, i.e., a separation of 
slow nuclei and fast electron dynamics \cite{TodorovTB1,TodorovTB2,Pantelides}.

There are ongoing discussions of run-away and possible water-wheel nuclei-dynamics 
effects\footnote{Including an electron-phonon coupling 
will obviously provide a modeling of Joule heating of the lattice and, 
hence, of a friction mechanism. However, it also opens up for both possible topological 
effects \protect\cite{ForcesWithPhonFriction} and frequency-selective, current-induced 
spontaneous emission as well as stimulated transitions, for example, as discussed in
Ref.~\protect\onlinecite{CurrentExcitation}.}
\cite{Verdozzi,Todorov1,Todorov2,Todorov3,ForcesWithPhonFriction} 
and of an inherent nonconservative nature \cite{Sorbello,Todorov2,Todorov3} of such electrostatic
forces. These discussions have recently been supplemented by works which also include
a coupling to a phonon bath \cite{vonOppen,ForcesWithPhonFriction}.
Typically, it is the electron behavior expressed by (\ref{eq:esForceFirstDef}) which are 
seen as the source for a possible structural instability in tunneling systems. Several
papers (for example, Ref.~\onlinecite{ForcesWithPhonFriction}) state that the nonconservative 
behavior express a \textit{nonadiabatic} property (a view that is consistent with the
here-presented finding that the exact steady-state thermodynamics forces must be 
explicitly conservative).  Inclusion of phonons tends to stabilize 
the dynamics \cite{ForcesWithPhonFriction,vonOppen}.
The papers on nonconservative electrostatic forces discuss an effect beyond Joule heating, an effect which 
should arise also in the absence of a possible current-induced coupling to vibrations.
The latter effect is clearly also important,  for example, as has very 
recently been illustrated \cite{vonOppen} in a study of a current-induced limit-cycle behavior 
produced when focusing the many-body interaction to an electron-vibrational coupling.
Overall, this development, a finding of a nonconservative character in the forces which are 
widely used for implementing nonequilibrium relaxations, accentuates the need for discussing
what forces are best suited to describe working nonequilibrium devices.  

\textit{This paper concentrates} the discussion of forces and relaxations on the 
\textit{strictly adiabatic behavior}, arising when there is an infinite ratio of the ion to electron masses. 
For this discussion I am finding the the exact thermodynamics forces are not only variational and 
explicitly conservative but also express an inherent adiabatic nature (always propelling the system 
to a state without any thermodynamics excitation). I find that the thermodynamics forces are suited 
for use in a nonequilibrium Born-Oppenheimer approximation.

It is, of course, clear that additional nonconservative forces, like wind forces
will also apply when one can not ignore the mass and momentum of the 
electron in the thermodynamics analysis. Understanding the nature of non-adiabatic 
forces is also an important problem which, however, lies beyond the scope of this paper.

\subsection{Relation to equilibrium thermodynamics}

The demonstration of the variational character (appendix C) is essentially identical 
to Mermin's original analysis \cite{Mermin65} in his formulation of an 
equilibrium-thermodynamics DFT.  However, one difference deserves a separate 
discussion: the emerging (rather than fixed) expression for the nonequilibrium 
Gibbs free energy operator $\hat{Y}_{\rm col}(t)$ does not seem to 
provide any reductio ad absurdum argument. Rather, changes in the SP collision 
potential term $V_{\rm col}$ will make themselves felt both in the Hamiltonian 
and in the Gibbs free energy operator. 

Nevertheless, the present nonequilibrium thermodynamic account is developed in the 
continuum Caroli partition scheme \cite{Caroli,ContinuumCaroli} that was also
used in the formulation \cite{LSCDFT} of a LS collision DFT. Rather than a traditional
reduction ad absurdum argument \cite{Mermin65}, it is here a  time-dependent DFT-type of argument 
(from the appendix of Ref.~\onlinecite{LSCDFT} or from Ref.~\onlinecite{ZhengTDDFT}) 
that underpins a DFT SP formulation. This analysis ensures that one can treat both 
the nonequilibrium thermodynamic quantities and associated nonequilibrium density matrix solutions through 
universal electron-density functionals.  

\subsection{Refining the density functional description}

I note that many-body interactions in nonequilibrium tunneling are known to 
cause changes in the statistical distribution among many-body 
eigenstates \cite{NEQreform,exactKondo,NEQunified,
NEQaverageEvolution,Dutt} in tunneling. The magnitude and complexity of such changes 
significantly exceed those arising in equilibrium cases 
because the Fermi-liquid behavior degrades and genuine inelastic 
processes will proceed \cite{KondoPRL,RateEq,Phon,QCLdisc}.

The Friedel sum rule is, of course, also of central importance in 
correct calculations of such many-body interactions \cite{LangrethFriedel,LangerAmbegaokar}.
In addition, it is in nonequilibrium tunneling a much more complex task to 
determine what possible many-body states represent the lower- or 
higher-lying many-body configurations of the emerging distribution \cite{NEQreform}.
A full, exact GCE thermodynamical account is essential to decide which such many-body 
configurations must be excluded in an implementation of an adiabatic 
transformation of nuclei positions.  

\section{Summary and conclusion}

To derive a rigorous description of GCE thermodynamics in interacting nonequilibrium tunneling, 
this paper recasts the full QKA \cite{Langreth} of electron dynamics. It identifies
a variational thermodynamics grand potential in a theory that is closely related to the 
exact steady-state ReNQSM \cite{NEQreform} and to Mermin's analysis of equilibrium
GCE thermodynamics \cite{Mermin65}. In part, this formal result is made possible by 
a definition of an operator for the nonequilibrium Gibbs free energy under general nonequilibrium tunneling 
conditions.  The paper furthermore notes the existence of a uniqueness 
of density proof (from the closely related LS collision DFT \cite{LSCDFT}) and uses that 
proof, and the extremal nature of the nonequilibrium grand potential, to formulate rigorously a
SP thermodynamics theory cast in terms of universal density functionals. The
density functionals characterize the internal and Gibbs free energy variation.
The paper establishes the widely used DBT formulation \cite{Lang} as a 
lowest-order yet consistent approximation to the here-presented 
nonequilibrium thermodynamics DFT.

A nonempirical theory of nonequilibrium tunneling must be able to reliably describe the 
structure of a tunneling system under operating condition and, in particular, to 
determine transport-induced morphology changes.  This has been a challenge because 
in actual nonequilibrium quantum statistical mechanics calculations, the focus is typically on the electron behavior. 
For the standard (open-boundary and thus infinite) model framework of tunneling, I
observe that there is a need to include effects of the Gibbs free energy variation. For example, 
electron redistribution effects \cite{NEQreform,RateEq,Phon,QCLdisc,DoyonAndrei,Dutt} makes 
it relevant to investigate and determine thermodynamics forces.  The demonstration of a
generalized Hellmann-Feynman theorem ensures a conservative nature
of the thermodynamics forces.

This paper has thus taken a fundamentally different route than the standard force
discussion and has avoided an indeterminate Ehrenfest-type force evaluation.
This paper demonstrates that the here-defined exact thermodynamics forces 
are both useful and will always implement (with sufficiently slow deformation) 
an adiabatic nature in relaxations. The thermodynamics forces are also based on a 
robust physics principle (entropy maximization) and are suited for
use in a nonequilibrium Born-Oppenheimer approximation for interacting nonequilibrium tunneling.

\acknowledgments

The author is grateful and thankful to David C. Langreth (who, sadly, 
is no longer with us) for interest and for input far beyond 
Ref.~\onlinecite{LangrethLS}.  The author also gratefully 
acknowledges discussions with Selman P.~Hershfield, Nicola Lanat{\`a}, 
and Gerald D.~Mahan.  The research was supported by the Swedish Research 
Council (VR).

\appendix

\section{Noninteracting Green function analysis}

The emitter `L' and collector `R' are initially described by states $|\phi_{\lambda,L/R}\rangle$ 
(of SP energy $\epsilon_{\lambda,L/R}$) but have no overlap with each other or with states of 
the central region $|\phi_{\lambda,C}\rangle$.  The initial (disconnected) system is 
described by retarded Green functions
\begin{eqnarray}
	\hat{g}_r^d(\omega) & = & \sum_{J=L/R/C} \hat{g}_r^{dJ}(\omega),
\label{grddef}\\
\hat{g}_r^{dJ={\rm L/R/C}}(\omega) & = & 
\sum_{\lambda}\frac{
|\phi_{\lambda_J}\rangle\langle \phi_{\lambda_J}|} 
{\omega-\epsilon_{\lambda_J}+i\eta}.
\label{grJdef}
\end{eqnarray} 
From these one can determine the initial density of state as a trace
\begin{eqnarray}
	\mathcal{D}^d(\omega) & = & \sum_{J={\rm L/C/R}} \mathcal{D}^{dJ}(\omega),
\label{DEQdosdef}\\
\mathcal{D}^{d,{\rm J=L/R/C}}(\omega) & = & \frac{i}{\pi}[\hat{g}_r^{dJ}(\omega)-\hat{g}_a^{dJ}(\omega)],
\label{DEQspectral}
\end{eqnarray}
using the general analytic properties of the retarded and advanced Green functions, 
$\hat{g}_a(\omega)=\hat{g}_r^*(\omega)$. The density of state includes a factor of 
2 for spin. A specification of the electron-distribution or less-than Green function is
given in Sec.~VII.C.

I provide below a formal nonequilibrium Green function (or equivalently a SP LS solution) framework for 
calculations and analysis in noninteracting GCE thermodynamics.

\subsection{From noninteracting Green functions to nonequilibrium thermodynamics} 

For the emitter and collector we seek the \textit{partial} or \textit{projected} density of 
state changes (\ref{eq:deltaDshiftEvaluation}) as well as the corresponding integrated changes 
(\ref{eq:deltaNshiftEvaluationNEGF}).  The integrated density of state changes 
(\ref{eq:deltaNshiftEvaluationNEGF}) reflects the net effects of the 
infinitesimal level shifts ($\epsilon_{\lambda_P} \to \tilde{\epsilon}_{\lambda_P}$)
that arise with the adiabatic turn on of the tunneling and which are reflected 
in the SP LS solutions \cite{AlbertXIX3,Gellmann,DoniachCite} 
\begin{equation}
|\psi_{{\rm LS},\lambda_{P=L/R}}^{(0)}\rangle = [1+\hat{g}_r^{d}(\tilde{\epsilon}_{\lambda_P})
\hat{T}^{(0)}(\tilde{\epsilon}_{\lambda_P})]
| \phi_{\lambda_P}\rangle.
\end{equation}
Here $\hat{T}^{(0)}$ denotes the T matrix of the connected (but noninteracting) system.
In fact, the set of SP LS solution states $\{|\psi_{{\rm LS},\lambda_{P=L/R}}^{(0)}\rangle\}$, 
can be used to determine the full change in the density of state \cite{ZarambaKohnLS} 
as well as the lead projected density of state changes, (\ref{eq:deltaDshiftEvaluation})

The full density of state change (in the connected but noninteracting system)
\begin{eqnarray}
	\Delta \mathcal{D}_{\rm tot}(\omega)  & = & \Delta \mathcal{D}_{\rm L}(\omega) + \Delta \mathcal{D}_{\rm R}(\omega) 
\label{eq:fullDdefsum}\\
	& = & 
	\mathcal{D}^d(\omega) + 
\frac{i}{\pi} \hbox{Tr} \{\hat{g}^d_r(\omega) \hat{T}^{(0)}(\omega) \hat{g}^d_r(\omega) \},
\label{eq:fullDdef}
\end{eqnarray}
reflects the full S and T-matrix behavior [expressed in (\ref{eq:FormalFriedel})]. The Dyson equation \cite{Dyson} for 
the retarded Green function 
\begin{equation}
\hat{g}_r^{(0)}(\omega) - \hat{g}_r^d(\omega) = \hat{g}_r^d(\omega)H_{1,V}\hat{g}_r^{(0)}(\omega)
\label{deltagrres}
\end{equation}
determines the energy levels $\epsilon_{\lambda_P}$, formally by solving for the poles of the 
scattering-state retarded Green function \cite{Gellmann}, (\ref{gr0def}) and (\ref{gr0Pdef}).
In turn, this formal-scattering theory analysis uniquely determines of the tunneling-induced changes 
in the zero-temperature thermodynamics grand potential
\begin{eqnarray}
\Delta \Omega_{\rm LS} & = & 
\sum_{P={\rm L/R}} \int_{-\infty}^{\mu_P} (\omega-\mu_P)\Delta \mathcal{D}_P(\omega) 
\nonumber\\
& = & 
-\sum_{P=L/R} \int_{-\infty}^{\mu_P}\Delta \mathcal{N}_P(\omega).
\label{eq:FormalSPOmegaLS}
\end{eqnarray}

\subsection{Grand-canonical ensemble thermodynamics study of Cu(111) adsorbates}

The interactions of Cu(111) adsorbates provides a simple example of the role
Gibbs free energy in a GCE thermodynamics study (Section II.C). This is a system
where it is sufficient to consider adsorbate-induced s-wave scattering phase shifts \cite{FriedelInteractSurfTest,WeissReview}.
Here I provide some details on the thermodynamics behavior from the Green function calculations which underpin the results
for example, reported in Refs.~\onlinecite{HarrisFriedelInteract,HarrisFriedelTrioInteract,SlidingRings}.
The isotropic surface state has an approximative free-electron dispersion \cite{ExternalResponse} given by effective 
mass $m_{\rm eff}$.  At the same time, it is an example which requires formal scattering theory
because the MSS scattering off, for example, an adatoms must be described 
nonperturbatively \cite{FriedelInteractSurfTest} and the same therefore applies for 
the description of the MSS-mediated interaction \cite{HarrisFriedelInteract}.

I Introduce a MSS Fermi energy $\epsilon_{\rm F}$ to express the difference between chemical 
potential $\mu_F$ and the bottom of the MSS state (in a zero-temperature analysis).  
The unperturbed two-dimensional dynamics is expressed in terms of the MSS retarded Green functions
\begin{equation}
g_r^{\rm MSS}(d;\omega)=
\frac{m_{\text{eff}}}{\pi\hbar} K_0(-iq_{\omega} r+i\eta).
\label{TwoDGreenRes}
\end{equation}
Here $\hbar \omega\equiv \hbar^2 q_{\omega}^2/2m_{\rm eff}$
and $K_0$ denotes the two-dimensional zeroth-order Bessel function.  Since the experimental and theoretical
characterizations rests on a s-wave description of the adatom-MSS interaction, I also introduce the 
MSS T-matrix
\begin{equation}
T_{\text{1s}}(\omega) = \frac{i \hbar^2}{m_{\text{eff}}}
\left[e^{i2\delta_0(\omega)} - 1\right]
\label{Tmatrix}
\end{equation}
for multiple $\delta$-function scattering \cite{HarrisFriedelInteract}.
This energy-dependent T-matrix is given by the s-wave phase shift
\begin{equation}
\delta_0(\omega)=\hbox{arccot}\left[\ln(\omega/\epsilon_{\rm F})/\pi+
\cot(\delta_{\rm F})\right],
\label{SwaveForm}
\end{equation}
which, in turn, is entirely specified the experimentally observed Fermi-level (s-wave) 
phase shift, $\delta_{\rm F}\approx \pm \pi/2$.

The MSS-mediated interaction is given by evaluating the integrated change in the density of states, 
$\Delta \mathcal{D}^{\delta_{\rm F}}_{\rm MSS,2s}(d;\omega)$, which arise
when two adatoms approach from an infinite distance and to some finite separation $d$. 

Since this is an equilibrium problem, one can use Lloyds' formula \cite{Lloyds} to directly express the integrated 
changes in the density of state as the two s-wave scatters approach from an infinite separation to
some finite separation $d$:
\begin{equation}
\Delta \mathcal{N}^{\delta_{\rm F}}_{\rm MSS,2s}(d;\omega)
 =  -2\Im \ln \{1-[T^{\delta_{\rm F}}_0(\omega)g^{MSS}_r(\omega,d)]^2\}. 
\label{eq:IntDOSchangesMSSexplicit}
\end{equation}
Using also formal Fourier-transformation theory, for example, as formulated by Lighthill\cite{Lighthill}, it is straightforward to
obtain an analytical evaluation in the limit of asymptotic adatom separation, $q_{\rm F}d \gtrsim \pi$.  In particular, it 
follows that the asymptotical internal energy variation is entirely dominated by \cite{NEQreformWithTime}
\begin{eqnarray}
	\Delta N_{\rm MSS,2s}^{\delta_{\rm F}}(d;\omega) & \sim & 
\left(\frac{2}{\pi}\right)^2 \Re 
\left[\frac{e^{i2q_{\rm F} d \sqrt{y}}}{q_{\rm F} d \sqrt{y}} \times \right.\nonumber\\
& & \left. \frac{(\ln(y)/\pi +\cot(\delta_{\rm F})+i)^2}
{[1+(\ln(y)/\pi +\cot(\delta_{\rm F}))^2]^2}
\right]_{y=\omega/\epsilon_{\rm F}}.
\label{ULSsumExpandFriedel}
\end{eqnarray}

The leading-order GCE variation in internal energy $U_{\rm LS}^{\rm MSS,2s}(d;\delta_{\rm F})$
coincides exactly with the variation predicted for the Gibbs free energy 
$Y_{\rm LS}^{\rm MSS,2s}(d;\delta_{\rm F}) = \mu_{\rm F} \Delta N_{\rm 2s}(\epsilon_{\rm F};\delta_{\rm F})$.  
The asymptotic decay $1/d$ of either of the individual GCE interaction components
differs from the $1/d^2$ decay form derived in a previous perturbative canonical-ensemble
calculation \cite{FriedelLK} and with experimental observations \cite{FriedelInteractSurfTest}.

The full GCE thermodynamics account, calculating the GCE thermodynamics grand potential
for the MSS-mediated interaction interactions, has instead a cancellation of the $1/d$ 
terms and has a correct $1/d^2$ oscillatory decay as well as a phase in agreement 
with the experiments \cite{FriedelInteractSurfTest}. The analysis thus illustrates 
that including the Gibbs free energy variation is essential in formal scattering 
theory studies of the GCE thermodynamics behavior \cite{ZarambaKohnLS,ScatteringPhysisorption,HarrisFriedelInteract}.

\subsection{Thermodynamics versus electrostatic forces}

Here I explore consequences of \textit{an assumption} `A)' that the 
electron distribution, i.e., $g_<^{(0)}(\mathbf{r},\mathbf{r},\omega)$, remains 
entirely specified by the changes in the full local density of state,
\begin{equation}
	\mathcal{D}_{\rm tot}(\mathbf{r},\omega) \equiv -\frac{2}{\pi}\Im \langle \mathbf{r}| \hat{g}^{(0)}_r (\omega)| \mathbf{r} \rangle.
\label{eq:LDOStot}
\end{equation}
For a discussion of electrostatic forces it is relevant to also inspect the 
variation in the corresponding integral
\begin{equation}
	\mathcal{N}_{\rm tot}(\mu')\equiv 
	\int_{-\infty}^{\mu'} d\omega \int d\mathbf{r} \mathcal{D}_{\rm tot}(\mathbf{r},\omega).
	\label{eq:NtotDef}
\end{equation}
Assumption `A)' applies in equilibrium \cite{NEQQSM}. \textit{When} assumption `A)' applies it 
allows an important simplifications through a Lloyds' formula \cite{Lloyds} evaluation of the 
thermodynamics grand potential and causes the thermodynamics forces to reduce to electrostatic 
forces.  I stress that assumption `A)' is not expected to hold under nonequilibrium conditions
and that there is no argument that an equivalence of thermodynamics and electrostatic 
forces holds in general.

Lloyds's formula \cite{Lloyds} is a succinct formulation of the full 
density of states changes
\begin{equation}
	\Delta\mathcal{D}_{\rm tot}(\omega) = -\frac{\partial}{\partial \omega}
	\frac{2}{\pi}\Im \hbox{Tr} \{\ln [1-H_{1,V}g_r^d(\omega)]\}.
\label{DOSchangeFormal}
\end{equation}
arising, for example, with the adiabatic turn on of a tunneling term $H_1^{(0)}$.
Lloyds' formula rests on the simple observation that
\begin{equation}
	- \frac{\partial}{\partial \omega} \hat{g}_r(\omega)= 
	\hat{g}_r(\omega) \hat{g}_r(\omega).
	\label{lloydidea}
\end{equation}
and the possibility for using a cyclic permutation of operations when
evaluating a full trace.  Lloyds' 
formula (\ref{DOSchangeFormal}) leads directly to a compact evaluation 
of the integral of the full density of state changes $\Delta \mathcal{N}_{\rm tot}(\omega)$.

Lloyds' formula \cite{Lloyds} provides significant advantages in calculation of 
forces \textit{when} assumption `A)' is applicable.  Lloyds' formula (\ref{DOSchangeFormal}) can 
then be applied for an exact analysis of the \textit{additional} changes produced in 
a noninteracting system by an \textit{infinitesimal adjustment,} $H_{1,V} \to H_{1,V}+\delta V$.
Here I take $\delta V = \int d\mathbf{r} \delta v(\mathbf{r}) \hat{n}(\mathbf{r})$; 
generalizations to other cases are straightforward.  To lowest order in this 
infinitesimal additional perturbation $\delta V$ one obtains from (\ref{DOSchangeFormal}) 
\begin{equation}
	\delta \Delta\mathcal{N}_{\rm tot}(\omega) =  
	- \frac{1}{\pi}\hbox{Tr} \{ \delta V [- 2 \Im \hat{g}_r^{(0)}(\omega)]\}.
\label{deltaIntDOSchangeFormal}
\end{equation}

In an analysis of nonequilibrium conditions, I assume that $\mu_{\rm L}>\mu_{\rm R}$ and I
formulate assumption `A)' in terms of the contributions to the lead-specific less-than Green function
contributions (Sec. II.B and above).  In general, it holds for the noninteracting nonequilibrium system that
\begin{eqnarray}
	\langle \mathbf{r} | \hat{g}^{(0)}_< (\omega) | \mathbf{r} \rangle
	& = & \sum_{P={\rm L/R}} \langle \mathbf{r} | \hat{g}^{(0)P}_{<,\mu_P} (\omega)| \mathbf{r} \rangle
	\nonumber\\
	& = & \pi \sum_{P={\rm L/R}} \mathcal{D}_P (\mathbf{r},\omega) f_{\mu_P}(\omega),\\
	\mathcal{D}_{\rm tot}(\mathbf{r},\omega) & = & \sum_P \mathcal{D}_P (\mathbf{r},\omega).
\end{eqnarray}
where $f_{\mu_P}(\omega)$ denotes the Fermi-Dirac distribution function. 
Assumption `A)' can be expressed (for $P={\rm L/R}$)
\begin{eqnarray}
	\langle \mathbf{r} | \hat{g}^{(0)P}_{<,\mu_P} (\omega) | \mathbf{r} \rangle  =  \kappa_P \pi f_{\mu_P}(\omega) \mathcal{D}_{\rm tot} (\mathbf{r},\omega)
	\label{eq:asumptionINmath}
\end{eqnarray}
where $\kappa_{P={\rm L/R}}$ are would-be proportionality factors. Since the electron distribution for 
$\omega < \mu_{\rm R}$ is no different than for a system in equilibrium, we must have 
$\kappa_{\rm L}+\kappa_{\rm R}=1$ but this boundary conditions is not used in the following analysis.

\textit{If} assumption `A)' holds, one can use of Lloyds' formula \cite{Lloyds} 
in the following evaluation of the lead-specific contributions (for $P={\rm L/R}$)
to the noninteracting thermodynamics grand potential
\begin{eqnarray}
	\Omega_{P} & \to  & - 2 \int_{-\infty}^{\mu_{P}} d\mu' \int d\mathbf{r} 
	\int \frac{d\omega}{2\pi} \langle \mathbf{r} | \hat{g}^{(0)P}_{<,\mu'} (\omega) | \mathbf{r} \rangle
	\nonumber\\
	& = & -2 \kappa_P \int d\mu' \int d\mathbf{r} \int \frac{d\omega}{2\pi} \pi \mathcal{D}_{\rm tot}(\mathbf{r},\omega) f_{\mu'}(\omega)
	\nonumber\\
	& = & - \kappa_P \int d\mu' \mathcal{N}_{\rm tot}(\mu').
	\label{eq:OmLWansatz}
\end{eqnarray}
Again, subject to `A)' and thus the applicability of (\ref{eq:asumptionINmath}),
it follows that infinitesimal changes in the potential (relevant for forces)
corresponds to changes in the lead-specific thermodynamics grand potential terms ($P={\rm L/R}$):
\begin{eqnarray}
	\delta \Omega_{P} & \to  & \kappa_P \int_{-\infty}^{\mu_P} d\mu' \frac{1}{\pi} 
	[ -2\Im g_r^{(0)} (\mathbf{r},\mathbf{r},\omega=\mu') ] \delta v(\mathbf{r})
	\\
	& = & 2\int d\mathbf{r} \delta v(\mathbf{r}) \int \frac{d\omega}{2\pi} 
	\langle \mathbf{r} | \hat{g}_{<,\mu_P}^{(0)P}(\omega)| \mathbf{r} \rangle.
\end{eqnarray}
\textit{Assumption `A)' implies} for noninteracting-particle system that the exact 
and explicitly conservative thermodynamics forces reduce to the widely used electrostatic forces:
\begin{eqnarray}
	\frac{\partial \Omega_{\rm LS}}{\partial\mathbf{R}_i} \cdot \delta \mathbf{R}_i & \to & 2 \int d\mathbf{r} \delta v(\mathbf{r}) \int \frac{d\omega}{2\pi} 
	\langle \mathbf{r}  | \hat{g}_< ^{(0)} (\omega) | \mathbf{r} \rangle \nonumber \\
	& = & \langle \frac{\partial H^{(0)}}{\partial\mathbf{R}_i} \rangle_{\rm GCE} \cdot \delta \mathbf{R}_i.
	\label{eq:ThermVselectForce}
\end{eqnarray}

In summary, thermodynamics and electrostatic forces do coincide in equilibrium. 
There also exists arguments, summarized in Ref.~\onlinecite{Sorbello}, that
electrostatic forces captures the adiabatic part of the force description in
linear response.  Under nonequilibrium conditions, however, assumption `A)' is not expected 
to generally hold and there is then no argument that the thermodynamics and 
electrostatic forces should agree.  This conclusion is corroborated by the 
phase-shift analysis in Appendix B.

\section{Phase-shifts analysis of thermodynamics}

To provide a phase-shift analysis of the thermodynamics in nonequilibrium tunneling
it is advantageous to consider the state-specific contributions to the
general formulation of the Friedel sum rule (\ref{eq:FormalFriedel}). I emphasize
that the SP behavior expressed in the T-matrix reflects
\textit{mutually independent} dynamics and that the on-shell S matrix is 
directly linked to the on-shell T matrix
\begin{eqnarray}
	\mathcal{S} & = & 1 - 2\pi i T
	\\
	T & = & \delta(\omega - H_V^0) \mathcal{T}
\end{eqnarray}
In fact, the T- and S-matrix have identical eigenstates, here denoted $| j \rangle$. 

The S-matrix (for particles at energy $\omega$) have eigenvalues $\exp[2i \delta_j(\omega)]$ which, apart from the assumed spin-degeneracy, 
reflect an isomorphic mapping to the set of initial states in the leads, namely the $H_d$ SP eigenstates 
$|\lambda_{P={\rm L/R}}\rangle$.  One can, as illustrated in Fig.~1, define a unitary transformation 
with matrix elements 
\begin{equation}
	\boldsymbol{\alpha}_{j,\lambda_P} \equiv \langle j | \lambda_P \rangle,
\end{equation}
which expresses a representation change 
\begin{equation}
	\mathcal{S} \longrightarrow \boldsymbol{d}_\mathcal{S} = \boldsymbol{\alpha} 
	\cdot \mathcal{S} \cdot \boldsymbol{\alpha}^\dag.
\label{eq:Smatrixdiagolize}
\end{equation}
This transformation diagonalizes the noninteracting S-matrix,
\begin{equation}
	(\boldsymbol{d}_\mathcal{S})_{j,j'} = \delta_{j,j'}\; \exp[2i \delta_j(\omega)].
\label{eq:diagonalizedSmatrix}
\end{equation}

I observe in passing that the T matrix is already by itself a powerful tool for approximating
the dynamics even for the case of a nonpertubative interaction \cite{ZarambaKohnLS,HarrisFriedelInteract}.
One often obtains an excellent approximation by simply setting 
\begin{equation}
	\ln[\mathcal{S}(\omega)] \approx -2\pi i T(\omega),
\end{equation}
an observation which suggests that we can can use an S-matrix eigenstate analysis to
assign an effective phase shift to each an every initial state $|\lambda_{P={\rm L/R}}\rangle$.

The argument for a state-specific evaluation can be made formal by considering a repeated application of the
diagonalization (\ref{eq:Smatrixdiagolize}) in every term of an expansion of the logarithm in (\ref{eq:FormalFriedel}):
\begin{eqnarray}
	\ln[\mathcal{S}(\omega)] & = & \boldsymbol{\alpha} \ln[\boldsymbol{\delta}_S(\omega)] \boldsymbol{\alpha}^{\dag}
	\nonumber\\
	& = & \boldsymbol{\alpha} [2i\delta_j(\omega)] \boldsymbol{\alpha}^{\dag}.
\end{eqnarray}
Here $\boldsymbol{\alpha}^{\dag}|\lambda_P\rangle$ represents the state $|\lambda_P\rangle$ in the
S-matrix eigenstate representation. An S-matrix eigenstate analysis will therefore directly
give an evaluation of the state-specific density of state contribution
\begin{eqnarray}
	2i\pi [\delta(\omega-\tilde{\epsilon}_{\lambda_P}) 
	- \delta(\omega-{\epsilon}_{\lambda_P})] 
	& \leftrightarrow & \frac{\partial}{\partial \omega}
	\langle \lambda_P | \ln[\mathcal{S}(\omega)] \lambda_P \rangle \\
	& = & 2i \frac{\partial}{\partial\omega} \delta_{\lambda_P}(\omega).
	\label{eq:contribFormal}
\end{eqnarray}
The latter expression is given in terms of an effective (state-projected) phase shift 
\begin{equation}
	\delta_{\lambda_P}(\omega) \equiv \sum_j \delta_j(\omega) | \langle j |\lambda_P \rangle |^2.
	\label{eq:contribEval}
\end{equation}

It is interesting to link the above formal S-matrix description with the very general 
analysis of one-dimensional tunneling that is reported in Ref.~\onlinecite{Albert1D}.  That textbook analysis 
considers transmission in cases where the density of state and the group velocity are different
in the two leads. That SP analysis is relevant for conditions which generally result with the application 
of a bias, for example, in modeling a heterostructure-based
resonant-tunneling diode \cite{RahmanDavies,Phon,QCLdisc}. For a propagating state `u' ( `v') originating from the 
far left (right) the transmission coefficients and reflection coefficients are expressed in polar notation
\begin{eqnarray}
	t_{u/v} & = & |t_{u/v}|\exp[i\Theta^{u/v}_t],\\
	r_{u/v} & = & |r_{u/v}|\exp[i\Theta^{u/v}_r].
\end{eqnarray}
The analysis of the finite-bias SP dynamics is based on evaluations of the
Wronskian, i.e., with strict adherence to flux conservation, and identifies 
the general relations among the scattering phase shifts \cite{Albert1D}
\begin{eqnarray}
	\Theta^u_t & = & \Theta^v_t,\\
	\Theta^u_r & = & \pi - \Theta^v_r + 2\Theta_t^u.
\end{eqnarray}

Before proceeding, I observe that application of this textbook analysis formally requires a 
reinterpretation for use in the Caroli formulation.  This is because the textbook analysis assumes the 
initial states to be propagating, i.e., not strictly limited to either of the leads as applies for $| \lambda_P\rangle$.  
However, the adaptation is straightforward and leads in a one-dimensional system to a simple explicit 
identification of an effective phase shift for the left-lead and right-lead states (at energy $\omega)$:
\begin{eqnarray}
	\delta_L^{\rm eff} (\omega) & = & \Theta_r^u(\omega),\\
	\delta_R^{\rm eff} (\omega) & = & \Theta_r^v(\omega).
\end{eqnarray}
One can also define a total phase shift derivative
\begin{equation}
	\frac{\partial}{\partial \omega} \delta_{\rm tot}^{\rm eff} (\omega) \equiv \frac{\partial}{\partial \omega} \Theta_r^{u}(\omega) + \frac{\partial}{\partial \omega} \Theta_r^{v}(\omega) = 2 \frac{\partial}{\partial \omega} \Theta_t^{u}(\omega)
	\label{eq:totphsaeshiftrelation}
\end{equation}
that determines the full value of the integrated change in the density of state (\ref{eq:NtotLS:phaseEvaluation}).

The above-described textbook analysis \cite{Albert1D} of one-dimensional tunneling highlights the
cause of differences between thermodynamics and electrostatic forces which arise under nonequilibrium conditions.
Appendix A demonstrates that electrostatic forces result when (\ref{eq:NtotLS:phaseEvaluation}) may 
be used to characterize the GCE thermodynamics change of an infinitesimal coordinate change. Section II.B 
explains that a full nonequilibrium characterization of the thermodynamics grand potential change in the limit of $\mu_{\rm L} \gg 
\mu_{\rm R}$ is instead completely specified by $\delta _{\rm L}^{\rm eff}$. Different phase shifts variations characterize 
the two types of forces under nonequilibrium conditions.  For example, when $\mu_{\rm L} \gg \mu_{\rm R}$ it holds that thermodynamics
and electrostatic forces have different natures given by
\begin{eqnarray}
	F_{\mathbf{R}_i}^{\rm GCE} & \leftrightarrow  & \frac{\partial}{\partial \mathbf{R}_i} \delta_{\rm L}^{\rm eff}(\omega),\\  
	F_{\mathbf{R}_i}^{\rm es,GCE} & \leftrightarrow  & \frac{\partial}{\partial \mathbf{R}_i} \delta_{\rm tot}^{\rm eff}(\omega),  
\end{eqnarray}
respectively. I stress that in the one-dimensional model problem \cite{Albert1D} there is no general relation 
between the transmission phase shift $\delta_{\rm tot}^{\rm eff}(\omega)=\Theta_t^{u}(\omega)$ and the reflection phase shift 
$\delta_{\rm L}^{\rm eff}(\omega)=\Theta_r^u(\omega)$ and that, consequently, there does not exist a general argument
that the nonequilibrium electrostatic force agrees with the exact and explicitly conservative thermodynamics 
forces\footnote{The difference, $\Theta_r^u\neq \Theta_t^u$, is also important in a detailed analysis of tunneling
times, cf. reference \protect\cite{Hauge}.}.

\section{Mermin-type variational thermodynamics for nonequilibrium tunneling}

This appendix summarizes the proof of the variational
nature of the nonequilibrium thermodynamic grand potential and of the 
global-minimum property of the exact (solution) density 
matrix.  The proof applies for time-independent or time-dependent interacting 
tunneling systems under nonequilibrium conditions defined as a collision problem, Sec.~IV.
The argument is a straightforward generalization of Mermin's previous
equilibrium analysis \cite{Mermin65}.

The first observation is that it is sufficient to prove the 
extremal property for any given time and hence I shall
suppress all mentioning of subscripts `col', `sc', or `LS' 
(as well as possible temporal argument) in the following.
A more compressed formulation of the extremal properties of 
the nonequilibrium thermodynamic grand potential, the nonequilibrium variational property can thus
be formulated
\begin{equation}
\Omega[\hat{\rho}]> \Omega[\hat{\rho}_0]\equiv \Omega_0,
\label{eq:reOmegaExtremal}
\end{equation}
where (for a Hermitian $\hat{X}_0$)
\begin{eqnarray}
\hat{\rho}_0 &\equiv & \frac{e^{-\beta \hat{X}_0}}{\hbox{Tr}\{e^{-\beta \hat{X}_0}\}},
\label{eq:reRho0Def}\\
\Omega_0 & \equiv  & -\frac{1}{\beta}\ln(\hbox{Tr}\{e^{-\beta \hat{X}_0}\}),
\label{eq:reNEQGrandPotVal}\\
\Omega[\hat{\rho}] & \equiv & 
%\frac{
{\rm Tr}\{ \hat{\rho} [X_0 + \beta^{-1} \ln \hat{\rho}]\},
%}
%{{\rm Tr}\{\hat{\rho}\}}.
\label{eq:reNEQGrandPotFunct}
\end{eqnarray}
and where trial nonequilibrium density matrices 
\begin{equation}
\hat{\rho} =\frac{e^{-\beta\hat{X}}}{\hbox{Tr}\{e^{-\beta\hat{X}}\}}
\label{eq:classredef}
\end{equation}
are given by any Hermitian operator $\hat{X}$.

As in the equilibrium theory, it is advantageous to also define a 
set of intermediate density matrices
\begin{eqnarray}
\hat{\rho}_\lambda & \equiv & \frac{e^{-\beta\hat{X}_\lambda}}
{\hbox{Tr} \{e^{-\beta\hat{X}_\lambda}\}},
\label{eq:lambdaRhoDef}\\
\hat{X}_\lambda & \equiv &\hat{X}_0+\lambda \Delta,
\label{eq:lambdaXDef}
\end{eqnarray}
where $\Delta \equiv -(1/\beta) \ln(\hat{\rho})-\hat{X}_0$. This family
of possible nonequilibrium density matrices (\ref{eq:lambdaRhoDef}) form a line 
between $\hat{\rho}_0$ and $\hat{\rho}$ where each member formally represents
(in the notation introduced in Sec.~VI) a solution of the thermodynamic grand potential
\begin{eqnarray}
\Omega_{\hat{X}_\lambda} & \equiv &
\frac{\hbox{Tr} \{e^{-\beta \hat{X}_\lambda} (\hat{X}_\lambda + \beta^{-1} \ln \hat{\rho}_\lambda)\}}
{\hbox{Tr}\{e^{-\beta \hat{X}_\lambda}\}}\nonumber\\
& = & -\frac{1}{\beta}\ln (\hbox{Tr}\{e^{-\beta \hat{X}_\lambda}\})
\neq \Omega[\hat{\rho}_\lambda].
\end{eqnarray}
I note that, unlike in the equilibrium case, a separate argument (for example, V-representability)
would be needed to conclude that any given member, $\hat{\rho}_\lambda$ (of the line of 
density-matrices) corresponds to an actual density matrix solution for a physical nonequilibrium 
tunneling system.  However, such a representability assumption is not of any consequence 
for the proof and discussion of the extremal nature of the nonequilibrium thermodynamic
theory. The essential observation is that all of the intermediate nonequilibrium 
density matrices (\ref{eq:lambdaRhoDef}) are still specified in the
form given in Eq.~(\ref{eq:classredef}).

The central step in the proof and argument is a demonstration of 
the variational property, 
\begin{equation}
 \frac{\partial}{\partial \lambda} \Omega_{\hat{X}_\lambda}
= \hbox{Tr} \{\Delta \hat{\rho}_\lambda\},
\label{eq:shortGenHFtheorem}
\end{equation}
that is, a compact formulation of the generalized Hellmann-Feynman theorem 
(\ref{eq:GenHFtheorem}). Here the key observation is the general
operator identity \cite{Mermin65}
\begin{eqnarray}
\frac{\partial}{\partial \lambda} 
e^{-\beta (\hat{X} + \lambda \Delta)} 
&  = & - e^{-\beta (\hat{X}+\lambda \Delta)} \int_0^\beta d\beta' 
\Delta_\lambda (\beta')
\label{eq:HFlemma}\\
\Delta_\lambda (\beta') & \equiv & 
e^{\beta'(\hat{X}+\lambda \Delta)}\, \Delta \,
e^{-\beta (\hat{X}+\lambda \Delta)},
\end{eqnarray}
which, of course, applies irrespectively of the nature of the statistical 
ensemble, for nonequilibrium and equilibrium conditions alike.
Introducing also
$\langle \hat{O} \rangle_\lambda \equiv \hbox {Tr} 
\{ \hat{\rho}_\lambda \hat{O} \}$, one can thus formulate 
the parametric density matrix changes, 
\begin{equation}
\frac{\partial}{\partial \lambda} \hat{\rho}_\lambda = - \int_0^\beta d\beta'
\hat{\rho}_\lambda [\Delta_\lambda(\beta')  - \langle \Delta \rangle_\lambda].
\label{eq:betaprimerecast}
\end{equation}

Mermin also makes the observation that a cyclic permutation of operators
permits a significant simplification in the evaluation of expectation 
values, for example \cite{Mermin65},
\begin{equation}
\langle \Delta_\lambda(\beta') \rangle_{\lambda} =
\langle \Delta \rangle_{\lambda}.
\label{eq:betaprimeeval}
\end{equation}
Inserting this result in (\ref{eq:betaprimerecast}) completes an explicit demonstration of the
generalized Hellmann-Feynman result (\ref{eq:shortGenHFtheorem}).

Finally, from (\ref{eq:HFlemma}) also follows a proof for the extremal nature of
the thermodynamical grand potential. It proceeds by the reformulation of the 
potential derivative
\begin{eqnarray}
\frac{\partial}{\partial \lambda} \Omega[\hat{\rho}_\lambda] 
& = &  -\lambda \hbox{Tr} \{\Delta \frac{\partial}{\partial \lambda}
\hat{\rho}_\lambda\}
\label{MerminA6}\\
& = & 
\lambda
\int_0^{\beta} d\beta' | \langle \Delta_\lambda (\beta/2) - \langle 
\Delta\rangle_\lambda \rangle |^2.
\label{MerminA10}
\end{eqnarray}
As in equilibrium thermodynamics, this results is sufficient to ensure that 
the difference $\Omega[\hat{\rho}]-\Omega[\hat{\rho}_0]$ is positive and
cannot vanish except when $\hat{\rho}\equiv \hat{\rho}_0$. 

In summary, the operator analysis of Mermin \cite{Mermin65} is sufficient to ensure 
(not only for the equilibrium case but also for the present nonequilibrium thermodynamics) that the
thermodynamic grand potential is both variational and extremal at the exact 
solution $\hat{\rho}_0$.

%####################### APPENDIX D ###################################### 

\section{Linking the variational QKA of tunneling to LS collision DFT}

The details of the continuum Caroli partition scheme, section III, are 
designed to allow a combination of the LS collision DFT \cite{LSCDFT} with the present general nonequilibrium
thermodynamic account. To accomplish that, it is necessary to also
ensure that the choice of details in the adiabatic turn on (for the electron QKA)
is made such that it permits the LS collision DFT proof of uniqueness of density.
The underlying issue is purely technical, but for completeness, I include 
here a mathematical analysis.

While Ref.~\onlinecite{LSCDFT} never explicitly stated this,
the formal LS collision DFT proof of uniqueness of density does, in principle,
require that the adiabatic-turn-on factor $a_\eta(t)$ vanishes 
not only at $t\to -\infty$, but also for all times $t < t_0$ where $t_0 \to -\infty$.  
In a strict mathematical sense, this is necessary because the LS collision DFT proof implicitly 
assumes one can work with a Taylor expansion in the temporal evolution. With the
simple adiabatic turn on [$\tilde{a}_{\eta}(t)=\exp(\eta t)$] that we normally employ in 
the QKA, it follows that all Taylor-expansion coefficients of the temporal evolution vanish in
the limit $t\to -\infty$.  It is clear, that alone an extremely soft modification of the
adiabatic turn on is required to ensure a rigorous mathematical description in the LS collision DFT 
analysis. It is also clear (because of the entropy flow) that such an extremely soft modification of 
the adiabatic turn on can have no consequence for the tunneling dynamics at relevant times when a 
steady-state does emerge.  The argument can be made formal.

To be specific I investigate an adiabatic turn on which [instead of the standard LS or QKA form, 
$\tilde{a}_\eta(t)=\exp(\eta t)$] is identically zero until some far away time $t_0\to -\infty$. I assume that
$a_\eta(t>t_0)$ is real, monotonously increasing, and satisfies $0 \leq a_\eta(t) 
\leq 1$, with a bounded derivative $(d a_\eta/dt) \leq \eta a_\eta(t)$.
An explicit choice of slightly modified adiabatic turn on factors can be expressed
\begin{equation}
a_\eta(t) = \exp(\eta t) b_\eta(t; t^\eta_0, t^\eta_1)
\label{eq:modifiedAdiabatic}
\end{equation}
where I have introduced a monotonously increasing positive 
function $b_\eta(t; t^\eta_0,t_1^\eta)$ with
$b_\eta(t<t_0^\eta) \equiv 0$, $b_\eta(t>t_1^\eta) \equiv 1$ and 
$db_\eta(t)/dt \ll \eta$. I choose 
$t^{\eta}_1=-\eta^{-(\alpha+1)}$ (with $\alpha\to \infty$) and $t^\eta_0=t^\eta_1/\eta^2$.  
I note that this set of choices removes any possible $t=-\infty$ singularities in the LCS-DFT 
proof of uniqueness of density and I proceed to demonstrate that this modified adiabatic 
turn on still leaves all details of the formal LS solutions in the QKA 
unchanged (in the relevant limit $\eta \to 0^+$).

The important fact is that $da_\eta(t)/dt$ remains essentially specified by 
$\eta \exp(\eta t)$ since this factor, in turn, determines the denominator in the
formal many-body LS solution (\ref{eq:LSmbEquation}).  Focusing on the analysis
in Ref.~\onlinecite{Gellmann} I introduce
\begin{eqnarray}
\hat{f}_\xi(t) & \equiv & e^{-i(E_\xi-H+i\eta)t},\\
\hat{F}_\xi(t) & \equiv & \int_{-\infty}^t dt' \hat{f}_\xi(t')
\nonumber\\
& = & \frac{i}{E_\xi-H+i\eta}e^{-i(E_\xi-H+i\eta)t},
\end{eqnarray}
and note that the boundary conditions, i.e., the original choice of
adiabatic turn on $\exp(\eta t)$, enters the many-body LS 
solution through the formal evaluation \cite{LippSchwing,Gellmann} 
\begin{equation}
|\Psi^{(+)}_{\xi}\rangle = \eta \hat{F}_\xi(0)|\Phi_\xi\rangle.
\label{eq:LSeval}
\end{equation}
Using instead the present collision picture with slightly 
modified adiabatic turn on (\ref{eq:modifiedAdiabatic}), the boundary
conditions instead enters the formal many-body solution through 
\begin{eqnarray}
\int_{-\infty}^0 dt' \hat{f}_\xi(t') b_{\eta}(t') 
& = & \hat{F}_\xi(t=0) - \hat{B}_\xi(\eta)\\
\hat{B}_\xi(\eta) & = &
\int_{t_0^\eta}^{t_1^\eta} dt' \hat{F}_\xi(t') \frac{db_\eta}{dt}(t'),
\end{eqnarray}
as obtained by partial integration.  The resulting LS solution
therefore has the form
\begin{eqnarray}
|\bar{\Psi}^{(+)}_\xi \rangle & = & |\Psi^{(+)}_{\xi}\rangle 
-| \delta \Psi^{(+)}_{\xi}\rangle,\\
| \delta \Psi^{(+)}_{\xi}\rangle 
& = & \eta
\hat{B}_\xi(\eta) |\Phi_\xi\rangle.
\label{eq:modifiedLSequation}
\end{eqnarray}
However, the relative normalization of the additional term 
in the modified many-body LS solution (\ref{eq:modifiedLSequation}) 
is strongly bounded,
\begin{eqnarray}
\left\| | \delta \bar{\Psi}^{(+)}_{\xi}\rangle 
\right\| & \ll & \eta 
\int_{t_0^\eta}^{t_1^\eta} dt' \, 
\left\| \eta e^{\eta t'} \hat{F}_\xi(0) |\Phi_\xi \rangle
 \right\|,
\label{eq:LSmodNormIntgA}\\
& < & \mathcal{B}(\eta)
\left\| |\Psi^{(+)}_{\xi}\rangle 
\right\|, 
\label{eq:LSmodNormIntg}
\end{eqnarray}
by the (state-independent) evaluation
\begin{equation}
\mathcal{B}(\eta) \equiv \exp(\eta t_1^\eta) = \exp[-(1/\eta)^{\alpha}]  \to 0
\label{eq:modNormlimit}
\end{equation}
The choice $t_1^{\eta}=\eta^{-(\alpha+1)}$ with $\alpha\to\infty$ certainly 
ensures that $\mathcal{B}(\eta)$ and hence the modification (\ref{eq:modifiedLSequation}) 
of the LS solution become irrelevant in the limit $\eta \to 0^+$.

In summary, I find that the LS collision DFT analysis \cite{LSCDFT} can be 
adapted to provide an independent proof for uniqueness of density 
while leaving the formal LS solutions of the many-body collision problems
unchanged. This opens for a density functional formulation of the 
nonequilibrium thermodynamic quantities in a description which resembles the 
equilibrium case presented in Ref.~\onlinecite{Mermin65}. 

\section{Adiabatic changes in thermodynamics state functions} 

This appendix supplements the description of changes in the
thermodynamics variation subject to the assumption of adiabatic 
transformations. That is, I derive formal expression for the changes
in thermodynamics state function with coordinate transformations, in essence,
assuming that the evolving nonequilibrium density matrix is at all times
(and all evolving coordinates) given by the steady-state value.
This analysis serves, for example, to detail the GCE breakdown 
in the traditional Hellmann-Feynmann force evaluation. 

\textit{The key observation\/} is that the operator analysis of Mermin 
can be used to obtain the parametric derivatives of the unnormalized nonequilibrium solution density matrix,
$\exp[-\beta(H-\hat{Y}_{\rm LS})]$. For example, use of (\ref{eq:HFlemma}) and (\ref{eq:betaprimeeval}) 
directly leads to an evaluation of the entropy content,
\begin{equation}
\frac{\partial
\bar{S}_{\rm LS}
}{\partial \mathbf{R}_i} 
 = -\beta^2 
\langle
[(H-\hat{Y}_{\rm LS})-(\bar{U}_{\rm LS}-\bar{Y}_{\rm LS})]
\frac{\partial
(H-\hat{Y}_{\rm LS})
}{\partial \mathbf{R}_i}
\rangle_{\hat{\rho}_{\rm LS}},
\end{equation}
a form which is entirely specified by differences in internal energy and
Gibbs free energy operators (and differences in their values).

There is a corresponding formal simplification in evaluation of the expectation value of the 
Gibbs weighting factor, 
\begin{equation}
\frac{\partial}{\partial\mathbf{R}_i} (\bar{U}_{\rm LS} - \bar{Y}_{\rm LS}) 
\equiv 
\frac{\partial}{\partial \mathbf{R}_i} \left(\bar{\Omega}_{\rm LS}+\frac{1}{\beta}\bar{S}_{\rm LS}\right).
\label{eq:weightingDeterminFormal}
\end{equation}
Using a cyclic permutation of operators in the trace yields 
\begin{eqnarray}
\frac{1}{\beta^2}\frac{\partial
\bar{S}_{\rm LS}
}{\partial \mathbf{R}_i} 
 & = & 
(\bar{U}_{\rm LS}-\bar{Y}_{\rm LS})
\frac{\partial}
{\partial \mathbf{R}_i} 
\bar{\Omega}_{\rm LS} \nonumber\\
& & -\frac{1}{2} 
\langle
\frac{\partial}{\partial \mathbf{R}_i}(H-\hat{Y}_{\rm LS})^2
\rangle_{\hat{\rho}_{\rm LS}},
\label{eq:SderivRewrite}
\end{eqnarray}
and it follows that
\begin{eqnarray}
\frac{\partial}{\partial\mathbf{R}_i} (\bar{U}_{\rm LS} - \bar{Y}_{\rm LS}) 
& = & 
\left[1+\beta(\bar{U}_{\rm LS}-\bar{Y}_{\rm LS})\right]
\frac{\partial}{\partial \mathbf{R}_i} \bar{\Omega}_{\rm LS}
\nonumber\\
& & -\frac{\beta}{2}  
\langle \frac{\partial (H-\hat{Y}_{\rm LS})^2}{\partial \mathbf{R}_i} \rangle_{\hat{\rho}_{\rm LS}}.
\label{eq:weightingExplicit}
\end{eqnarray}

It is also possible to complete a formal evaluation of the derivatives of 
total-internal energy and Gibbs free energy (also when evaluated in isolation) thanks to the
nature \cite{NEQreform} of the Gibbs free energy or electron-redistribution operator.
In steady state tunneling the Gibbs free energy must, of course, also be time independent.
The time evolution is naturally formulated in terms of the Liouville operation
\begin{equation}
\mathcal{L}_H[\hat{Y}_{\rm LS}] \equiv [\hat{Y}_{\rm LS},H] \propto \eta (\hat{Y}_{\rm LS}-\hat{Y}_d)\to 0.
\end{equation}
The Gibbs free energy operator commutes with the interacting steady-state Hamiltonian
$H\equiv H(t=0)$ in steady-state tunneling problems. As pointed out by Hershfield \cite{NEQreform},
the role of $\hat{Y}_{\rm LS}$ is to redistribute the electrons among the many-body eigenstates of 
$H$. Using the commutator result, $[H,\hat{Y}_{\rm LS}]=0$, together 
with Mermin's observation, (\ref{eq:betaprimeeval}), leads to the formal evaluations
\begin{eqnarray}
\frac{\partial
\bar{U}_{\rm LS}
}{\partial \mathbf{R}_i} 
& = & 
\langle \frac{\partial H}
{\partial \mathbf{R}_i} \rangle_{\rho_{\rm LS}} +
\beta 
\langle (H-\bar{U}_{\rm LS}) 
\frac{\partial
(H-\hat{Y}_{\rm LS})
}{\partial \mathbf{R}_i} 
\rangle_{\hat{\rho}_{\rm LS}},
\\
\frac{\partial
\bar{Y}_{\rm LS}
}{\partial \mathbf{R}_i} 
& = & 
\langle \frac{\partial \hat{Y}_{\rm LS}}
{\partial \mathbf{R}_i} \rangle_{\rho_{\rm LS}} +
\beta 
\langle (\hat{Y}_{\rm LS}-\bar{Y}_{\rm LS}) 
\frac{\partial
(H-\hat{Y}_{\rm LS})
}{\partial \mathbf{R}_i} 
\rangle_{\hat{\rho}_{\rm LS}},
\end{eqnarray}
of the total internal energy and Gibbs free energy derivatives, respectively.

Finally I note that the reformulation of the derivative of the expectation value of
the Gibbs weighting factor (\ref{eq:weightingExplicit})
can also be used to obtain an alternative expression of the here-defined thermodynamic forces,
\begin{equation}
\mathbf{F}^{\rm GCE}_{{\rm LS}, \mathbf{R}_i} 
= - 
\frac
{\frac{\partial}{\partial \mathbf{R}_i} \langle H-\hat{Y}_{\rm LS}\rangle_{\hat{\rho}_{\rm LS}}
+(\beta/2) \langle \frac{\partial}{\partial \mathbf{R}_i}(H-\hat{Y}_{\rm LS} )^2\rangle_{\hat{\rho}_{\rm LS}}}
{1 + \beta (\bar{U}_{\rm LS}-\bar{Y}_{\rm LS}) }.
\label{eq:AlternativeForceExpression}
\end{equation}

\end{document}